\documentclass[a4paper,11pt]{article}
\pdfoutput=1 % if your are submitting a pdflatex (i.e. if you have
\usepackage{jheppub} % for details on the use of the package, please
\usepackage[T1]{fontenc} % if needed
\usepackage{slashed}
\usepackage{subfigure}
\usepackage[usenames,dvipsnames]{xcolor}
\usepackage{enumitem}

\usepackage{marginnote}
\usepackage[normalem]{ulem}
\usepackage{orcidlink}

\DeclareMathOperator{\Li}{Li}
\newcommand{\cJ}{\mathcal{J}}
\newcommand{\Lqcd}{\Lambda_\text{QCD}}
\newcommand{\eq}[1]{Eq.~\eqref{eq:#1}}

\renewcommand{\sec}[1]{Sec.~\ref{sec:#1}}

\newcommand{\fig}[1]{Fig.~\ref{fig:#1}}

\newcommand{\tab}[1]{Table~\ref{tab:#1}}
\newcommand{\nn}{\nonumber}

\title{\boldmath  Precision DIS thrust predictions for HERA and EIC}

\author[a]{June-Haak Ee\orcidlink{0000-0002-4699-2100},}
\author[*,b,c]{Daekyoung Kang\orcidlink{0000-0002-9145-6913},\note[*]{Corresponding author.}}
\author[a]{Christopher Lee\orcidlink{0000-0003-2385-7536},}
\author[d,e]{Iain W. Stewart\orcidlink{0000-0003-0248-0979}}

\affiliation[a]{Theoretical Division, Los Alamos National Laboratory, PO Box 1663, MS B283, \\ Los Alamos, NM 87545, USA}
\affiliation[b]{Key Laboratory of Nuclear Physics and Ion-beam Application (MOE) and Institute of Modern Physics,
Fudan University,
Shanghai 200433, China}
\affiliation[c]{Department of Physics, Korea University, Seoul 02841, Korea}
\affiliation[d]{Center for Theoretical Physics,  Massachusetts Institute of 
Technology, \\Cambridge, MA  02139, USA}
\affiliation[e]{University of Vienna, Faculty of Physics, Boltzmanngasse 5, 
        A-1090 Wien, Austria}

\emailAdd{jhee@lanl.gov}
\emailAdd{dkang@fudan.edu.cn}
\emailAdd{clee@lanl.gov}
\emailAdd{iains@mit.edu}

\abstract{We present predictions for the DIS 1-jettiness event shape $\tau_1^b$, or DIS thrust, using the framework of Soft Collinear Effective Theory (SCET) for factorization, resummation of large logarithms, and rigorous treatment of nonperturbative power corrections, matched to fixed-order QCD away from the resummation region. Our predictions reach next-to-next-to-next-to-leading-logarithmic (N$^3$LL) accuracy in resummed perturbation theory, matched to $\mathcal{O}(\alpha_s^2)$ fixed-order QCD calculations obtained using the program \texttt{NLOJet++}. We include a rigorous treatment of hadronization corrections, which are universal across different event shapes and kinematic variables $x$ and $Q$ at leading power, and supplement them with a systematic scheme to remove $\mathcal{O}(\Lambda_\textrm{QCD})$ renormalon ambiguities in their definition. The framework of SCET allows us to connect smoothly the nonperturbative, resummation, and fixed-order regions, whose relative importance varies with $x$ and $Q$, and to rigorously estimate theoretical uncertainties, across a broad range of $x$ and $Q$  covering existing experimental results from HERA as well as expected new measurements from the upcoming Electron-Ion-Collider (EIC). Our predictions will serve as an important benchmark for the EIC program, enabling the precise determination of the QCD strong coupling $\alpha_s$ and the universal nonperturbative first moment parameter $\Omega_1$.}

\begin{document} 
\preprint{
\vspace{-0.75cm}

\setcounter{tocdepth}{2}

\begin{flushright} 
LA-UR-25-23190\\
MIT-CTP 5856
\end{flushright} \vspace*{-0.75cm}
}

\maketitle

\flushbottom

\section{Introduction}
\label{sec:intro}

Since the dawn of quantum chromodynamics (QCD), deep inelastic scattering (DIS) has served as a powerful and precise microscope, revealing the internal structure of the proton and the dynamics of its constituents.
From the existence of quarks themselves, to the distributions of all parton flavors, and to new states of matter, DIS probes a wide range of physics inside the tiny proton. In fact, this process forms the basis for many upcoming studies at the Electron Ion Collider (EIC), which will push the frontiers of precision in each of these and other aspects of proton and hadron structure and QCD itself \cite{AbdulKhalek:2021gbh}. 

In this paper, we explore how the measurements of the final-state hadron distributions in DIS can be used to determine the strong coupling $\alpha_s$ to high precision, together with the size of the leading nonperturbative effects. Indeed, many such determinations have been performed to date \cite{ParticleDataGroup:2024cfk}; global fits of DIS structure functions to next-to-next-to-leading-order (NNLO) QCD calculations have yielded values of $\alpha_s(m_Z)$ at the one- to a few-percent level of precision, while jet measurements at HERA have yielded determinations at a few to ten-percent level of precision. Jet measurements are particularly sensitive to hadronization and serve as crucial tests for hadronization parameters and models. 

Observables called hadronic event shapes \cite{Dasgupta:2003iq} are in principle an outstanding candidate for the high-precision determination of $\alpha_s$ due to the potential or actual availability of high-accuracy predictions in perturbation theory, with powerful handles on the universality and scaling of the dominant hadronization effects. Their definitions are typically simple, focusing on global measures of QCD radiation that shape the distribution of final-state hadrons and provide sensitivity to $\alpha_s(\mu)$ over a wide spectrum of energy scales $\mu$. For $e^+e^-$ event shapes this program has reached impressive levels with predictions of event shapes like thrust \cite{Abbate:2010xh}, $C$-parameter \cite{Hoang:2014wka,Hoang:2015gta} and heavy jet mass  \cite{Chien:2010kc,Benitez-Rathgeb:2024ylc,Benitez:2025vsp} to  N$^3$LL resummed order matched to NNLO [$\mathcal{O}(\alpha_s^3)$] fixed-order predictions. Some of these have led to determinations of $\alpha_s(M_Z)$ included in the PDG~\cite{Workman:2022ynf}, e.g. from a global fit of data on energy-energy correlators (EEC) to NNLL+NNLO predictions \cite{Kardos:2018kqj}, and  determinations from thrust \cite{Abbate:2010xh,Gehrmann:2012sc} and $C$-parameter \cite{Hoang:2015hka}. 
Determinations based on first principle QCD treatments of power corrections, rather than Monte Carlo (MC) models, tend to yield noticeably lower values (e.g., \cite{Abbate:2010xh,Gehrmann:2012sc,Hoang:2015hka}), creating some tension with the world average. 
In recent years, significant discussion has focused on improving the estimation of uncertainties from nonperturbative power corrections, particularly in the far tail (multi-jet) of event shape distributions \cite{Luisoni:2020efy,Caola:2021kzt,Caola:2022vea,Nason:2023asn}. 
There has also been work that questions the use resummation of large logarithms for event shapes~\cite{Nason:2023asn}. 
Ongoing research continues to address various questions in $e^+e^-$ determinations of $\alpha_s(m_Z)$,  see Refs.~\cite{Nason:2023asn,Bell:2023dqs,Benitez-Rathgeb:2024ylc,Benitez:2024nav,Nason:2025qbx}.

Independent sensitivity to fundamental QCD ingredients makes it compelling to apply similar techniques of resummation and rigorous treatment of power corrections to an entirely different experimental environment with different systematics---DIS in $ep$ collisions. DIS event shapes exhibit a rich dependence on parton distribution functions, $\alpha_s(M_Z)$, and hadronization.
Event shapes in DIS analogous to those defined in $e^+e^-$ collisions have been computed and proposed since the turn of the millennium (e.g. \cite{Antonelli:1999kx,Dasgupta:2001eq,Dasgupta:2003iq}). Later, several DIS event shapes in the category of $N$-jettiness \cite{Stewart:2010tn} (specifically, 1-jettiness) were proposed and computed to various levels of accuracy. In this paper we focus on one of these versions of DIS 1-jettiness, namely, $\tau_1^b$ in the notation of \cite{Kang:2013nha,Kang:2014qba},
which is also equivalent to the DIS thrust $\tau_Q$ defined in \cite{Antonelli:1999kx}. We will give the detailed definition below. Similar to $e^+e^-$ thrust \cite{Brandt:1964sa,Farhi:1977sg} it groups particles into two regions defined by symmetric hemispheres in the Breit frame, where the proton and virtual photon form a collision axis. The fact that it is entirely determined by only measuring the ``current'' hemisphere (away from the proton-beam remnant direction) and does not require use of a jet algorithm, makes it experimentally attractive.
For this reason the HERA H1 collaboration recently produced a measurement of the $\tau_1^b$ event shape  \cite{H1:2024aze} over a wide range of $x,Q$, to which we will compare our predictions near the end of this paper.

$\tau_1^b$ differs from other versions of 1-jettiness such as $\tau_1^a$ defined in \cite{Kang:2013nha} or $\tau_1$ in \cite{Kang:2012zr}, which define 1-jettiness regions and reference axes defined by certain  algorithms such as jet and jettiness algorithms \cite{Chu:2022jgs}. These versions have a slightly simpler factorization theorem in the resummation region \cite{Kang:2012zr,Kang:2013nha}, but can be somewhat more complicated to measure experimentally than $\tau_1^b$.  In this paper we bring the perturbative accuracy of predictions for $\tau_1^b$ to the N$^3$LL order of resummed accuracy, matched to NLO [$\mathcal{O}(\alpha_s^2)$] predictions from the program \texttt{NLOJet++} \cite{Nagy:2001xb, Nagy:2003tz}. 
For historical reference, $\tau_1^a$ along with $\tau_1^b$ were computed to NNLL resummed accuracy in \cite{Kang:2012zr,Kang:2013nha} and LO [$\mathcal{O}(\alpha_s)$] fixed-order accuracy in \cite{Kang:2013lga,Kang:2014qba}, and later to N$^3$LL resummed + LO [$\mathcal{O}(\alpha_s)$] fixed-order accuracy in \cite{Kang:2015swk,Kang:2015moa,AbdulKhalek:2021gbh}.
$\tau_1$ had also been computed to NNLL + LO [$\mathcal{O}(\alpha_s)$] accuracy in \cite{Kang:2013lga}, and together with $\tau_1^a$ was recently brought up to N$^3$LL + NLO [$\mathcal{O}(\alpha_s^2)$] accuracy in \cite{Cao:2024ota}.

In addition to the perturbative accuracy of the predictions, as we have alluded to above, a rigorous estimate of the size of nonperturbative effects due to hadronization is also essential for a reliable determination of $\alpha_s$. 
In the region, $\Lqcd/Q\ll \tau_1^b \ll 1$, mostly dijet events contribute, highly-accurate resummed perturbative predictions are relevant, and the leading nonperturbative effect can be implemented as a power correction that is essentially a shift of the perturbative distribution by a constant \cite{Dokshitzer:1995zt,Dokshitzer:1995qm} depending on a universal nonperturbative parameter $\Omega_1$,
\begin{equation}
\label{eq:shift}
\frac{d\sigma}{d\tau_1^b}(\tau_1^b) = \frac{d\sigma_\text{PT}}{d\tau_1^b}\biggl(\tau_1^b - \frac{c_\tau \Omega_1}{Q}\biggr)\,,
\end{equation}
where $c_\tau = 2$. 
The power of this relation lies in the universality of $\Omega_1$, which is independent of DIS variables $x$ and $Q$, as well as the specific observable, for a broad class of event shapes that includes $\tau_1^b$. For different event shapes, only the constant coefficient $c_\tau$ varies \cite{Lee:2006fn}, 
as long as they fall into the categories with the same hadron mass corrections~\cite{Mateu:2012nk}.
The distribution \eq{shift} can be expressed as
a convolution of the perturbative distribution with a shape function, whose first moment is $\Omega_1$. 
In our work, we ensure that the shape function remains universal in both $x$ and $Q$, in such a way that the dependence on the universal parameter $\Omega_1$ remains manifest. It is necessary to take care of renormalon ambiguities in the definition of $\Omega_1$ and to define it in a scheme where this ambiguity and corresponding counterpart in the perturbative cross section are canceled manifestly. We implement one such scheme,  the $R$-gap scheme \cite{Hoang:2008fs, Jain:2008gb, Hoang:2008yj}, in this paper.

Complementing the theoretical developments presented in this paper, a significant technical achievement involves the development of enhanced computational frameworks for calculating the $\tau_1^b$ cross section. Our collaboration created two independent codebases implementing resummation and renormalon subtraction, which demonstrate excellent consistency, showing discrepancies better than 0.1\%  across the relevant regions of the $\tau_1^b$ distribution.
By leveraging parallelization and memory optimization techniques, our codes achieve significant performance improvements---approximately 10 times faster than the ones we used in previous works \cite{Kang:2014qba, Kang:2015swk,AbdulKhalek:2021gbh}.
Further details on the numerical implementations are provided in the relevant sections of the paper. 

This paper is organized as follows. Section~\ref{sec:definition} provides a brief overview of the kinematics of DIS in the Breit frame, defines the 1-jettiness event shape $\tau_1^b$, and discusses its key features.
Section~\ref{sec:formalism} presents the factorization formula for the $\tau_1^b$ 
distribution and details the fixed-order components of the factorization theorem up to $\mathcal{O}(\alpha_s^2)$. It also discusses the resummation of large logarithms, the treatment of nonperturbative effects, and the subtraction of $\mathcal{O}(\Lambda_\textrm{QCD})$ renormalons. Section~\ref{sec:nonsing} focuses on the computation of the nonsingular contributions at $\mathcal{O}(\alpha_s)$ and $\mathcal{O}(\alpha_s^2)$ using \texttt{NLOJet++}. We  conduct multiple validation tests of the \texttt{NLOJet++} results and describe how these nonsingular contributions are incorporated into the overall theoretical predictions.
Section~\ref{sec:profile} discusses the implementation of the profile functions to perform renormalization group (RG) evolution of the components in the factorization theorem. 
Section~\ref{sec:results} presents our theoretical predictions for the $\tau_1^b$ distributions, and illustrates the perturbative convergence after resummation and $\mathcal{O}(\Lambda_\textrm{QCD})$ renormalon subtraction.
Our predictions are compared to recent HERA H1 measurements \cite{H1:2024aze} for selected values of $x$ and $Q$. Furthermore, we analyze the sensitivity of our theoretical predictions to the DIS variables $x$, $Q$, the strong coupling $\alpha_s$, the nonperturbative parameter $\Omega_1$, and the parton distribution functions (PDFs). The paper concludes in Sec.~\ref{sec:conclusion}.
Several appendices provide technical details to support the main text.

\section{Definitions and kinematics} 
\label{sec:definition}

In DIS, an incoming electron $(\ell)$ collides with a proton $(p)$, and after the scattering, the outgoing electron scatters off and the proton is broken up
into a multi-particle state $(X)$,
%---------------
\begin{equation}
\label{eq:DIS}
%---------------
\ell(k)+ p(P)\to \ell(k')+X(p_X)\,,
%---------------
\end{equation}
%---------------
where $k$ and $k'$ are the initial and final momenta of the electron, 
$P$ is the initial proton momentum, and $p_X=\sum_{i} p_i$ is the total momentum of the hadronic final state $X$.
\begin{figure}
    \centering
    \includegraphics[width=0.725\linewidth]{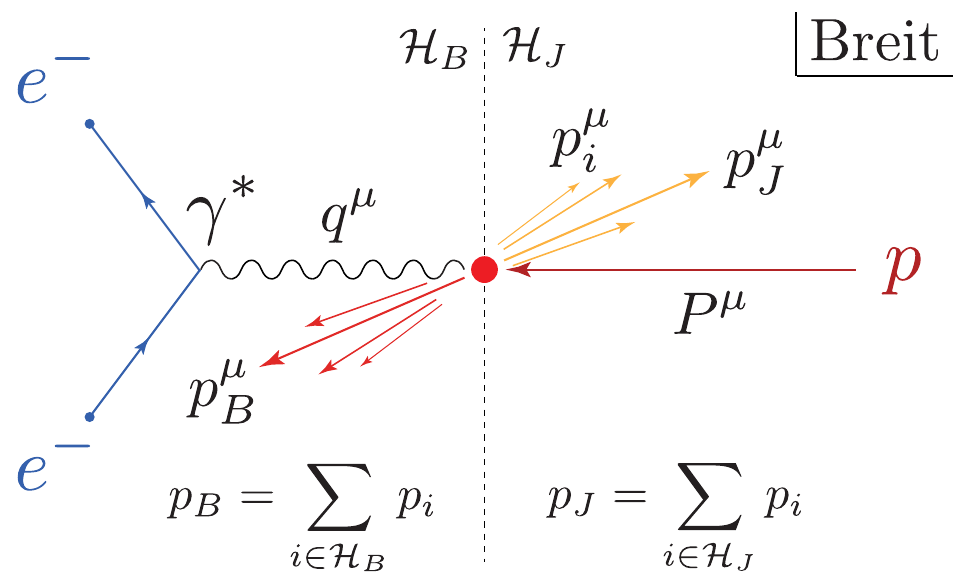}
    \vspace{-1em}
    \caption{Two-jet like DIS event in the Breit frame. $q^\mu$ and $P^\mu$ are the incoming momenta of the off-shell photon and proton, respectively, and $p_J$ and $p_B$ are the sum of the momenta in the jet and beam hemispheres, ${\cal H}_J$ and ${\cal H}_B$, respectively. }
    \label{fig:DIS-kinematics}
\end{figure}
In the course of this scattering, the off-shell photon with momentum $q$ is exchanged:
%---------------
\begin{equation}
%---------------
q = k-k'\,.
%---------------
\end{equation}
%---------------
In the following, we assume that the masses of the electron and proton
are negligible compared to the off-shell photon energy:
%---------------
\begin{equation}
\label{eq:massless-condition}
%---------------
P^2\approx0\,,
\quad
k^2\approx0\,,
\quad
k'^2\approx0\,.
%---------------
\end{equation}
%---------------
Since $k$ and $k'$ are light-like, $q$ is space-like:
%---------------
\begin{equation}
%---------------
q^2 = (k-k')^2 = -2k\cdot k'=-2E E'(1-\hat{\mathbf{k}}\cdot\hat{\mathbf{k}}')\le 0\,,
%---------------
\end{equation}
%---------------
where $E$ ($\hat{\mathbf{k}}$) and $E'$ ($\hat{\mathbf{k}}'$) are the energies (the Euclidean unit vectors)
of $k$ and $k'$, respectively.
Consequently, we define the following positive quantity 
%---------------
\begin{equation}
\label{def:Q}
%---------------
Q^2 = -q^2\,. 
%---------------
\end{equation}
%---------------
We define the usual dimensionless Bjorken scaling variable $x$ and the 
dimensionless variable $y$ as
%---------------
\begin{equation}
\label{def:xy}
%---------------
x = \frac{Q^2}{2P\cdot q}\,,
\quad
y = \frac{2P\cdot q}{2P\cdot k}\,.
%---------------
\end{equation}
%---------------
The three variables $Q$, $x$, and $y$ are related to one another via the kinematic constraint
$Q^2 = sxy$, where $s = (P+k)^2$ is the total invariant mass of the incoming states.
In the massless limit [\eq{massless-condition}], the kinematically allowed regions for $x$ and $y$ are given by
%---------------
\begin{equation}
\label{eq:xy-kinematically-allows}
%---------------
\frac{Q^2}{s}\le x \le 1
\quad\textrm{and}\quad
\frac{Q^2}{s}\le y \le 1\,.
%---------------
\end{equation}
%---------------

In this paper, we consider a variant of DIS 1-jettiness, $\tau_1^b$, whose formal definition is \cite{Kang:2013nha}
\begin{equation}
\label{eq:tau1b}
    \tau_1^b \equiv \frac{2}{Q^2}\sum_{i\in X}
    \min \left\{q_B^b\cdot p_i,\, q_J^b\cdot p_i\right\},
\end{equation}
where the sum runs over all particles $i$ in the final state $X$. The minimum operator separates the final-state particles with momenta $p_i$ into one of two regions, depending on which reference vector $q_{B,J}$ they are closer to, measured by the dot product.  The two reference four-vectors are defined by
\begin{equation}
\label{eq:qBqJ}
    {q_B^b} = x P,
    \qquad
    {q_J^b} = q + x P.
\end{equation}
This is a specific case of the ``1-jettiness'' event shape $\tau_1$ \cite{Stewart:2010tn}, where the value of $\tau_1$ becomes small when the final-state hadrons are collimated along the jet direction $q_J$ and initial-state radiation along the beam direction $q_B$. 
Other choices of reference vectors or measures are available \cite{Jouttenus:2011wh}. The alternative choices were considered in \cite{Kang:2013nha,Kang:2013wca,Chu:2022jgs}. The choice of $\tau_1^b$ in \eq{tau1b} is so named because it groups final-state hadrons into perfectly symmetric hemispheres in the \emph{Breit} frame: the beam hemisphere $\mathcal{H}_B$ when $q_B^b\cdot p_i < q_J^b\cdot p_i$, and  the jet hemisphere $\mathcal{H}_J$ otherwise. 
Although it is natural to work in the Breit frame, note that $\tau_1^b$ is Lorentz invariant by definition, so it can be computed in any frame with identical results. Unlike some other choices, \eq{tau1b} does not rely on the use of any jet algorithm to identify the outgoing jet or beam/jet regions, with reference vectors \eq{qBqJ} entirely determined by the lepton and incoming proton kinematics.

\begin{figure}
    \centering
    \includegraphics[width=0.65\linewidth]{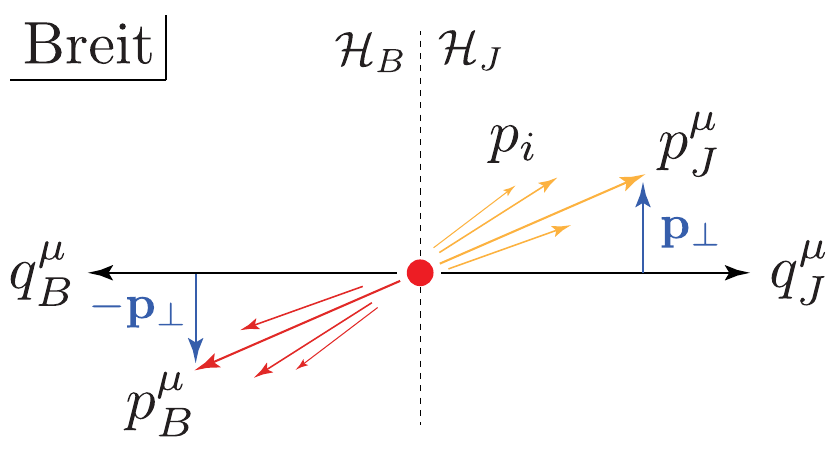}
    \vspace{-1em}
    \caption{The jet and beam momenta $p_J$ and $p_B$, and the reference vectors $q_J^b$ and $q_B^b$ for $\tau_1^b$ in the Breit frame. Note that the reference vectors are not necessarily aligned with the physical jet and beam momenta, with offsets given by the transverse momentum $\pm \mathbf{p}_\perp$. }
    \label{fig:jet-beam-vectors}
\end{figure}

In the Breit frame, the virtual photon with momentum $q$ and the proton with momentum $P$ are aligned along the $z$ axis (\fig{DIS-kinematics}). In this frame, $q$ and $P$ are given by
\begin{equation}
\label{eq:qBreit}
    q \stackrel{\text{Breit}}{=} Q(0, 0, 0, 1),
    \quad
    P \stackrel{\text{Breit}}{=} \frac{Q}{2x}(1,0,0,-1),
\end{equation}
where we used Eq.~\eqref{def:xy}. 
If we define the two light-like vectors along $z$ axis as 
%---------------
\begin{equation}
%---------------
n_z \stackrel{\textrm{Breit}}{=} (1,0,0,1)\,,
\quad
\bar{n}_z\stackrel{\textrm{Breit}}{=}(1,0,0,-1)\,, 
%---------------
\end{equation}
%---------------
then $q$ and $P$ in the Breit frame can be expressed in terms of $n_z$ and $\bar{n}_z$ as follows:
%---------------
\begin{equation}
\label{eq:q-P-in-Breit}
%---------------
q \stackrel{\textrm{Breit}}{=} \frac{Q}{2}(n_z-\bar{n}_z)\,,
\quad
P \stackrel{\textrm{Breit}}{=} \frac{Q}{2x}\bar{n}_z\,. 
%---------------
\end{equation}
%---------------
Similarly, the two reference vectors in the definition of $\tau_1^b$ in Eq.~\eqref{eq:tau1b} can be written in the Breit frame as (\fig{jet-beam-vectors})
\begin{equation}
    {q_B^b} \stackrel{\text{Breit}}{=} \frac{Q}{2}\bar{n}_z,
    \quad
    {q_J^b} \stackrel{\text{Breit}}{=} \frac{Q}{2}n_z.
\end{equation}
The outgoing jet and beam remnants are not necessarily aligned exactly along $q_{J,B}$ but can be offset from them due to transverse momentum $\mathbf{p}_\perp$ of radiation from the initial proton beam and corresponding momentum recoil. This is a feature of $\tau_1^b$ and not a bug. The vector $q_J^b = q+xP$ in \eq{qBqJ} is in fact the outgoing jet momentum in the Born approximation, and deviations from it are a measure of the recoil effect of initial-state radiation. This is in contrast to $\tau_1^a$ in \cite{Kang:2013nha}, which always re-aligns $q_J$ to be along the measured jet axis.

By making use of the momentum conservation $p_B = p_X - p_J$ where $p_{B} = \sum_{i\in \mathcal{H}_{B}} p_i$ and $p_{J} = \sum_{i\in \mathcal{H}_{J}} p_i$ in the Breit frame, one can show \cite{Kang:2013nha} that $\tau_1^b$ agrees with the classical DIS thrust variable $\tau_Q$ \cite{Antonelli:1999kx}:
\begin{equation}\label{eq:DIS-thrust}
    \tau_1^b  \stackrel{\text{Breit}}{=} 1-\frac{2}{Q}\sum_{i\in \mathcal{H}_J}(p_i)_z = \tau_Q.
\end{equation}
In this form, we observe that $\tau_1^b=\tau_Q$ can, in principle, be measured entirely in terms of particles within the jet hemisphere. This property allows for measurements free from remnant fragmentation, making it advantageous for experimental analysis. 
Additionally, $\tau_1^b$ is a global observable, meaning it is free of non-global logarithms \cite{Dasgupta:2002dc}, which enables computations with high theoretical accuracy.
Furthermore, there is a physical upper limit on the possible values of $\tau_1^b$, imposed by kinematic constraints:
the $z$ component of the jet momentum $(p_J)_z$ must be positive, while the $z$-component of the beam momentum is negative in the Breit frame  \cite{Kang:2014qba}:
\begin{equation}
\label{eq:def_taubmax}
\tau^b_\textrm{max}
=
\begin{cases}
1,&\textrm{for $x<\frac{1}{2}$},
\\[1ex]
\displaystyle
\frac{1-x}{x},&\textrm{for $x\ge\frac{1}{2}$}.
\end{cases}
\end{equation}

A distinctive feature of the $\tau_1^b$ distribution, unlike thrust in $e^+e^-$ collisions,  
is that one hemisphere ($\mathcal{H}_J$) can be completely empty, devoid of final-state particles, as long as $x<1/2$. This possibility was observed as early as Ref.~\cite{Streng:1979pv}.
For such configurations, the measured value of $\tau_1^b$ will be exactly 1 \cite{Kang:2014qba}, thus our variable can be used to identify such events. This will appear in the $\tau_1^b$ distribution as a peak near $\tau_1^b=1$. 
We will show that this singular behavior as $\tau_1^b \to 1$ is not captured by the SCET factorization formula,  
as it cannot be described by two-jet events.  
In our analysis, this effect is accounted for by full QCD fixed-order calculations.
At one-loop order, this manifests as a delta function, $\delta(1-\tau_1^b)$ \cite{Kang:2014qba}.  
At two-loop order, numerical results from \texttt{NLOJet++} reveal that this singular behavior is smeared towards lower values of $\tau_1^b$.
As we will review in the later Sections, this behavior is observed in experimental data and agrees well with these theoretical calculations.

\section{Formalism}
\label{sec:formalism}
\subsection{Overview}
The shape of the $\tau_1^b$ distribution, like other 2-jet event shapes, varies significantly across its domain, organized naturally into three regions: peak, tail, and far-tail, each described accurately by different but smoothly connected theoretical methods. These regimes are characterized by the power counting:
\begin{align}
\begin{split}
\textrm{peak region}: & \quad 
\tau_1^b \sim \Lambda_\textrm{QCD}/Q \ll 1,
\\
\textrm{tail region}: & \quad 
\Lambda_\textrm{QCD}/Q \ll \tau_1^b \ll 1,
\\
\textrm{far-tail region}: & \quad
\tau_1^b \sim 1.
\end{split}
\end{align}
In the peak region, the soft scale approaches $\Lambda_\textrm{QCD}$, and a nonperturbative shape function is required to model the soft physics. The peak region is dominated by physics at a nonperturbative scale and will require a nonperturbative model. The tail region exhibits large logarithms of $\tau_1^b$ in perturbation theory which will require resummation, and the nonperturbative effects obey an OPE allowing their leading contribution to be described by a universal shift parameter. In the far-tail region, which is populated by multijet events, fixed-order perturbation theory provides an accurate description. In \sec{nonsing} and \sec{profile} we will tie together predictions in each region smoothly.

After putting predictions for these regions together, our final expression for the $\tau_1^b$ distribution will be given by

%---------------
\begin{align}
\label{eq:k-int-cumulant}
%---------------
\sigma(\tau_1^b)
&=
\int dk\,
\sigma_\textrm{PT}\left(\tau_1^b-\frac{k}{Q}\right)
\left[
e^{-2\delta(R,\mu_S)(d/dk)}
F
\left(
k-2\Delta(R,\mu_S)
\right)
\right],
%---------------
\end{align}
%---------------
where the perturbative cross section $\sigma_\textrm{PT}$ is convolved with a shape function $F$.
The function~$F$ captures nonperturbative soft radiation, peaking around $k \sim \Lambda_{\textrm{QCD}}$ and 
falling off faster than any power (such as exponentially)
at larger~$k$.
The perturbative cross section $\sigma_\textrm{PT}$ computed in the $\overline{\textrm{MS}}$ scheme has an $\mathcal{O}(\Lambda_\textrm{QCD})$ renormalon ambiguity. 
To remove this ambiguity, the exponential factor preceding the shape function $F$ is introduced as a renormalon subtraction.
Here, $\delta(R,\mu_S)$ cancels the leading renormalon ambiguity order-by-order in $\alpha_s$,  
and $\Delta$ is the gap parameter that accounts for the physical hadronic threshold.
These elements are further discussed in Sec.~\ref{sec:shape-renormalon}.

In our notation, $\sigma$ and $\sigma_\textrm{PT}$ can represent either the differential or cumulative $\tau_1^b$ distribution.
\begin{equation}\label{eq:cumulant-def}
\sigma(\tau_1^b) = \frac{d\sigma}{d\tau_1^b}
\quad
\textrm{or}
\quad
\sigma(\tau_1^b) = \sigma_c(\tau_1^b) = \int_0^{\tau_1^b} d\tau \frac{d\sigma}{d\tau}.
\end{equation}
For simplicity, we suppress notation that these are also differential in $x$ and $Q^2$.

The perturbative cross section is expressed as the sum of the singular and nonsingular contributions:
\begin{equation}
\label{eq:sing-plug-ns}
\sigma_\textrm{PT}(\tau_1^b; \mu_H, \mu_J, \mu_B, \mu_S, \mu_\textrm{ns})
=
\sigma_\textrm{PT}^\textrm{s}(\tau_1^b; \mu_H, \mu_J, \mu_B, \mu_S)
+
\sigma_\textrm{PT}^\textrm{ns}(\tau_1^b; \mu_\textrm{ns}).
\end{equation}
The singular contribution $\sigma_{\textrm{PT}}^\textrm{s}$ contains logarithmic terms of the form $\log^n\tau_1^b$ for the cumulative and $(\log^{n-1}\tau_1^b)/\tau_1^b$ for the differential. These logarithms are captured and resummed 
using the SCET framework \cite{Bauer:2000ew,Bauer:2000yr,Bauer:2001ct,Bauer:2001yt,Bauer:2002nz} at leading power. The nonsingular contribution $\sigma_{\textrm{PT}}^\textrm{ns}$ includes 
terms that are suppressed by one or more powers of $\tau_1^b$.
It accounts for the  
fixed-order QCD distribution with the most singular terms subtracted to avoid double counting.  
The scales $\mu_{H,J,B,S}$ are associated with hard, jet, beam, and soft functions in the factorization discussed in the next subsection. The scale $\mu_\textrm{ns}$ corresponds to the nonsingular part.  
Their choices are discussed in Sec.~\ref{sec:profile}.

We compute the differential cross section by differentiating the cumulative distribution,  
$\sigma_c(\tau_1^b)$, with respect to $\tau_1^b$. 
In this process, we ensure that the $\tau_1^b$-dependent scales, $\mu_i(\tau_1^b)$,  
are the same in both terms in the numerator in the equation below to avoid spurious contributions \cite{Abbate:2010xh}:
\begin{equation}\label{eq:cumulant-to-differential}
\frac{d\sigma}{d\tau_1^b}
= \lim_{\epsilon\to 0}
\frac{\sigma_c(\tau_1^b +\epsilon, \mu_i(\tau_1^b))-\sigma_c(\tau_1^b -\epsilon, \mu_i(\tau_1^b))}{2\epsilon}.
\end{equation}
Equation~(\ref{eq:cumulant-to-differential}) applies to the singular contributions.  
For the nonsingular contribution, analytic results are currently only available at $\mathcal{O}(\alpha_s)$ \cite{Kang:2014qba}.  
We use the differential results from the known analytic expression at $\mathcal{O}(\alpha_s)$ \cite{Kang:2014qba},  
and the numerical calculations from \texttt{NLOJet++} at $\mathcal{O}(\alpha_s^2)$.\footnote{The numerical results from \texttt{NLOJet++} are provided as histograms for the differential cross section, making it natural to work directly with the differential cross section. Additionally, the numerical results from \texttt{NLOJet++} are highly noisy at small $\tau_1^b$, rendering the computation of the cumulative distribution integrated starting from $\tau_1^b=0$ unreliable.}

Due to the absence of $Z^0$-boson contributions in the fixed-order full QCD results at $\mathcal{O}(\alpha_s)$ \cite{Kang:2014qba} and at $\mathcal{O}(\alpha_s^2)$ in \texttt{NLOJet++}, our results include only virtual photon exchange diagrams for the nonsingular contributions. However, the singular contributions from $Z^0$-boson exchange are included through the SCET factorization formula \cite{Kang:2013nha}.

We follow the standard order counting as used in Refs.~\cite{Kang:2013nha,Almeida:2014uva,Bell:2018gce}.
The fixed-order $\mathcal{O}(\alpha_s)$ contributions are called LO, and the fixed-order $\mathcal{O}(\alpha_s^2)$ contributions are called NLO, as the tree-level fixed-order contributions at $\mathcal{O}(\alpha_s^0)$ are given by the delta distribution  $\delta(\tau_1^b)$.  Table~\ref{tab:order} shows the accuracy of the fixed-order ingredients matched with the summation orders of logs (NLL,~NNLL, N$^3$LL).

\begin{table}
    \centering
    \begin{tabular}{c|c|c|c|c|c|c|c}
        \hline
        \hline
        \rule{0pt}{2.8ex}
        Accuracy & $\Gamma_\textrm{cusp}[\alpha_s]$ & $\gamma_{H,B,J,S}[\alpha_s]$ & $\beta[\alpha_s]$ & $\{H,B,J,S\}[\alpha_s]$ & Nonsingular & $\gamma_\Delta^{\mu,R}$ & $\delta$ \\[0.6ex]
         \hline
         \hline
          \rule{0pt}{2.8ex}
        NLL     & $\alpha_s^2$ & $\alpha_s$ & $\alpha_s^2$ & $\alpha_s^0$ & - & $\alpha_s$ & - \\[0.5ex]
        NNLL    & $\alpha_s^3$ & $\alpha_s^2$ & $\alpha_s^3$ & $\alpha_s$ & $\alpha_s$ & $\alpha_s^2$ & $\alpha_s$\\[0.5ex]
        N$^3$LL & $\alpha_s^4$ & $\alpha_s^3$ & $\alpha_s^4$ & $\alpha_s^2$ & $\alpha_s^2$ & $\alpha_s^3$ & $\alpha_s^2$ \\[0.5ex]
         \hline
         \hline
    \end{tabular}
    \caption{Ingredients included at N$^k$LL accuracies.
    $\Gamma_\textrm{cusp}$ and $\gamma_{H,B,J,S}$ are the cusp and non-cusp anomalous dimensions for the hard, beam, jet, and soft functions, respectively. $\beta$ is the QCD beta function, $\gamma_\Delta^{\mu,R}$ are the anomalous dimensions for the $R$-gap parameters, and $\delta$ refers to the renormalon subtraction terms in the same scheme.}
    \label{tab:order}
\end{table}

\subsection{Factorization}

Within the framework of SCET, the dijet cross section is factorized into hard, collinear, and soft components.
The factorization theorem for the $\tau_1^b$ distribution in dijet limit, as derived in Ref.~\cite{Kang:2013nha}, is given by
\begin{align} \label{eq:tau1b-FT}
\begin{split}
\frac{d\sigma^\textrm{s}_\textrm{PT}}{d\tau_1^b}
={}&
\sigma_0^b
\int dt_J dt_B dk_S \,
\delta\left(\tau_1^b - \frac{t_J+t_B}{Q^2}-\frac{k_S}{Q}\right)
S_\textrm{PT}(k_S,\mu)
\\
&
\times
\sum_q
\int d^2 \mathbf{p}_\perp
J_q(t_J-\mathbf{p}_\perp^2,\mu)
\left[
H_q(y,Q^2,\mu)
\mathcal{B}_q(t_B,x,\mathbf{p}_\perp^2,\mu)
+
\left(
q\leftrightarrow \bar{q}
\right)
\right].
\end{split}
\end{align}
In this equation, $\sigma_0^b$ represents the Born-level cross section, defined as
\begin{equation}\label{eq:born-crosssection}
    \sigma_0^b\equiv \frac{d\sigma_0^b}{dx dQ^2} = \frac{2\pi \alpha^2_\textrm{em}}{Q^4}[(1-y)^2+1],
\end{equation}
with $\alpha_\textrm{em}$ being the electromagnetic coupling constant at the scale $Q$.
The delta function in Eq.~\eqref{eq:tau1b-FT} serves as a measurement function for $\tau_1^b$. It enforces the kinematic constraint that links the beam, jet, and soft momenta contributions to $\tau_1^b$ in the factorization formula, and ensures that the sum of their contributions equals the measured $\tau_1^b$. The soft function $S_\textrm{PT}$ describes the perturbative soft radiation exchanged between the particles in the two hemispheres, while the jet function $J_q$ handles the collinear radiation within the jet. The jet function is independent of quark flavor in the massless limit. The beam function $\mathcal{B}_{q(\bar{q})}$ encodes the transverse-momentum-dependent (TMD) initial-state radiation (ISR) from the incoming quark (anti-quark). 
Note that in the factorization theorem in Eq.~\eqref{eq:tau1b-FT}, only quark jet and beam functions contribute, as the contributions from the gluon operators vanish due to quark-number conservation, vector current conservation, and the assumption of massless leptons \cite{Stewart:2009yx, Kang:2013nha}. As a result, the gluon beam and jet functions are absent from the factorization theorem, and the gluon PDF contributes only indirectly through its contribution to the quark beam function. 
The hard functions, $H_{q(\bar{q})}$, describe the hard scattering contributions in DIS for both virtual $\gamma$ and $Z^0$ exchange
and are expressed in terms of vector and axial-vector Wilson coefficients, as well as leptonic factors. 

The $\mathbf{p}_\perp^2$ dependence in the beam and jet functions arises because the reference vectors, $q_{J,B}$, used to define $\tau_1^b$ are not necessarily aligned with the physical jet and beam momenta $p_{J,B}$ (see Fig.~\ref{fig:jet-beam-vectors}). In the Breit frame, where initial states are aligned along the $z$-axis, the exchanged gauge boson momentum $q$ in \eq{qBreit} has $\mathbf{q}_\perp=0$, and energy-momentum conservation ensures that the transverse momenta of the jet and beam are balanced. Therefore, in our factorization formula, we sum over all possible configurations in $\mathbf{p}_\perp$. However, the range of $\mathbf{p}_\perp^2$ is bounded as $0\le \mathbf{p}_\perp^2 \le t_J$, as the jet function vanishes for the negative arguments. Additionally, the transverse-momentum-dependent beam function imposes a constraint $0\le \mathbf{p}_\perp^2 \le t_B(1-z)/z$ to ensure the positivity of the invariant mass of the ISR jet \cite{Jain:2011iu, Gaunt:2014xxa}.
To manage the $\mathbf{p}_\perp^2$ dependence, we shift variables, $t_J\to t_J + \mathbf{p}_\perp^2$ and $t_B\to t_B - \mathbf{p}_\perp^2$, such that the $\mathbf{p}_\perp^2$ integration is confined to the beam function. This leads to the simplified factorization formula:
\begin{align} \label{eq:tau1b-FT-2}
\begin{split}
\frac{d\sigma^\textrm{s}_\textrm{PT}}{d\tau_1^b}
={}&
\sigma_0^b
\int dt_J dt_B dk_S \,
\delta\left(\tau_1^b - \frac{t_J+t_B}{Q^2}-\frac{k_S}{Q}\right)
S_\textrm{PT}(k_S,\mu)
\\
&
\times
\sum_q
J_q(t_J,\mu)
\left[
H_q(y,Q^2,\mu)
{B}_q(t_B,x,\mu)
+
\left(
q\leftrightarrow \bar{q}
\right)
\right],
\end{split}
\end{align}
where the integrated beam function ${B}_q$ is defined as
\begin{equation}\label{eq:integrated-beam}
{B}_{i}(t_B,x,\mu)
\equiv
\int d^2 \mathbf{p}_\perp 
\mathcal{B}_i(t_B-\mathbf{p}_\perp^2,x,\mathbf{p}_\perp^2,\mu).
\end{equation}
Note that this differs from the beam function relevant for $\tau_1^a$, which is a function of the same variables $t_B$, $x$, and $\mu$ \cite{Kang:2013nha}, which is also called the ordinary beam function $B_i$ \cite{Stewart:2009yx}. At the risk of some confusion, for the sake of simpler notation in this paper, we will use $B_i$ to refer to our particular projection of the TMD beam function onto $t_B$ in \eq{integrated-beam}. 

\begin{figure}
    \centering
    \includegraphics[width=0.6\linewidth]{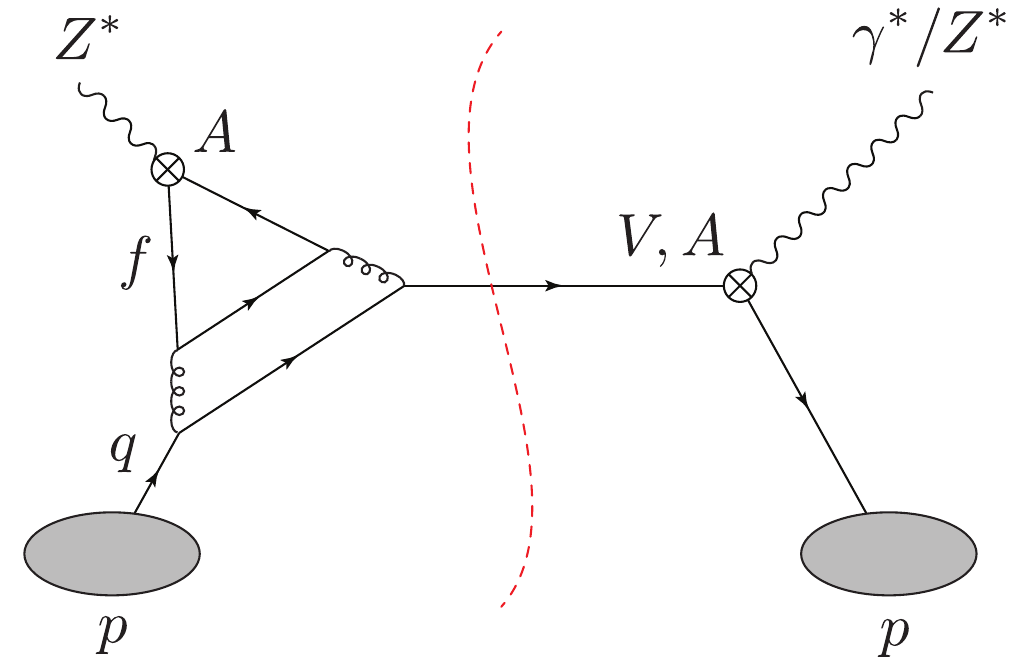}
    \vspace{-1em}
    \caption{The triangle anomaly graph contributing to the flavor-singlet part of matching coefficient $C_{Afq}$ in \eq{CVA}. The conjugate mirror graph is not shown.}
    \label{fig:anomaly}
\end{figure}

We now analyze each component of the factorization formula in Eq.~\eqref{eq:tau1b-FT-2}.
The hard functions, $H_q$ and $H_{\bar{q}}$, are given by \cite{Kang:2013nha}
\begin{align}
\label{eq:hard}
\begin{split}
H_{q,\bar q}(y,Q^2,\mu) 
=& \sum_{f,f'}
\big[
\left(C_{Vfq}^* C_{Vf'q} L_{gff'}^{VV} + C_{Afq}^* C_{Af'q} L_{gff'}^{AA} \right) 
\\
&
\quad\quad
\mp r(y) 
\left( C_{Vfq}^* C_{Af'q} L_{\epsilon ff'}^{VA}  + C_{Afq}^* C_{Vf'q} L_{\epsilon ff'}^{AV}\right) 
\big],
\end{split}
\end{align}
where the upper sign is for $H_q$ and the lower sign is for $H_{\bar{q}}$. The term $r(y)$ is defined as
\begin{equation}
r(y) \equiv \frac{y(2-y)}{1+(1-y)^2}\,.
\end{equation}
The vector and axial Wilson coefficients are given as
\begin{equation}\label{eq:CVA}
C_{Vfq} = \delta_{fq} C(q^2,\mu^2)\,,\quad C_{Afq} = \delta_{fq}C(q^2,\mu^2) 
+ C^\textrm{sing}_{fq}(q^2)\,,
\end{equation}
where the flavor-diagonal component for both terms is captured by $C(q^2,\mu^2)$, and the axial term includes an additional ``flavor singlet'' contribution $C^\textrm{sing}$, arising from the axial anomaly diagrams. As illustrated in \fig{anomaly}, the triangle anomaly graph contributes to the flavor-singlet part of $C_{Afq}$, starting at $\mathcal{O}(\alpha_s^2)$.
The leptonic factors $L_{gff'}^{VV}$, $L_{gff'}^{AA}$, $L_{\epsilon ff'}^{AV}$, and $L_{\epsilon f'f}^{VA}$ are given in Ref.~\cite{Kang:2013nha}:
\begin{align}
\label{eq:leptonic}
\begin{split}
L_{gff'}^{VV} &= Q_f Q_{f'} - \frac{(Q_f v_{f'} + v_f Q_{f'})v_e }{1+m_Z^2/Q^2}
+ \frac{ v_f v_{f'}(v_e^2 + a_e^2)}{(1+m_Z^2/Q^2)^2},
\\
L_{gff'}^{AA} &= \frac{a_f a_{f'} (v_e^2 + a_e^2)}{(1+m_Z^2/Q^2)^2},
\\
%L_{gff'}^{AV} &= L_{g f'f}^{VA} = \frac{a_f}{1+m_Z^2/Q^2}\left[ Q_{f'} v_e -  \frac{v_{f'}(v_e^2+a_e^2)}{1+m_Z^2/Q^2}\right]  \,,\nn \\
L_{\epsilon ff'}^{AV} &= L_{\epsilon f'f}^{VA} =
\frac{a_f a_e}{1+m_Z^2/Q^2}\left( Q_{f'} - \frac{2v_{f'}v_e}{1+m_Z^2/Q^2}\right).
\end{split}
\end{align}
The formulas for the flavor-diagonal contribution $C(q^2,\mu^2)$ and the flavor-singlet contribution $C_{fq}^\text{sing}(q^2)$ in \eq{CVA} are given in Appendix~\ref{app:fixed-order-coeff}. 
%and the formula for the flavor-singlet contributions are given below.
Plugging Eq.~\eqref{eq:leptonic} into Eq.~\eqref{eq:hard}, and retaining only the contributions that survive at $\mathcal{O}(\alpha_s^2)$, we obtain
\begin{equation}\label{eq:hard-before-simp}
H_{q,\bar q}(y,Q^2,\mu) = H(Q^2,\mu) L_{q,\bar{q}}(y,Q^2)  + H^\textrm{sing}_{q,\bar q}(y,Q^2,\mu)\,,
\end{equation}
where we define $H(Q^2,\mu) \equiv |C(q^2=-Q^2,\mu)|^2$ and
$L_{q,\bar{q}}(y,Q^2) \equiv L_{gqq}^{VV} + L_{gqq}^{AA} \mp 2 r(y) L_{ \epsilon qq}^{VA}$,
and
\begin{equation}
\label{eq:Hsing}
H^\textrm{sing}_{q,\bar q} = \sum_{f} ({C^\textrm{sing}_{fq}}^* + C^\textrm{sing}_{fq}) \big[ L_{gfq}^{AA} \mp r(y) L^{AV}_{\epsilon f q}\bigr].
\end{equation}
The hard function in Eq.~\eqref{eq:hard-before-simp} is expanded as a power series in $\alpha_s$ and the corresponding coefficients are provided in Appendix~\ref{app:fixed-order-coeff}.

The perturbative soft function $S_\textrm{PT}(k_S,\mu)$ in Eq.~\eqref{eq:tau1b-FT} is the symmetric projection of the double-differential soft function 
\cite{Kang:2013nha}:
\begin{equation}
    S_\textrm{PT}(k_S,\mu)=\int dk_S^J dk_S^B\,
    \delta(k_S - k_S^J - k_S^B) S_\text{hemi}(k_S^J, k_S^B, \mu),
\end{equation}
where $S_\text{hemi}(k_S^J, k_S^B, \mu)$ refers to the hemisphere soft function depending on hemisphere momenta $k_s^J$ add $k_S^B$, which has been computed up to 1-loop order in Ref.~\cite{Fleming:2007xt} and 2-loop order in Refs.~\cite{Kelley:2011ng, Monni:2011gb, Hornig:2011iu}.
Although these soft functions were originally computed for $e^+e^-$ dijet event shapes, it has been shown that the soft functions for DIS 1-jettiness and $pp$ 0-jettiness are equivalent  to the $e^+e^-$ soft function, at least to $\mathcal{O}(\alpha_s^2)$, which is the order relevant for our calculation \cite{Kang:2015moa, Boughezal:2015eha}. Expressing the soft function in terms of the plus distributions, we have
%---------------
\begin{align}\label{eq:s-hemi-orig}
%---------------
S_\textrm{PT}(k_S,\mu) 
= 
\frac{1}{\mu}
\sum_{m=-1} S_m(\alpha_s)
\mathcal{L}_m(k_S/\mu),
%---------------
\end{align}
%---------------
where the plus distribution, denoted as $\mathcal{L}_m$,
is defined by
%---------------
\begin{align}\label{eq:plus_equation}
\begin{split}
%---------------
\mathcal{L}_{m}(u)
=
\begin{cases}
\displaystyle
\delta(u), & \textrm{for $m=-1$,}
\\[1.5ex]
\displaystyle
\left[
\frac{\theta(u)\ln^m(u)}{u}
\right]_+, & \textrm{for $m>-1$,}
\end{cases}
%---------------
\end{split}
\end{align}
%---------------
where the symbol $[\quad]_+$ representing plus distribution is defined in Appendix~\ref{app:fixed-order-coeff}.
The coefficient functions $S_m(\alpha_s)$ up to $\mathcal{O}(\alpha_s^2)$ are also listed in Appendix~\ref{app:fixed-order-coeff}.

The jet function is also expressed in the same way as \cite{Kang:2013nha}:
%---------------
\begin{align}\label{eq:exp-jet-function}
J(t,\mu)
=
\frac{1}{\mu^2}
\sum_{m=-1} J_m(\alpha_s)
\mathcal{L}_m(t/\mu^2).
\end{align}
%---------------
The plus distributions $\mathcal{L}_m(t/\mu^2)$ are defined as in Eq.~\eqref{eq:plus_equation}, and the coefficient functions $J_m(\alpha_s)$ are listed in Appendix~\ref{app:fixed-order-coeff}.

The transverse-momentum-dependent beam function, $\mathcal{B}_i$, for $i=q,\bar{q},g$, is factorized into the PDF $f_j$ and its perturbative matching coefficient $\mathcal{I}_{ij}$ as follows \cite{Jain:2011iu}:
\begin{equation}\label{eq:transverse-dep-beam}
    \mathcal{B}_i(t,x,\mathbf{p}_\perp^2,\mu)
    =
    \sum_{j}
    \int_x^1 \frac{dz}{z}
    \mathcal{I}_{ij}(t,z,\mathbf{p}_\perp^2, \mu)f_j(x/z,\mu),
\end{equation}
where the index $j$ sums over the parton flavors in the proton. The beam functions are matched onto PDFs via the operator product expansion (OPE), up to power corrections of order $\Lambda_\textrm{QCD}^2/t$ and $\Lambda_\textrm{QCD}^2/\mathbf{p}_\perp^2$, assuming $\mathbf{p}_\perp^2$ is large enough to be a perturbative scale \cite{Boussarie:2023izj}. The coefficient functions $\mathcal{I}_{ij}$ are the radiative kernels describing the ISR from the parent parton $j$ until the parton $i$ enters the hard interaction.  
As discussed below Eq.~\eqref{eq:tau1b-FT}, in our factorization theorem, only $\mathcal{B}_i$ for $i=q,\bar{q}$ contribute. 
In this analysis, we use the PDF set NNPDF4.0 at NNLO, determined at $\alpha_s(m_Z) = 0.118$ \cite{NNPDF:2021njg}. To study the event shape at a different $\alpha_s(m_Z)$ value, the corresponding PDF set must be used.\footnote{NNPDF4.0 PDF sets are available for $\alpha_s(m_Z) = 0.116,\, 0.117,\, 0.1175,\, 0.118,\, 0.1185,\, 0.119,\, 0.120.$}
The tree-level and one-loop expressions for $\mathcal{I}_{ij}$ are given in \cite{Jain:2011iu} and the two-loop terms are given in \cite{Gaunt:2014xxa}. 

For the $\tau_1^b$ distribution, we need the integrated beam function defined in Eq.~\eqref{eq:integrated-beam} which can be written as
\begin{equation}\label{eq:integrated-beam-2}
{B}_i(t,x,\mu)
=
\sum_{j}
\int_x^1 \frac{dz}{z}
\mathcal{J}_{ij}(t,z, \mu)f_j(x/z,\mu),
\end{equation}
where the new kernel $\mathcal{J}$ is connected to $\mathcal{I}$ through the following relation: 
%---------------
\begin{equation}
\label{eq:Jij-Iij-relation}
%---------------
\mathcal{J}_{ij}(t,z,\mu)
=
\int d^2 \mathbf{k}_\perp
\mathcal{I}_{ij}(t-\mathbf{k}_\perp^2,z,\mathbf{k}_\perp^2,\mu)
=
\pi t
\int_{1-\frac{1-z}{z}}^1 dy\,
\mathcal{I}_{ij}(ty,z,t(1-y),\mu).
%---------------
\end{equation}
%---------------
Here, the variable $y$ is introduced through $y = 1-\mathbf{k}_\perp^2/t$,
and $\mathbf{k}_\perp^2$ is constrained by $0\le \mathbf{k}_\perp^2 \le t(1-z)/z$ to ensure positivity of the ISR jet's invariant mass. 

We can express ${B}_i$ in the same form as the soft and jet functions
in Eqs.~\eqref{eq:s-hemi-orig} and \eqref{eq:exp-jet-function} as
\begin{align}\label{eq:Bi-plus}
{B}_i(t,x,\mu)
=
\frac{1}{\mu^2}
\sum_{m=-1} 
{B}_{i,m}(x,\mu;\alpha_s)
\mathcal{L}_m(t/\mu^2),
\end{align}
where
\begin{equation}\label{eq:Bim-Jijm}
{B}_{i,m}(x,\mu;\alpha_s)
=
\sum_{j}
\int_x^1 \frac{dz}{z}
\mathcal{J}_{ij,m}(z;\alpha_s)
f_j(x/z,\mu).
\end{equation}
One of the most computationally intensive aspects of the numerical calculation involves evaluating the integrated beam function, which requires convolving PDFs with the kernel $\mathcal{J}_{ij}$ for all possible parton pairs $i$ and $j$. To improve efficiency, we precompute the numerical coefficients ${B}_{i,m}(x,\mu;\alpha_s)$ on a two-dimensional grid in $x$ and $\mu$, then apply bicubic spline interpolation to accelerate the computations. This interpolation method fits cubic polynomials within each two-dimensional grid cell, ensuring smooth and accurate approximations.\footnote{Our grid for the interpolated beam function spans the $Q$--$x$ plane linearly with a resolution of $99 \times 499$, covering the range $2~\textrm{GeV} \le Q \le 100~\textrm{GeV}$ (99 points) and $0 < x < 1$ (499 points).}
The explicit expressions of $\mathcal{J}_{ij,m}(z;\alpha_s)$ are listed in Appendix~\ref{app:fixed-order-coeff}.

\subsection{Resummation to $\text{N}^3\text{LL}$}
The DIS thrust distribution contains large logarithms of $\tau_1^b$, which grow significant as $\tau_1^b\to 0$. To keep the perturbative expansion and theoretical uncertainty under control, these logarithms must be resummed to all orders in $\alpha_s$. In this work, although we perform the resummation to N$^3$LL accuracy, the resummation formulae in this subsection remain valid for higher orders. 
The resummation is achieved through the RG evolution of the hard, jet, beam, and soft functions in Eq.~\eqref{eq:tau1b-FT}, as outlined in Appendix~D of Ref.~\cite{Kang:2013nha}:
%---------------
\begin{align}
\label{eq:RG-simple}
\begin{split}
%---------------
H_q(y,Q^2,\mu)
&=
H_q(y,Q^2,\mu_H)
U_H(Q^2,\mu_H,\mu),
\\
{B}_q(t,x,\mu)
&=
\int dt'
{B}_q(t-t',x,\mu_B)
U_{B_q}(t',\mu_B,\mu),
\\
J_q(t,\mu)
&=
\int dt'
J_q(t-t',x,\mu_J)
U_{J_q}(t',\mu_J,\mu),
\\
S_\textrm{PT}(k,\mu)
&=
\int dk' 
S_\textrm{PT}(k-k',\mu_S)
U_S^2(k',\mu_S,\mu).
%---------------
\end{split}
\end{align}
%---------------
At any resummation order, dependence on the intermediate renormalization scale $\mu$ cancels, ensuring that the final cross section is independent of this choice. The evolution factors used in this resummation follow the conventions as in Ref.~\cite{Kang:2013nha}, with the additional terms required for N$^3$LL accuracy provided in Appendix~\ref{app:N3LL-evolution}.

In Eq.~\eqref{eq:RG-simple}, the fixed-order soft, beam, and jet functions, along with their evolution factors, are expressed in terms of the plus distributions [Eqs.~\eqref{eq:s-hemi-orig}, \eqref{eq:exp-jet-function}, \eqref{eq:Bi-plus}, and \eqref{eq:evol-summary}]. To evaluate the convolution integrals among these functions in Eq.~\eqref{eq:tau1b-FT}, we rescale the plus distributions in the fixed-order functions as follows: 
%---------------
\begin{align}
\label{eq:each-functions-rescale-2}
\begin{split}
%---------------
S(k_S,\mu_S) 
&=
\frac{1}{\xi}
\sum_{m=-1} S_m\left(\frac{\xi}{\mu_S}\right)
\mathcal{L}_m\left(\frac{k_S}{\xi}\right),
\\
{B}_i(t_B, x,\mu_B)
&=
\frac{1}{\xi Q}
\sum_{m=-1}
{B}_{i,m}\left(x,\mu_B,\frac{\xi Q}{\mu_B^2}\right)
\mathcal{L}_m\left(\frac{t_B}{\xi Q}\right),
\\
J(t_J,\mu_J)
&=
\frac{1}{\xi Q}
\sum_{m=-1}
{J}_m\left(\frac{\xi Q}{\mu_J^2}\right)
\mathcal{L}_m\left(\frac{t_J}{\xi Q}\right).
%---------------
\end{split}
\end{align}
%---------------
The coefficients $G_m(\lambda)$ for $G = \{S, B_i, J\}$ are derived from the original forms in 
Eqs.~\eqref{eq:s-hemi-orig}, \eqref{eq:exp-jet-function} and \eqref{eq:Bi-plus}, using the rescaling identity for plus distributions:
\begin{equation}\label{eq:rescale}
\lambda \mathcal{L}_n(\lambda u) = \sum_{k=0}^n
\binom{n}{k}\ln^k\lambda\, \mathcal{L}_{n-k}(u) 
+
\frac{\ln^{n+1}\lambda}{n+1}\delta(u).
\end{equation}
The rescaled coefficients are given by
\begin{align}
\begin{split}
G_{-1}(\lambda) &= G_{-1} + \sum_{k=0}^\infty 
G_k \frac{\ln^{k+1}\lambda}{k+1},
\\
G_\ell(\lambda) &= \sum_{k=0}^\infty
\frac{(\ell+k)!}{\ell!\,k!} G_{\ell+k} \ln^{k}\lambda.
\end{split}
\end{align}
These rescalings are independent of the choice of $\xi$, allowing us to choose any $\xi$ that simplifies the calculation.

After rescaling the functions as in Eq.~\eqref{eq:each-functions-rescale-2} and performing the plus distribution integrals using the identities in Eq.~(E8) of Ref.~\cite{Kang:2013nha}, we obtain the formula for the resummed singular cross section:
%---------------
\begin{align}\label{eq:singular-after-resummation}
%---------------
\sigma^\textrm{s}_\textrm{PT}(\tau_1^b)
={}&
\sigma_0^b
\frac{e^{\mathcal{K}-\gamma_\textrm{E}\Omega}}
{\Gamma(1+\Omega)}
\left(\frac{Q}{\mu_H}\right)^{\eta_H}
\left(\frac{\xi Q}{\mu_B^2}\right)^{\eta_{B}}
\left(\frac{\xi Q}{\mu_J^2}\right)^{\eta_{J}}
\left(\frac{\xi}{\mu_S}\right)^{2\eta_S}
\left(\frac{Q}{\xi}\right)
\nonumber
\\
&
\times
\sum_q
\bigg[
H_{q}(y,Q^2,\mu_H)
\sum_{ \substack{ m_1,m_2, \\ m_3=-1 } }
% \sum_{m_1,m_2,m_3=-1}
{J}_{m_1}\left(\frac{\xi Q}{\mu_J^2}\right)
{B}_{q,m_2}\left(x,\mu_B,\frac{\xi Q}{\mu_B^2}\right)
S_{m_3}\left(\frac{\xi}{\mu_S}\right)
\nonumber
\\
&
\quad
\times
\sum_{\ell_1= -1}^{m_1+m_2+1}
\sum_{\ell_2= -1}^{\ell_1+m_3+1}
\sum_{\ell_3=-1}^{\ell_2+1}
V_{\ell_1}^{m_1m_2} 
V_{\ell_2}^{\ell_1 m_3}
V_{\ell_3}^{\ell_2}(\Omega)
G\left(\frac{\tau_1^b Q}{\xi}\right)
+
(q\leftrightarrow \bar{q})
\bigg],
%---------------
\end{align}
%---------------
% \begin{align}\label{eq:singular-after-resummation}
% %---------------
% \sigma^\textrm{s}_\textrm{PT}
% ={}&
% \sigma_0^b
% \frac{e^{\mathcal{K}-\gamma_\textrm{E}\Omega}}
% {\Gamma(1+\Omega)}
% \left(\frac{Q}{\mu_H}\right)^{\eta_H}
% \left(\frac{\xi Q}{\mu_B^2}\right)^{\eta_{B}}
% \left(\frac{\xi Q}{\mu_J^2}\right)^{\eta_{J}}
% \left(\frac{\xi}{\mu_S}\right)^{2\eta_S}
% \left(\frac{Q}{\xi}\right)
% \nonumber
%  \\
% &
% \times
% \sum_{\{n_i\}}
% \frac{[\alpha_s(\mu_H)]^{n_0}
% [\alpha_s(\mu_J)]^{n_1}
% [\alpha_s(\mu_B)]^{n_2}
% [\alpha_s(\mu_S)]^{n_3}}{(4\pi)^{n_0+n_1+n_2+n_3}}
% \nonumber
% \\
% &
% \times
% \sum_q
% \bigg[
% H^{(n_0)}_{q}(y,Q^2,\mu_H)
% \sum_{m_1,m_2,m_3=-1}
% {J}_{m_1}^{(n_1)}\left(\frac{\xi Q}{\mu_J^2}\right)
% {B}_{q,m_2}^{(n_2)}\left(x,\mu_B,\frac{\xi Q}{\mu_B^2}\right)
% S_{m_3}^{(n_3)}\left(\frac{\xi}{\mu_S}\right)
% \nonumber
% \\
% &
% \quad
% \times
% \sum_{\ell_1= -1}^{m_1+m_2+1}
% \sum_{\ell_2= -1}^{\ell_1+m_3+1}
% \sum_{\ell_3=-1}^{\ell_2+1}
% V_{\ell_1}^{m_1m_2} 
% V_{\ell_2}^{\ell_1 m_3}
% V_{\ell_3}^{\ell_2}(\Omega)
% G\left(\frac{\tau_1^b Q}{\xi}\right)
% +
% (q\leftrightarrow \bar{q})
% \bigg],
% %---------------
% \end{align}
% %---------------
where the function $G\left(\frac{\tau_1^b Q}{\xi}\right)$
depends on whether the cross section is differential or cumulative:
\begin{align}
\label{eq:G_tau_sing}
G\left(\frac{\tau_1^b Q}{\xi}\right)
\equiv
\begin{cases}
\displaystyle
\mathcal{L}_{\ell_3}^{\Omega}\left(\frac{\tau_1^b Q}{\xi}\right)
&\textrm{for}~
\sigma^\textrm{s}_\textrm{PT}
=
\displaystyle
\frac{d\sigma^\textrm{s}_\textrm{PT}}{d\tau_1^b}
,
\\[2ex]
\displaystyle
G_{\ell_3}^{\Omega}\left(\frac{\tau_1^b Q}{\xi}\right)
&\textrm{for}~
\sigma^\textrm{s}_\textrm{PT}
=
\sigma_{\textrm{PT},c}^{\textrm{s}}.
\end{cases}
\end{align}
Here, $\frac{d\sigma^\textrm{s}_\textrm{PT}}{d\tau_1^b}$ is the differential singular cross section 
and $\sigma_{\textrm{PT},c}^{\textrm{s}}$ is the cumulative singular cross section defined in Eq.~\eqref{eq:cumulant-def}.
The evolution kernels $\mathcal{K}$ and $\Omega$, which resum large logarithms, are defined as
\begin{align}
\begin{split}
\mathcal{K} &\equiv K_H(\mu_H,\mu) + K_J(\mu_J,\mu) + K_B(\mu_B,\mu) + K_S(\mu_S,\mu),
\\
\Omega &\equiv \eta_J(\mu_J,\mu) + \eta_B(\mu_B,\mu) + 2\eta_S(\mu_S,\mu).
\end{split}
\end{align}
The explicit forms of $K_i$ and $\eta_i$ can be found in Appendix~D of Ref.~\cite{Kang:2013nha}, with the N$^3$LL extensions provided in Eq.~\eqref{eq:N3LL-extention-func}. The coefficients $V_k^n(\Omega)$ and $V_{k}^{mn}$, which arise from convolutions of the plus distributions with evolution factors, are given in Appendix~E of Ref.~\cite{Kang:2013nha}. 

In Eq.~\eqref{eq:singular-after-resummation}, the explicit $\tau_1^b$ dependence enters only through
$G\left(\frac{\tau_1^b Q}{\xi}\right)$. For the differential singular cross section, it is given by
%---------------
\begin{equation}
%---------------
\mathcal{L}_n^a(u)
=
\begin{cases}
\displaystyle
\delta(u),& \textrm{for $n=-1$,}
\\[2ex]
\displaystyle
\left[
\frac{\theta(u)\log^n u}{u^{1-a}}
\right]_+,& \textrm{for $n>-1$.}
\end{cases}
%---------------
\end{equation}
%---------------
For the cumulative singular cross section, 
$G^\Omega_{\ell_3}$ is given by
\begin{equation}
G^\Omega_{\ell_3}\left(\frac{\tau_1^b Q}{\xi}\right)
\equiv
\left(\frac{Q}{\xi}\right)
\int_0^{\tau_1^b}
d\tau_1
\mathcal{L}_{\ell_3}^{\Omega}\left(\frac{\tau_1 Q}{\xi}\right)
=
\int_0^{\frac{\tau_1^b Q}{\xi}}
du\,
\mathcal{L}_{\ell_3}^{\Omega}\left(u\right).
\end{equation}
For $\ell_3=-1$, we have $G^\Omega_{-1}\left(u\right)=1$, and for $\ell_3>-1$,
\begin{equation}\label{eq:def-of-G}
G^\Omega_{\ell_3}\left(u\right)
=
\frac{\gamma(1+\ell_3, -\Omega\log u)}{(-\Omega)^{1+\ell_3}},
\end{equation}
where $\gamma(s,t)$ is the lower incomplete gamma function:
\begin{equation}
\gamma(s,t) = \int_0^t du\, u^{s-1} e^{-u}. 
\end{equation}
Into the result for the resummed singular distribution in Eq.~\eqref{eq:singular-after-resummation}, we insert the appropriate $\tau_1^b$-dependent scales $\mu_i$ that minimize logs in the fixed-order functions and allow for smooth transitions to both nonperturbative (peak) and fixed-order (far-tail) regions of the distribution, and $R$ from the $R$-gap scheme after performing the $k$ convolution in Eq.~\eqref{eq:k-int-cumulant}. We will explain our exact choices for these scales in \sec{profile}.

\subsection{Power corrections and renormalon subtraction}
\label{sec:shape-renormalon}
In this subsection, we provide an overview of the key concepts related to nonperturbative power corrections and the treatment of renormalon subtractions for the $\tau_1^b$ distributions. 

To account for the nonperturbative soft radiation and hadronization occurring at the scale of $\Lambda_\textrm{QCD}$, we convolve the nonperturbative shape function $F(k)$ with the perturbative soft function $S_\textrm{PT}$ \cite{Ligeti:2008ac, Hoang:2007vb, Korchemsky:2000kp}:
\begin{equation}\label{eq:shape-intro}
S(k,\mu_S)
=
\int dk' S_\textrm{PT}(k-k',\mu_S) F(k').
\end{equation}
The shape function $F(k)$, which captures the nonperturbative physics, is peaked around $k\sim \Lambda_\textrm{QCD}$ and falls off exponentially for larger values of $k$. Plugging  Eq.~\eqref{eq:shape-intro} into the factorization formula in Eq.~(\ref{eq:tau1b-FT-2}), we get the following expression for the convolved cross section: 
%---------------
\begin{equation}
\label{eq:power-correction}
%---------------
\sigma(\tau_1^b)
=
\int dk\,
\sigma_\textrm{PT}
\left(
\tau_1^b
-
\frac{k}{Q}
\right)
F(k),
%---------------
\end{equation}
%---------------
where $\sigma(\tau_1^b)$ refers to either the differential or cumulative cross section, and
$\sigma_\textrm{PT}$ is the cross section computed with the perturbative soft function.

For the nonperturbative shape function $F(k)$, 
we adopt the parametrization proposed in Ref.~\cite{Ligeti:2008ac}, where $F(k)$ is expanded as a sum of basis functions:
%---------------
\begin{equation}
\label{eq:model-NP-soft}
%---------------
F(k) = 
\frac{1}{\lambda}
\left[
\sum_{n=0}^N
c_n 
f_n
\left(\frac{k}{\lambda}\right)
\right]^2\,,
%---------------
\end{equation}
%---------------
with basis functions $f_n$ given by:
\begin{equation}\label{eq:basis_function}
f_n(u) = 8\sqrt{\frac{2u^3(2n+1)}{3}}
e^{-2u} P_n
\left[
1-2\left(1+4u+8u^2+\frac{32}{3}u^3\right)e^{-4u}
\right],
\end{equation}
where $P_n$ represents the Legendre polynomials, defined by
\begin{equation}
P_n(u) = \frac{1}{2^n n!}
\frac{d^n}{du^n}
(u^2-1)^n.
\end{equation}
These basis functions are orthonormal on the interval $(0,\infty)$,
\begin{equation}
\int_0^\infty du\, f_m(u) f_n(u) = \delta_{mn}.
\end{equation}
Consequently, the normalization condition of the shape function, $\int dk \, F(k)=1$, imposes the constraint $\sum_{i}c_i^2 = 1$.
The characteristic scale
$\lambda$, typically of order $\mathcal{O}(\Lambda_\textrm{QCD})$,
is an additional parameter if the sum is truncated at a finite $N$.
In general, the parameters $c_i$ and $\lambda$
should be determined by fitting to experimental data
in the peak region.
Our shape function, which encapsulates the nonperturbative physics, is designed to be universal across $x$ and $Q$. 

In the region where $ Q\tau_1^b \gg \Lambda_\textrm{QCD}$, the shape function $F$ can be expanded for $k\gg \Lambda_\textrm{QCD}$ in terms of nonperturbative matrix elements of operators \cite{Mateu:2012nk}:
\begin{equation}
F(k) = \delta(k) - \delta'(k) \bar{\Omega}_1
+
\mathcal{O}\left(\frac{\alpha_s\Lambda_\textrm{QCD}}{k^2}\right)
+
\mathcal{O}\left(\frac{\Lambda_\textrm{QCD}^2}{k^3}\right).
\end{equation}
This leads to the OPE for the cross section Eq.~\eqref{eq:power-correction}
\cite{Kang:2013nha},
%---------------
\begin{equation}
\label{eq:OPE_shape}
%---------------
\sigma(\tau_1^b)
=
\left[
\sigma_\textrm{PT}(\tau_1^b)
-
\frac{2\bar\Omega_1}{Q}
\frac{d}{d\tau_1^b}\sigma(\tau_1^b)
\right]
\left[
1+\mathcal{O}\left(\frac{\alpha_s \Lambda_\textrm{QCD}}{Q\tau_1^b}\right)
+\mathcal{O}\left(\frac{\Lambda_\textrm{QCD}^2}{Q^2{\tau_1^b}^2}\right)
\right],
%---------------
\end{equation}
%---------------
where the first moment of the shape function, $\bar{\Omega}_1$, is defined as
\cite{Abbate:2010xh}
%---------------
\begin{equation}
\label{eq:no-gap}
%---------------
2\bar{\Omega}_1
=
\int dk\, k F(k).
%---------------
\end{equation}
%---------------
In the OPE expression Eq.~\eqref{eq:OPE_shape}, the first set of power corrections arises from perturbative effects to the leading power level, while the second set reflects purely nonperturbative corrections at subleading order. This result confirms that, in the tail region, the distribution experiences a shift $\tau_1^b\to \tau_1^b - 2\bar{\Omega}_1/Q$.
Notably, this first moment $\bar{\Omega}_1$ is a nonperturbative matrix element defined in $\overline{\textrm{MS}}$ scheme
\cite{Abbate:2010xh,Kang:2013nha}.

However, it is well-known that the partonic soft function $S_\textrm{PT}(k,\mu)$, calculated perturbatively in $\overline{\textrm{MS}}$ scheme, suffers from an $\mathcal{O}(\Lambda_\textrm{QCD})$ renormalon ambiguity~\cite{Hoang:2007vb}. This arises because the partonic threshold at $k=0$ in $S_\textrm{PT}(k,\mu)$ differs from the physical hadronic threshold. Furthermore, the nonperturbative matrix element $\bar{\Omega}_1$ also contains an $\mathcal{O}(\Lambda_\textrm{QCD})$ renormalon. To reduce numerical instabilities, we subtract these $\mathcal{O}(\Lambda_\textrm{QCD})$ renormalon order-by-order in $\alpha_s$ from both the perturbative soft function and the first moment $\bar\Omega_1$. 

To manage this ambiguity, we introduce a gap parameter $\bar\Delta$, representing the minimum hadronic energy deposit \cite{Hoang:2007vb}. The shape function 
which builds in this gap parameter is
%---------------
\begin{equation}
%---------------
F(k)
\to 
F(k-2\bar\Delta).
%---------------
\end{equation}
%---------------
Using this gapped shape function, 
we can rewrite the relation in 
Eq.~(\ref{eq:no-gap}) as 
%---------------
\begin{equation}
\label{eq:relation-omega1-delta}
%---------------
\int dk\, k F(k-2\bar\Delta)
=
2\bar\Delta
+
\int_{0}^\infty dk\, k
F(k)
=
2\bar{\Omega}_1.
%---------------
\end{equation}
%---------------
Here, $\bar\Delta$ carries the renormalon 
ambiguity in $\bar{\Omega}_1$. Then,
we define a renormalon-free definition of $\Omega_1$ by splitting $\bar\Delta$ into a nonperturbative
component ${\Delta}(R,\mu)$, free from the 
$\mathcal{O}(\Lambda_\textrm{QCD})$ renormalon ambiguity, and a perturbative series $\delta(R,\mu)$ with the same renormalon ambiguity as $\bar\Omega_1$:
%---------------
\begin{equation}\label{eq:Delta-sep}
%---------------
\bar\Delta = {\Delta}(R,\mu_S)
+
\delta(R,\mu_S).
%---------------
\end{equation}
%---------------
The new scale dependence on $R$ will arise due to a scheme choice needed to define the condition for making $S_\text{PT}$ and the gap renormalon-free; this will become clear in our chosen method that we describe below. The scale $\mu_S$ is inherited from perturbative renormalization of $S_\text{PT}$ itself.
Thus, Eq.~\eqref{eq:relation-omega1-delta} gives the renormalon-subtracted definition of $\Omega_1$ as follows:
%---------------
\begin{equation}
\label{eq:renormalon-free-omega}
%---------------
2\Omega_1(R,\mu_S)
=
\int dk\, k F(k- 2{\Delta}(R,\mu_S))
=
2{\Delta}(R,\mu_S) + 
\int dk\, k F(k),
%---------------
\end{equation}
%---------------
where $\Omega_1(R,\mu_S)$ is now renormalon-free. Its scheme
conversion formula from $\overline{\textrm{MS}}$ to the new scheme is given by
%---------------
\begin{equation}
%---------------
\Omega_1(R,\mu_S) = \bar{\Omega}_1 - \delta(R,\mu_S). 
%---------------
\end{equation}
%---------------
$\Omega_1(R,\mu_S)$ depends on the subtraction series $\delta(R,\mu_S)$, which defines the scheme.
In this work, we use the $R$-gap scheme adopted in Refs.~\cite{Hoang:2008fs, Jain:2008gb, Hoang:2008yj, Abbate:2010xh} where
%---------------
\begin{equation}
%---------------
\delta(R,\mu_S)
=
\frac{R}{2}
e^{\gamma_\textrm{E}}
\frac{d}{d\log(ix)}
[\log S_\textrm{PT}(x,\mu_S)]
\big|_{x=(iRe^{\gamma_\textrm{E}})^{-1}}.
%---------------
\end{equation}
%---------------
Here $S_\textrm{PT}(x,\mu_S)$ is the position space perturbative soft function, and
$R$ is a cutoff parameter used to remove the infrared renormalon. 

With these $R$-gap scheme parameters, we can rewrite the soft function in Eq.~\eqref{eq:shape-intro} in a form that is free from the renormalon ambiguities:
\begin{align}\label{eq:shape-function-after-R-gap}
\begin{split}
S(k,\mu_S)
&=
\int dk' 
S_\textrm{PT}\left(k-k',\mu_S\right)
F(k'-2\bar\Delta)
\\
&
=
\int dk' 
\left[
e^{-2\delta(R,\mu_S)(\partial/\partial k)}
S_\textrm{PT}
\left(k-k',\mu_S\right)
\right]
F(k'-2{\Delta}(R,\mu_S)).
\end{split}
\end{align}
Here, in the first line, we introduce the gap parameter $\bar\Delta$,
and in the second line, we use the decomposition of $\bar\Delta$ from Eq.~\eqref{eq:Delta-sep}, shifting $k'$ by $2\delta(R,\mu_S)$. This subtraction is expressed using an exponential operator, written in terms of a derivative with respect to $k$, which performs the perturbative subtraction on the $S_\textrm{PT}(k,\mu)$ computed in $\overline{\textrm{MS}}$ scheme. 

The perturbative series for the subtraction terms $\delta(R,\mu)$ is given by
%---------------
\begin{equation}
%---------------
\delta(R,\mu)
=
Re^{\gamma_\textrm{E}}
\sum_{i=1}^\infty
\left[\frac{\alpha_s(\mu)}{4\pi}\right]^i
\delta_i(R,\mu).
%---------------
\end{equation}
%---------------
We use the one- and two-loop coefficients for the renormalon subtractions $\delta_i$, given by \cite{Hoang:2008fs, Abbate:2010xh}:\footnote{It is worth noting that in Eq.~(56) of \cite{Hoang:2008fs}, the factor $Re^{\gamma_\textrm{E}}$ is omitted. }
%---------------
\begin{align}
\begin{split}
%---------------
\delta_1(R,\mu_S)
={}&
-8C_FL_R
\\
\delta_2(R,\mu_S)
={}&
C_A C_F
\left[-\frac{808}{27}-\frac{22}{9}\pi^2+28\zeta_3+
\left(
-\frac{536}{9}+\frac{8}{3}\pi^2
\right)L_R
-\frac{88}{3}L_R^2
\right]
\\
&
+
C_F T_F n_f
\left(
\frac{224}{27}+\frac{8}{9}\pi^2
+\frac{160}{9}L_R +\frac{32}{3}L_R^2
\right),
%---------------
\end{split}
\end{align}
%---------------
where $\zeta_3$ (Ap\'ery's constant) is defined as the value of the Riemann zeta function at argument $3$, and $L_R \equiv \log(\mu_S/R)$. 

It is appropriate for the nonperturbative cutoff parameter $R$ to be around 1~GeV, which corresponds to the nonperturbative scale, ensuring $\Omega_1\sim \Lambda_\textrm{QCD}$. However, in the tail region, where $\mu_S\sim Q\tau_1^b\gg 1~\textrm{GeV}$, the logarithm $L_R=\log(\mu_S/R)$ can become large. 
To avoid large logarithmic contributions in the subtraction terms $\delta_i(R,\mu_S)$, we should choose $R\sim \mu_S$ making the cutoff scale $R$ dependent on $\tau_1^b$, similar to $\mu_S$. 

This creates a conflict between the criteria $R\sim 1~\textrm{GeV}$ (for nonperturbative effects) and $R\sim \mu_S$ (to minimize large logarithms). To address this conflict and sum the large logarithms while keeping $\Delta(R,\mu_S\sim R)$ free of renormalon ambiguities, we use $R$-evolution \cite{Hoang:2008yj,Hoang:2008fs}. In this approach, ${\Delta}(R,\mu)$ follow the following $R$ and $\mu$ evolution equations:
%---------------
\begin{align}
\label{eq:RG-for-del}
\begin{split}
%---------------
R \frac{d}{dR}
{\Delta(R,R)} &= 
-R\sum_{n=0}^\infty \gamma_n^R
\left(
\frac{\alpha_s(R)}{4\pi}
\right)^{n+1},
\\
\mu \frac{d}{d\mu}
{\Delta(R,\mu)}
&=
2 R e^{\gamma_\textrm{E}}
\sum_{n=0}^\infty 
\Gamma_n^\textrm{cusp}
\left(
\frac{\alpha_s(\mu)}{4\pi}
\right)^{n+1}.
%---------------
\end{split}
\end{align}
%---------------
The anomalous dimension for the $\mu$ evolution is given by the cusp anomalous dimension as $\gamma_\Delta^{\mu} =-2e^{\gamma_\textrm{E}}\Gamma^\textrm{cusp}[\alpha_s]$
and the anomalous dimension coefficients for the $R$ evolution, $\gamma_n^R$, are provided in Appendix~\ref{app:renormalon-sub}.

The solution for Eq.~\eqref{eq:RG-for-del} is
%---------------
\begin{align}
\label{eq:evolv-of-bar-del}
\begin{split}
%---------------
\Delta(R,\mu)
={}&
\Delta(R_\Delta,\mu_\Delta)
+2R_\Delta e^{\gamma_\textrm{E}}
\eta_{\Gamma}(\mu_\Delta, R_\Delta)
+
D[\alpha_s(R),\alpha_s(R_\Delta)]
+
2Re^{\gamma_\textrm{E}}
\eta_{\Gamma}(R,\mu).
%---------------
\end{split}
\end{align} 
%---------------
Here, the first term represents the parameter $\Delta(R_\Delta,\mu_\Delta)$ at the reference scales $\mu=\mu_\Delta$ and $R=R_\Delta$, the second term accounts for the $\mu$-evolution from $\mu=\mu_\Delta$ to $\mu=R_\Delta$, and the third term accounts for the $R$-evolution from $(\mu,R)=(R_\Delta,R_\Delta)$ to $(R,R)$, and the last term accounts for the $\mu$-evolution from $\mu=R$ to $\mu$.
The evolution factor $\eta_\Gamma$ is given in Eq.~\eqref{eq:N3LL-extention-func}, and the $R$-evolution factor, $D[\alpha_s(R),\alpha_s(R_\Delta)]$, is given in Appendix~\ref{app:renormalon-sub}. 
Note that the $R$ evolution factor vanishes 
at lowest order (NLL in double log counting),
and is nonvanishing from NNLL. 
The RGE solution for $\Delta(R,\mu_S)$ yields a similar solution for a running $\Omega_1(R,\mu_S)$ through Eq.~(\ref{eq:renormalon-free-omega}). 

In this work, we use the input parameters $\Delta(R_\Delta,\mu_\Delta)=0.05~\textrm{GeV}$ and $\Omega_1(R_\Delta, \mu_\Delta) = 0.5~\textrm{GeV}$ at the reference scales $R_\Delta=\mu_\Delta = 2~\textrm{GeV}$. These values are illustrative but reasonable for the purposes of our analysis.\footnote{It is important to note that the values of $\Omega_1$ and $\Delta$ for DIS may differ from those for $e^+e^-$ collisions (i.e. \cite{Abbate:2010xh}), and should ultimately be determined by comparing theoretical predictions with experimental data.}
We adopt the simplest implementation of the shape function, setting $c_0=1$ in Eq.~\eqref{eq:model-NP-soft}. 
With an appropriate choice of $\lambda$, this accurately describes the tail region, where only the parameter $\Omega_1$ is relevant. 
However, it is important to note that the higher-order coefficients in the shape function, $c_{i>0}$, can be added to improve the description in the peak region. A demonstration of the impact of these higher coefficients is provided in Appendix~\ref{app:with_c2}.

The parameter $\lambda$ is then determined from Eq.~(\ref{eq:renormalon-free-omega}) as
%---------------
\begin{align}
\label{eq:omega1-c0-1}
%---------------
\lambda = 2\left[\Omega_1(R,\mu_S) - {\Delta}(R,\mu_S)\right]
=
2\left[\Omega_1(R_\Delta,\mu_\Delta) - {\Delta}(R_\Delta,\mu_\Delta)\right].
%---------------
\end{align}
%---------------
Note that the dependence on $R$ and $\mu_S$ cancels between $\Omega_1$ and $\Delta$, allowing $\lambda$ to be determined directly from their values at the reference scales. With $c_0=1$, all higher moments $\Omega_{n>1}$ can then be expressed as functions of $\Omega_1(R,\mu_S)$ and $\Delta(R,\mu_S)$. For example, the second moment $\Omega_2$ is given by
\begin{align}\label{eq:2nd-moment}
\begin{split}
\Omega_2(R,\mu_S) &\equiv
\int dk\left(\frac{k}{2}\right)^2 F(k-2\Delta(R,\mu_S))
\\
&=
\frac{[\Delta(R,\mu_S)]^2-2\Delta(R,\mu_S)\Omega_1(R,\mu_S)+5[\Omega_1(R,\mu_S)]^2}{4}.
\end{split}
\end{align}
However, as shown in Appendix~\ref{app:with_c2}, $\Omega_2$ exhibits a strong correlation with the higher-order coefficient $c_2$. Therefore, variations in $\Omega_2$ can be introduced by setting $c_2$ to a nonzero value while keeping $\Omega_1$ fixed. 

\begin{figure}
    \centering
    \includegraphics[width=0.5\linewidth]{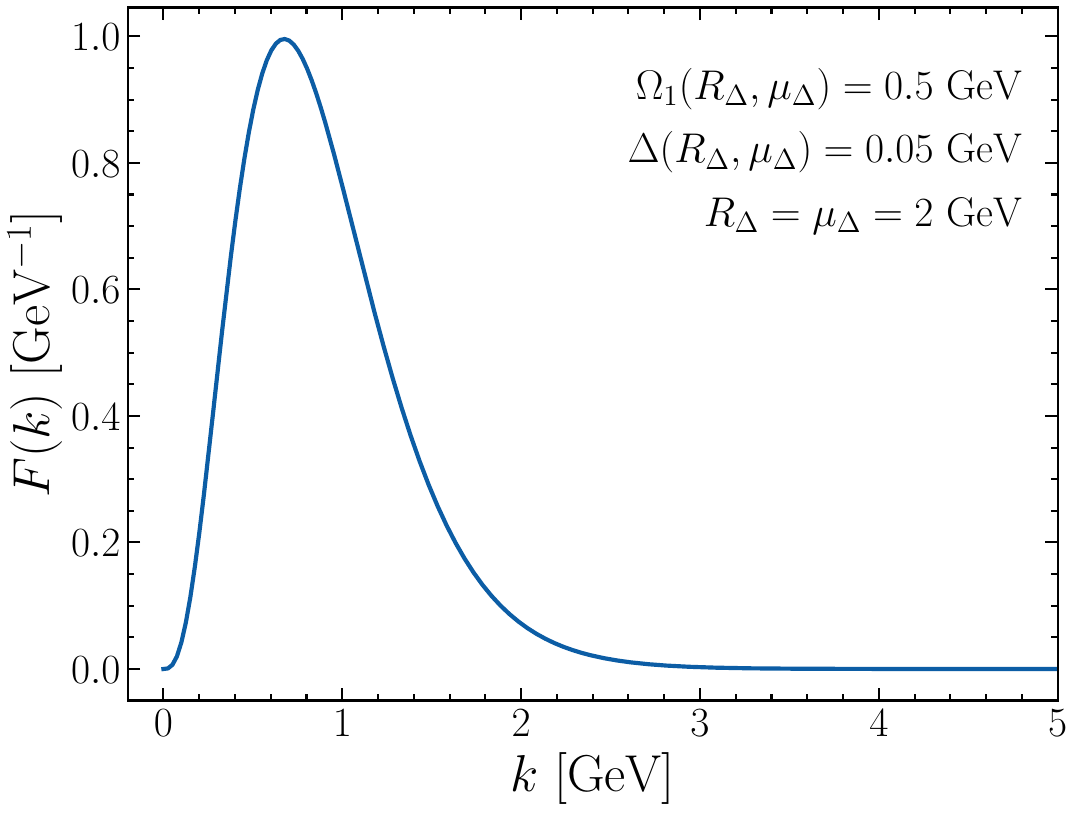}
    \vspace{-1em}
    \caption{An illustrative shape function $F(k)$ with $c_0=1$ and $c_{i\ge 1}=0$.
    This function can be fit from small $\tau_1^b$ experimental data, but for the purpose of the analysis here we keep it fixed. }
    \label{fig:fk}
\end{figure}

This implementation correctly keeps the effect of the first moment $\Omega_1$, that is, the translation of the distribution in the tail region [Eq.~\eqref{eq:OPE_shape}]. One can compare the form of the soft shape function employed in \cite{Cao:2024ota}, where the relation of its parameters to the single universal parameter controlling the first moment is not so immediately manifest, though the universality relations can still be implemented there.
Fig.~\ref{fig:fk} illustrates the shape function used in this analysis, with parameter $\lambda = 0.9~\textrm{GeV}$. 
To obtain the results from our codes that do not include nonperturbative corrections, we can simply turn them off by setting $F(k) = \delta(k)$ and $\Delta=\delta=0$. Correspondingly, with this setting, we find that $\Omega_1 = 0$ from Eq.~(\ref{eq:renormalon-free-omega}). 
The functional form of $R(\tau_1^b)$ used in this work is discussed in Sec.~\ref{sec:profile}.

\subsection{Final formula for singular distribution}

Incorporating the renormalon subtraction scheme encapsulated in \eq{shape-function-after-R-gap} and the $R$-evolution in \eq{evolv-of-bar-del} into  the cumulative $\tau_1^b$ distribution in \eq{singular-after-resummation}, we obtain our
final form  
of the singular distribution with renormalon subtraction:
%---------------
\begin{align}\label{eq:singular-cumulant-after-renormalon}
%---------------
&
\sigma_{c}^\textrm{s}(\tau_1^b)
\nonumber 
\\
={}&
\sigma_0^b
\frac{e^{\mathcal{K}-\gamma_\textrm{E}\Omega}}
{\Gamma(1+\Omega)}
\left(\frac{Q}{\mu_H}\right)^{\eta_H}
\left(\frac{\xi(\tau_1^b)Q}{\mu_B^2}\right)^{\eta_{B}}
\left(\frac{\xi(\tau_1^b)Q}{\mu_J^2}\right)^{\eta_{J}}
\left(\frac{\xi(\tau_1^b)}{\mu_S}\right)^{2\eta_S}
\nonumber 
\\
&
\times
\sum_q
\bigg[
H_{q}(y,Q^2,\mu_H)
%\sum_{m_1,m_2,m_3=-1}
\sum_{ \substack{ m_1,m_2, \\ m_3=-1 } }
{J}_{m_1}\left(\frac{\xi(\tau_1^b)Q}{\mu_J^2}\right)
{B}_{q,m_2}\left(x,\mu_B,\frac{\xi(\tau_1^b) Q}{\mu_B^2}\right)
S_{m_3}\left(\frac{\xi(\tau_1^b)}{\mu_S}\right)
\nonumber
\\
&
\quad
\times
\sum_{\ell_1= -1}^{m_1+m_2+1}
\sum_{\ell_2= -1}^{\ell_1+m_3+1}
\sum_{\ell_3=-1}^{\ell_2+1}
V_{\ell_1}^{m_1m_2} 
V_{\ell_2}^{\ell_1 m_3}
V_{\ell_3}^{\ell_2}(\Omega)
I_{\ell_3}^{\Omega}
(\tau_1^b)
+
(q\leftrightarrow \bar{q})
\bigg].
%---------------
\end{align}
%---------------
% %---------------
% \begin{align}\label{eq:singular-cumulant-after-renormalon}
% %---------------
% &
% \sigma_{c}^\textrm{s}(\tau_1^b)
% \nonumber 
% \\
% ={}&
% \sigma_0^b
% \frac{e^{\mathcal{K}-\gamma_\textrm{E}\Omega}}
% {\Gamma(1+\Omega)}
% \left(\frac{Q}{\mu_H}\right)^{\eta_H}
% \left(\frac{\xi(\tau_1^b)Q}{\mu_B^2}\right)^{\eta_{B}}
% \left(\frac{\xi(\tau_1^b)Q}{\mu_J^2}\right)^{\eta_{J}}
% \left(\frac{\xi(\tau_1^b)}{\mu_S}\right)^{2\eta_S}
% \nonumber 
%  \\
% &
% \times
% \sum_{\{n_i\}}
% \frac{[\alpha_s(\mu_H)]^{n_0}
% [\alpha_s(\mu_J)]^{n_1}
% [\alpha_s(\mu_B)]^{n_2}
% [\alpha_s(\mu_S)]^{n_3}}{(4\pi)^{n_0+n_1+n_2+n_3}}
% \nonumber
% \\
% &
% \times
% \sum_q
% \bigg[
% H^{(n_0)}_{q}(y,Q^2,\mu_H)
% \sum_{m_1,m_2,m_3=-1}
% {J}_{m_1}^{(n_1)}\left(\frac{\xi(\tau_1^b)Q}{\mu_J^2}\right)
% {B}_{q,m_2}^{(n_2)}\left(x,\mu_B,\frac{\xi(\tau_1^b) Q}{\mu_B^2}\right)
% S_{m_3}^{(n_3)}\left(\frac{\xi(\tau_1^b)}{\mu_S}\right)
% \nonumber
% \\
% &
% \quad
% \times
% \sum_{\ell_1= -1}^{m_1+m_2+1}
% \sum_{\ell_2= -1}^{\ell_1+m_3+1}
% \sum_{\ell_3=-1}^{\ell_2+1}
% V_{\ell_1}^{m_1m_2} 
% V_{\ell_2}^{\ell_1 m_3}
% V_{\ell_3}^{\ell_2}(\Omega)
% I_{\ell_3}^{\Omega}\left[\xi(\tau_1^b)\right]
% +
% (q\leftrightarrow \bar{q})
% \bigg].
% %---------------
% \end{align}
% %---------------
Here, our parameter choice $\xi(\tau_1^b) \equiv \tau_1^b Q-2\Delta(R,\mu_S)$ simplifies the rescaling of the plus distributions in the perturbative and nonperturbative functions, as it naturally aligns with the gap subtraction in the renormalon formalism. This choice allows for efficient convolution with the shape function while incorporating the renormalon subtraction as follows:\footnote{In obtaining this form from Eq.~\eqref{eq:shape-function-after-R-gap}, we implicitly performed integration by parts with respect to $k$, to move all the differential operators effecting the renormalon subtraction onto the shape function $F$ instead of the perturbative soft function. The boundary terms vanish because the shape function decays asymptotically: it follows a power-law behavior as $k\to 0$ and an exponential behavior as $k\to \infty$.}
%---------------
\begin{equation}
%---------------
I_{\ell_3}^{\Omega}(\tau_1^b)
=
\xi(\tau_1^b)
\int_0^1 du\,
G_{\ell_3}^{\Omega}
\left(u\right)
\left\{
\exp\left[{\frac{2\delta(R,\mu_S)}
{\xi(\tau_1^b)}\frac{d}{du}}\right]
F
\left[
\xi(\tau_1^b)(1-u)
\right]
\right\},
%---------------
\end{equation}
%---------------
where $G_\ell^\Omega(u)$ is defined in Eq.~\eqref{eq:def-of-G}. 
To properly carry out the renormalon subtraction, it is necessary to expand both the SCET fixed-order functions and the renormalon subtraction terms in $\delta(R,\mu_S)$ perturbatively in $\alpha_s$ and truncate them consistently to the order of accuracy at which the cross section is compute (see Refs.~\cite{Abbate:2010xh,Bell:2023dqs}). 

To derive the differential $\tau_1^b$ singular distribution, we apply Eq.~\eqref{eq:cumulant-to-differential}, which relates the cumulative cross section to its differential counterpart through the finite difference method. For the numerical analysis of the resummed, renormalon-subtracted singular contributions, we developed the two independent implementations within our collaboration. The first, an updated \texttt{Mathematica} \cite{Mathematica} code, builds upon previous works \cite{Kang:2014qba, Kang:2015swk}. The second, a newly developed \texttt{Python} code, incorporates parallelization and memory optimization techniques, achieving a tenfold speed improvements over the \texttt{Mathematica} version.\footnote{For example, the \texttt{Python} code can generate the full $\tau_1^b$ distributions (with 422 bins), including renormalon subtractions, in under 10 minutes on \textit{Apple M3 Max} with 16 physical cores.} The two codes demonstrate excellent agreement, with differences in the $\tau_1^b$ singular distributions at the level of 0.1\% in all relevant regions of $\tau_1^b$ where we work.

\section{Nonsingular distribution}
\label{sec:nonsing}

The singular contributions discussed in the previous section constitute the dominant contributions to the description of dijet-like events,  
where $\tau_1^b$ lies in the tail region,  
$\Lambda_\textrm{QCD}/Q \ll \tau_1^b \ll 1$. There are further contributions multiplied by additional powers of $\tau_1^b$, which are power corrections when $\tau_1^b\ll 1$, but are the same size as the dijet contributions when $\tau_1^b\sim 1$. We call these additional terms ``nonsingular''.  
We determine these nonsingular contributions using fixed-order cross sections computed up to NLO [$\mathcal{O}(\alpha_s^2)$] accuracy. 
By combining the singular contribution derived in the previous section with the nonsingular contribution discussed here, as introduced in \eq{sing-plug-ns}, we obtain the complete N$^3$LL + NLO [$\mathcal{O}(\alpha_s^2)$] prediction, accurate across the full $\tau_1^b$ range. 

The nonsingular cross section is obtained by subtracting the fixed-order singular contribution $d\sigma^\textrm{s, fixed}_\textrm{PT}$, from the fixed-order QCD cross section $d\sigma^\textrm{QCD}_\textrm{PT}$:
\begin{equation}
\label{eq:def-of-nonsingular}
\frac{d\sigma^\textrm{ns}_\textrm{PT}}{d\tau_1^b}
\equiv
\frac{d\sigma^\textrm{QCD}_\textrm{PT}}{d\tau_1^b}
-
\frac{d\sigma^\textrm{s, fixed}_\textrm{PT}}{d\tau_1^b}.
\end{equation}
We take the analytic nonsingular cross section at $\mathcal{O}(\alpha_s)$  
from Ref.~\cite{Kang:2014qba}.  
At $\mathcal{O}(\alpha_s^2)$, we compute the QCD distribution using \texttt{NLOJet++} \cite{Nagy:2001xb, Nagy:2003tz}, which implements the Catani-Seymour dipole subtraction method \cite{Catani:1996vz}.
To ensure numerical accuracy,  
we validate the \texttt{NLOJet++} results at $\mathcal{O}(\alpha_s)$ by comparing them to the analytic result  
from Ref.~\cite{Kang:2014qba}. In the remainder of this section, any time we refer to the QCD, singular, or non-singular cross section, we mean its fixed-order (not resummed) expansion.

The most computationally intensive aspect of the analysis is the MC integration required to obtain the nonsingular cross section using \texttt{NLOJet++}. As \texttt{NLOJet++} is not natively parallelized, a single run with $10^9$ NLO events takes approximately 24 hours on a single-core CPU. High-quality nonsingular distributions typically require at least $10^{12}$ MC events, amounting to around 24,000 CPU hours in total. 
To handle this computational demand, all numerical runs of \texttt{NLOJet++} were executed on a high-performance-computing (HPC) cluster, ensuring the precision necessary for our analysis.

\begin{figure}
    \centering
    \vspace{-1em}
    \includegraphics[width=1\linewidth]{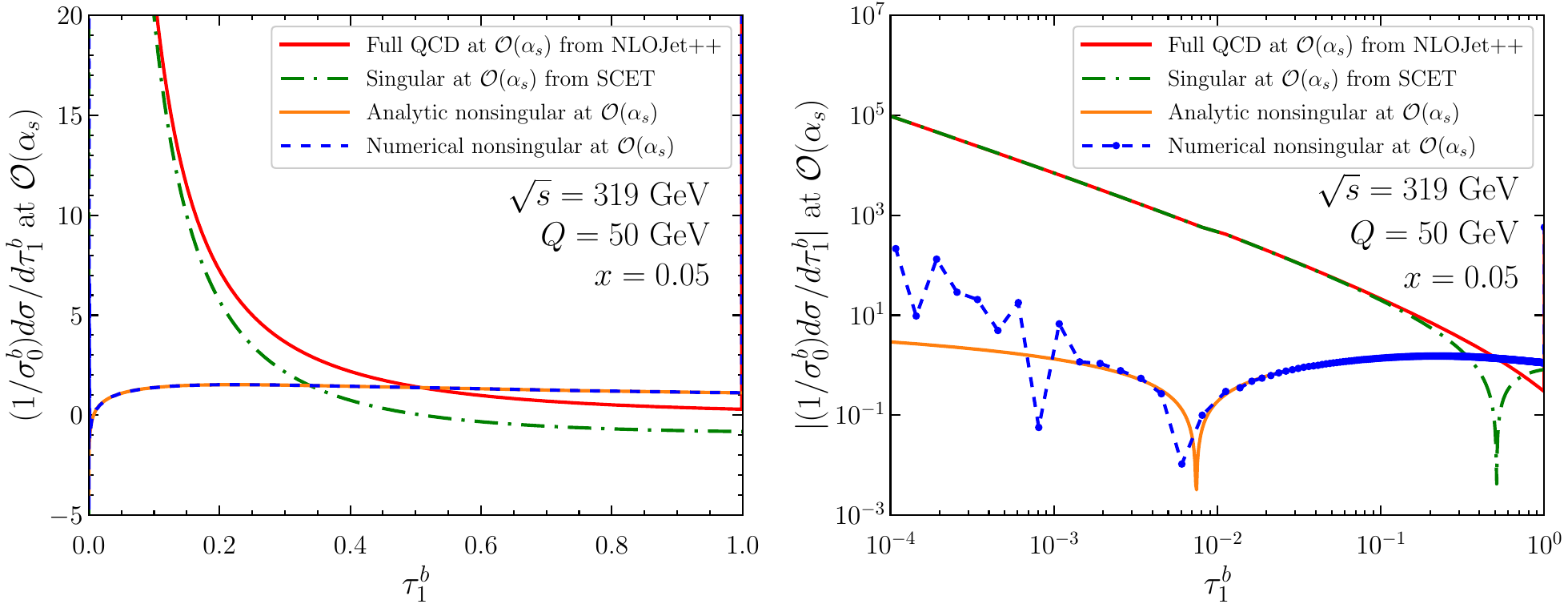}
    \vspace{-2em}
    \caption{Components of the $\mathcal{O}(\alpha_s)$ fixed-order cross section at $\sqrt{s}=319~\textrm{GeV}$, $Q=50~\textrm{GeV}$, $x=0.05$ at $\mu=Q$.
    The left plot shows the cross sections in linear scale for $\tau_1^b$, while the right plot displays the absolute values of the cross sections in logarithmic scale.
%\\
}
\hspace{0.2cm}
    \label{fig:ns_LO}
    \centering
    \includegraphics[width=1\linewidth]{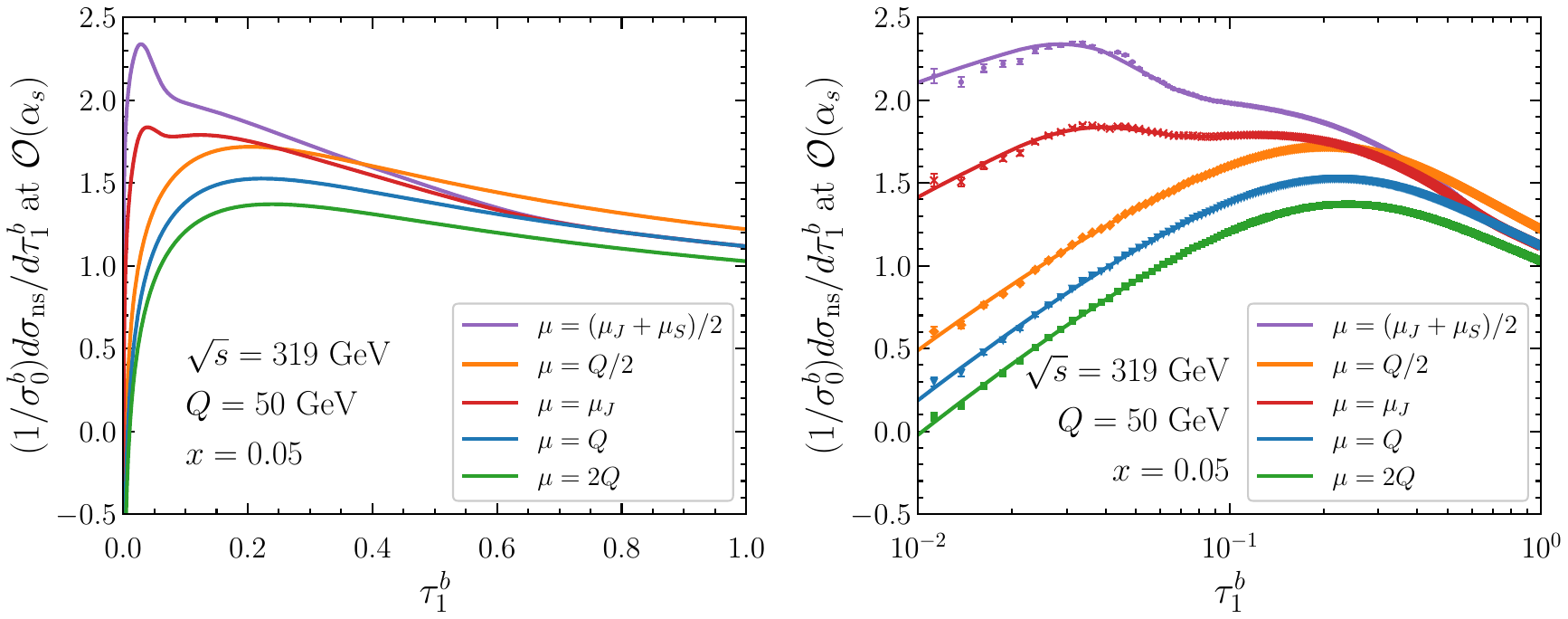}
    \vspace{-2em}
    \caption{Nonsingular $\mathcal{O}(\alpha_s)$ contributions to the cross section, reconstructed for five scale choices, with $\sqrt{s}=319~\textrm{GeV}$, $Q=50~\textrm{GeV}$, $x=0.05$. The left plot shows results with a linear scale for $\tau_1^b$, while the right plot shows the log scale. Points with error bars represent numerical results from \texttt{NLOJet++}. Solid lines represents the analytic nonsingular cross sections from Ref.~\cite{Kang:2014qba}.}
    \label{fig:ns_LO_scale_variations}
\end{figure}

In Fig.~\ref{fig:ns_LO}, we display the $\mathcal{O}(\alpha_s)$ terms of the full QCD cross section, the singular cross section from SCET, and the nonsingular cross section in linear (left) and logarithmic (right) scales as a function of $\tau_1^b$. These results are based on $10^{12}$ Born-level MC events with the infrared cutoff $10^{-10}$. As expected, the singular cross section converges with the full QCD cross section as $\tau_1^b\to 0$, and the nonsingular results from \texttt{NLOJet++} match the analytic results well for $\tau_1^b>0.01$.
Note that for $\tau_1^b \gtrsim 0.5$, the fixed-order singular cross section becomes negative, highlighting the necessity of including the nonsingular contribution at and beyond this point.\footnote{This indicates the breakdown of the two-jet-like description for events with $\tau_1^b> 0.5$.}

To estimate the perturbative uncertainties in the nonsingular cross sections  (discussed in Sec.~\ref{sec:profile}), we consider the five scale variations for $\mu$.  Fig.~\ref{fig:ns_LO_scale_variations} presents these  variations at $\mathcal{O}(\alpha_s)$, comparing numerical and analytic results. We observe strong agreement for $\tau_1^b>0.01$, but small numerical instabilities begin to arise at smaller values. 
Such instabilities are not present in the analytic results, so we use the analytic results at $\mathcal{O}(\alpha_s)$ in our code.

\begin{figure}
    \centering
    \vspace{-1em}
    \includegraphics[width=1\linewidth]{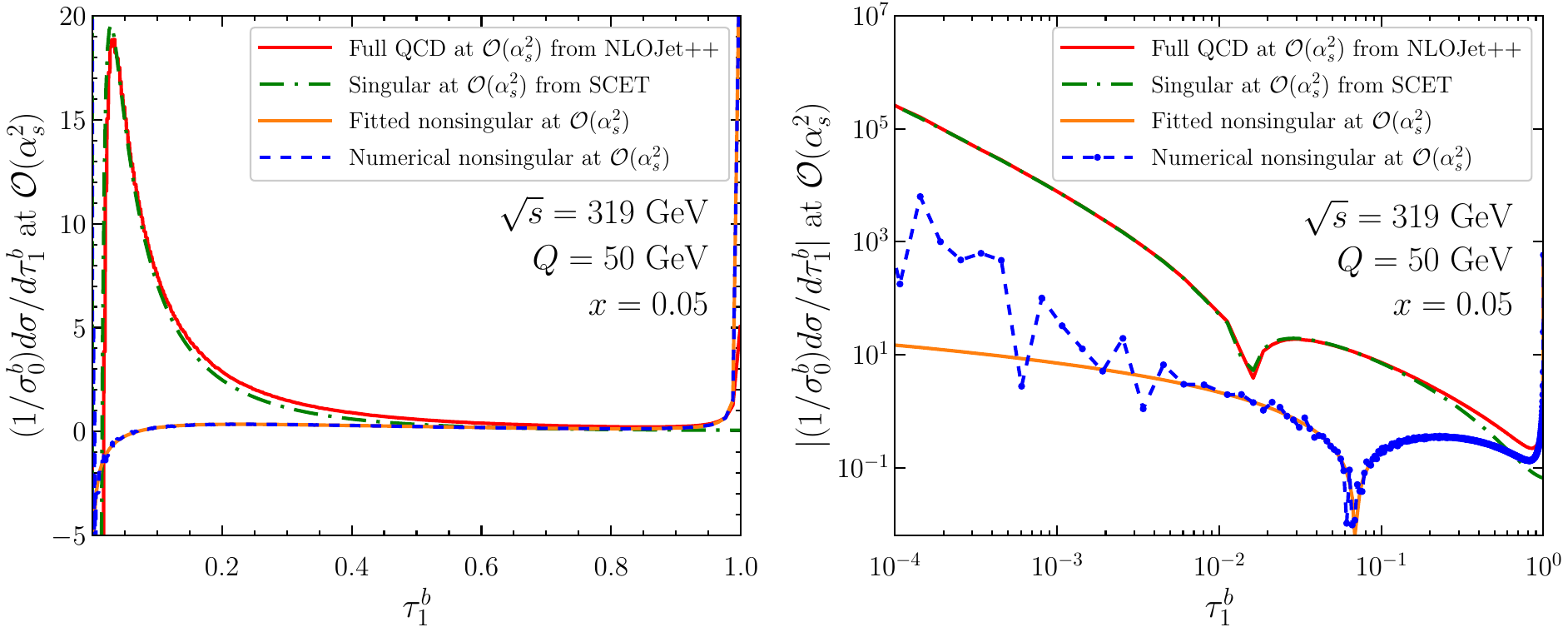}
    \vspace{-2em}
    \caption{Components of the $\mathcal{O}(\alpha_s^2)$ fixed-order cross section at $\sqrt{s}=319~\textrm{GeV}$, $Q=50~\textrm{GeV}$, $x=0.05$ at $\mu=Q$.
    The left plot shows linear scales in $\tau_1^b$, while the right plot displays the  absolute values of the cross sections in logarithmic scale. 
%\\
    }
\hspace{0.2cm}
    \label{fig:ns_NLO}
    \centering
    \includegraphics[width=1\linewidth]{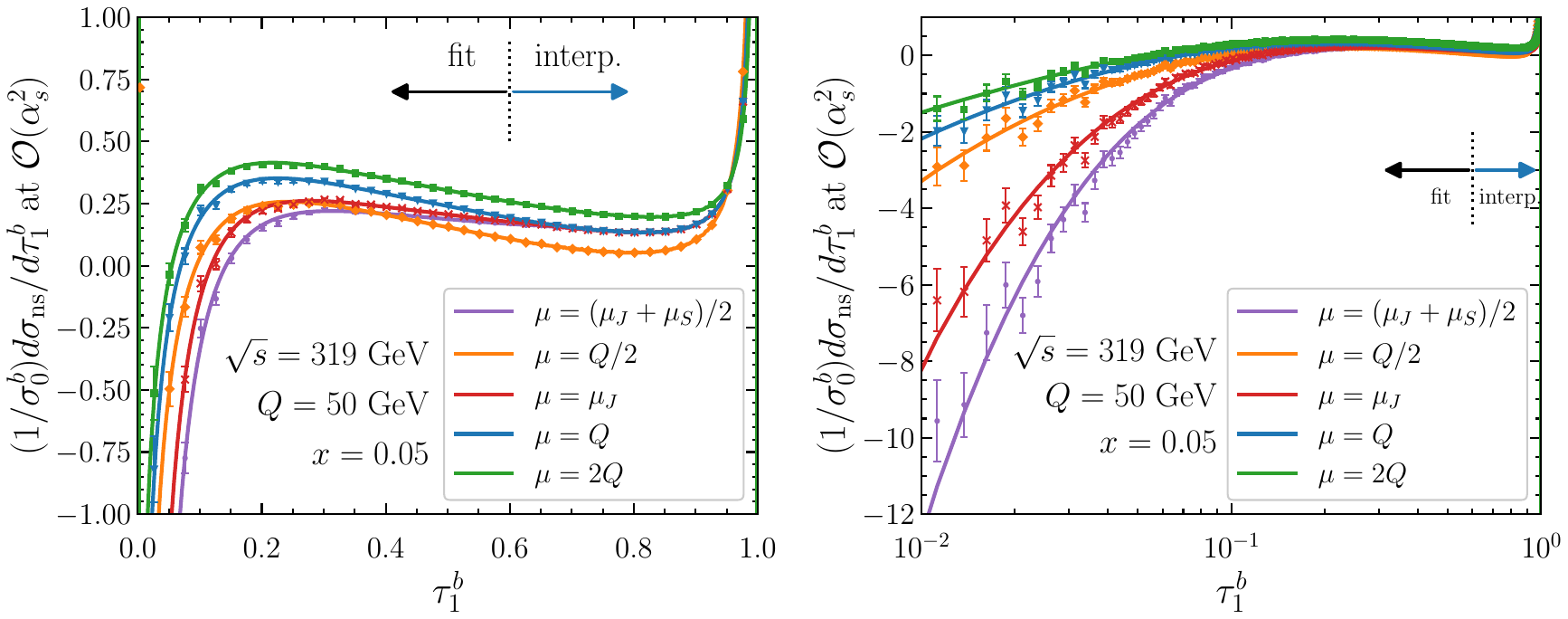}
    \vspace{-2em}
    \caption{Numerical and fit/interpolation results for the $\mathcal{O}(\alpha_s^2)$ non-singular cross section, reconstructed for five scale choices, with $\sqrt{s}=319~\textrm{GeV}$, $Q=50~\textrm{GeV}$, $x=0.05$.
    The left plot shows results with a linear scale for $\tau_1^b$, while the right plot shows the log scale. Points with error bars represent numerical results from \texttt{NLOJet++}.
    The black and blue arrows indicate the regions where the fitting and interpolation were performed, respectively. Solid lines represents the fit/interpolation nonsingular cross sections as described in Eq.~\eqref{eq:ns_interp_fit}. When the results at $\mathcal{O}(\alpha_s)$ and $\mathcal{O}(\alpha_s^2)$ are combined, there is a cancellation that reduces the overall scale dependence for the complete cross section, compared to what is shown here.}
    \label{fig:ns_NLO_scale_variations}
\end{figure}
In Fig.~\ref{fig:ns_NLO}, we present the $\mathcal{O}(\alpha_s^2)$ full QCD, singular, and nonsingular results in linear (left) and logarithmic (right) scales. Since numerical errors increase significantly as $\tau_1^b$ decreases,  
we use fitted results, which we discuss shortly.
These results are based on $3.6\times 10^{12}$ NLO Monte-Carlo events, all generated with the same infrared cutoff.
As $\tau_1^b\to 0$, the singular cross section from SCET asymptotically agrees with the full QCD results from \texttt{NLOJet++}.
The scale variations of the $\mathcal{O}(\alpha_s^2)$ nonsingular cross sections are depicted in Fig.~\ref{fig:ns_NLO_scale_variations}. The error bars represent MC uncertainties from \texttt{NLOJet++}. In order to smooth out the noise, we fit the data, and the solid lines show the results of this fit across the five scale variations.

At ${\cal O}(\alpha_s^2)$ the asymptotic behaviors of the nonsingular cross sections as $\tau_1^b\ll 1$ can be described by the following ansatz: \cite{Abbate:2010xh}
\begin{equation}\label{eq:fit_formula}
\frac{d\sigma^\textrm{ns}}{d\tau_1^b}\bigg|_{\textrm{fit},\:\alpha_s^2}
=
a_0 + a_1 \log \tau_1^b + a_2  \log^2 \tau_1^b + a_3  \log^3 \tau_1^b + b_3 \tau_1^b \log^3 \tau_1^b,
\end{equation}
which contains five fit parameters. To determine the optimal fit for the noisy \texttt{NLOJet++} results, we tested multiple models using the Akaike information criterion (AIC) and cross-validation (CV) \cite{Akaike1998, 3327fbe9-39c0-3916-9425-5d6fb128f17c}. The five-parameter ansatz was chosen as it best balances underfitting and overfitting.

We perform weighted fits using the function in Eq.~\eqref{eq:fit_formula}, with the weights based on the MC uncertainties from \texttt{NLOJet++} results.
Because of the significant MC uncertainties for small $\tau_1^b$, the fits are largely insensitive to the choice of the lower limit of the fit region. Therefore, the fit region is defined as $\tau_1^b\in (0, \tau_\textrm{upper})$, with $\tau_\textrm{upper}$ as the upper limit, and this parameter is treated as a hyperparameter of the fit. The upper limit, $\tau_\textrm{upper}$, should not be too large because the asymptotic form of the nonsingular cross section cannot capture its behavior as $\tau_1^b\to 1$. At the same time, $\tau_\textrm{upper}$ should not be too small, as the numerical results for $\tau_1^b>0.1$ are more reliable and have smaller uncertainties. 
Based on these criteria and the AIC and CV results, we choose  
$\tau_\textrm{upper} = 0.65\tau_\textrm{max}^b$  
as the optimal hyperparameter for the five-parameter model in Eq.~\eqref{eq:fit_formula}, where $\tau_\textrm{max}^b$ is defined in Eq.~\eqref{eq:def_taubmax}.

Due to the noisy numerical data at small $ \tau _1^b$ and the reliability of \texttt{NLOJet++} at larger $ \tau_1^b$, we implement the following approach for the $\mathcal{O}(\alpha_s^2)$ nonsingular cross section. For $\tau_1^b > \tau_\textrm{upper}$, we use the interpolated \texttt{NLOJet++} results, where Monte Carlo uncertainties are negligible. For $\tau_1^b < \tau_\textrm {upper}$, we instead employ the fitted results described earlier to avoid noise from direct numerical data. These two regions are smoothly connected using a transition function $f(z,z_0,\epsilon_0)$:
\begin{equation}\label{eq:ns_interp_fit}
\frac{d\sigma^\textrm{ns}}{d\tau_1^b}\bigg|_{\alpha_s^2}
=
\left[1-f(\tau_1^b, \tau_\textrm{upper}, \epsilon_0)\right]
\frac{d\sigma^\textrm{ns}}{d\tau_1^b}\bigg|_{\textrm{fit},\:\alpha_s^2}
+
f(\tau_1^b, \tau_\textrm{upper}, \epsilon_0)
\frac{d\sigma^\textrm{ns}}{d\tau_1^b}\bigg|_{\textrm{interpolation},\:\alpha_s^2},
\end{equation}
where the transition function is defined as
\begin{equation}\label{eq:smooth-function}
f(z, z_0, \epsilon_0) =\frac{1}{1+e^{-(z-z_0)/\epsilon_0}}.
\end{equation}
We set $\epsilon_0=0.02$ to ensure a smooth transition, with minimal effect around $\tau_\textrm{upper}$.
As shown in Fig.~\ref{fig:ns_NLO_scale_variations}, the fitted function seamlessly joins with the interpolated results, effectively capturing the trends in the noisy data.

To validate these results for the fit/interpolation functions describing the nonsingular $\tau_1^b$ distributions, we verify whether the resulting sum of singular and nonsingular distributions integrate to the correct total QCD cross section, as described in \cite{Hoang:2008fs, Bell:2018gce}.
In general, the full QCD $\tau_1^b$ cross sections at fixed renormalization/factorization scales can be expressed as\footnote{If the renormalization/factorization scales vary with $\tau_1^b$ (i.e., $\mu=\mu(\tau_1^b)$), the coefficients $A$ and $B$ also become dependent on $\tau_1^b$, rendering Eqs.~\eqref{eq:hatsig_QCD} and \eqref{eq:hatsig_sing_fixed} below invalid. In this analysis, we  compute $r_c$ at fixed scales $\mu=Q$, $Q/2$, and $2Q$ and check the validity of the nonsingular cross sections. }
\begin{equation}\label{eq:full_QCD_fixed}
\frac{d\sigma_\textrm{PT}^\textrm{QCD}}{d\tau_1^b}
=
A\delta(\tau_1^b)
+
\left[B(\tau_1^b)\right]_+
+
r(\tau_1^b),
\end{equation}
where $A$ is the constant coefficient of the delta function, $B$ is a singular function written in terms of plus distributions, and $r$ is the remainder function that can have at most integrable singularities as $\tau_1^b\to 0$. The SCET factorization formula  captures the singular behavior of the full QCD cross sections. Therefore, we can separate the terms in Eq.~\eqref{eq:full_QCD_fixed} according to the definition in Eq.~\eqref{eq:def-of-nonsingular} as follows:
\begin{equation}\label{eq:asso_sing_fixed_ns_fixed}
\frac{d\sigma^\textrm{s,fixed}}{d\tau_1^b}
=A\delta(\tau_1^b)
+
\left[B(\tau_1^b)\right]_+,
\quad
\frac{d\sigma^\textrm{ns}}{d\tau_1^b}
= r(\tau_1^b).
\end{equation}
Integrating the QCD cross section and the singular cross section in $\tau_1^b$ from 0 to 1 gives
\begin{align}
\label{eq:hatsig_QCD}
\sigma_{\textrm{PT}}^\textrm{QCD}
&\equiv
\int_0^1 d\tau_1^b
\frac{d\sigma_\textrm{PT}^\textrm{QCD}}{d\tau_1^b}
=
A + r_c,
\\
\label{eq:hatsig_sing_fixed}
\sigma_{\textrm{PT}}^\textrm{s,fixed}
&\equiv
\int_0^1 d\tau_1^b
\frac{d\sigma_{\textrm{PT}}^\textrm{s,fixed}}{d\tau_1^b}
= A,
\end{align}
where we define the total integral of the remainder function $r(\tau_1^b)$ as
\begin{equation}
r_c \equiv \int_0^1 d\tau_1^b\, r(\tau_1^b),
\end{equation}
and apply the property of the plus distribution
\begin{equation}\label{eq:plus-int-identity}
\int_0^1 d\tau_1^b
\left[B(\tau_1^b)\right]_+ = 0.
\end{equation}
The coefficient $A$ is derived from the known fixed-order expressions for the hard, jet, beam, and soft functions in SCET, and the total fixed-order cross section in Eq.~\eqref{eq:hatsig_QCD} can be calculated using known two-loop analytic expressions \cite{Furmanski:1981cw}. In this work, we use the \texttt{QCDNUM} package~\cite{Botje:2010ay}, which provides routines for computing massless structure functions in unpolarized scattering up to $\mathcal{O}(\alpha_s^2)$.\footnote{Note that these analytic expressions must be integrated numerically over the same PDF sets.} From this, $r_c$ can be written as 
\begin{equation}
\label{eq:rc1-from-analytic}
r_c = \sigma_{\textrm{PT}}^\textrm{QCD} - \sigma_{\textrm{PT}}^\textrm{s,fixed},
\end{equation}
which allows us to determine $r_c$ without needing to explicitly calculate the remainder function, i.e., the differential nonsingular cross section in Eq.~\eqref{eq:asso_sing_fixed_ns_fixed}. 

From the numerical results of \texttt{NLOJet++}, we obtain the full QCD cross section for $\tau_1^b>0$ as
\begin{equation}\label{eq:QCD-from-NLOJET++}
\frac{d\sigma^\textrm{QCD}_\textrm{PT}}{d\tau_1^b}\Big|^\texttt{NLOJet++}_{\tau_1^b>0}
=
B(\tau_1^b)
+
r(\tau_1^b).
\end{equation}
From the known analytic form of the singular cross section for $\tau_1^b>0$, we have
\begin{equation}
\label{eq:singular-dist}
\frac{d\sigma^\textrm{s,fixed}_\textrm{PT}}{d\tau_1^b}\Big|_{\tau_1^b>0} = B(\tau_1^b),
\end{equation}
where we used the plus-distribution property 
$\left[B(\tau_1^b)\right]_+  = B(\tau_1^b)$ when $\tau_1^b>0$.
Integrating the difference of Eqs.~\eqref{eq:QCD-from-NLOJET++} and \eqref{eq:singular-dist} over $\tau_1^b$ from a small value $\epsilon>0$ to $1$, we obtain
\begin{equation}\label{eq:rcep}
r_{I}(\epsilon) \equiv
\int_\epsilon^1 d\tau_1^b\,r(\tau_1^b) = 
\int_\epsilon^1 
d\tau_1^b
\left[
\frac{d\sigma^\textrm{QCD}_\textrm{PT}}{d\tau_1^b}\Big|^\texttt{NLOJet++}_{\tau_1^b>0}
-
\frac{d\sigma^\textrm{s,fixed}_\textrm{PT}}{d\tau_1^b}\Big|_{\tau_1^b>0}
\right],
\end{equation}
and
\begin{equation}\label{eq:num-rc1}
r_c = 
\lim_{\epsilon\to0}
r_I(\epsilon).
\end{equation}
The two representations of $r_c$ in Eqs.~\eqref{eq:rc1-from-analytic}
and \eqref{eq:num-rc1} should converge as $\epsilon\to0$, since the remainder function $r(\tau_1^b)$ has integrable singularities as $\tau_1^b\to 0$. 
Therefore, comparing these two expressions in Eqs.~\eqref{eq:rc1-from-analytic} and \eqref{eq:num-rc1} serves as a nontrivial check of the numerical results from \texttt{NLOJet++}. 
In practice, due to the numerical instabilities at small $\tau_1^b$, we investigate  whether $r_I(\epsilon)$ develops a plateau before numerical errors become severe at very small $\epsilon$. The value of this plateau should align with the analytical result from Eq.~\eqref{eq:rc1-from-analytic}. 

To illustrate this procedure, we perform the check of $r_c$ in Eq.~\eqref{eq:num-rc1} for the $\tau_1^b$ cross section at $\sqrt{s}=319$ GeV, $Q=50$ GeV, and $x=0.05$.
The total cross sections are expressed in terms of the hadronic structure functions $F_1$ and $F_L$ as follows:
\begin{equation}
\sigma_{\textrm{PT}}^\textrm{QCD}
=
2
\left\{
F_1(x, Q^2) + \frac{1-y}{x\left[1+(1-y)^2\right]} F_L(x,Q^2)
\right\}.
\end{equation}
Using the numerical results for $F_1(x, Q^2)$ and $F_L(x,Q^2)$ from \texttt{QCDNUM}, 
we compute the fixed-order full QCD cross sections up to $\mathcal{O}(\alpha_s^2)$. 
At this $\sqrt{s}$, $Q$, and $x$, and at the scale $\mu=Q$, the full QCD cross sections in Eq.~\eqref{eq:hatsig_QCD} are
\begin{subequations}
\begin{equation}
\sigma_{\textrm{PT}}^\textrm{QCD}
=
\sigma_{\textrm{PT}}^\textrm{QCD}\big|_{\mathcal{O}(\alpha_s^0)}
+
\sigma_{\textrm{PT}}^\textrm{QCD}\big|_{\mathcal{O}(\alpha_s^1)}
+
\sigma_{\textrm{PT}}^\textrm{QCD}\big|_{\mathcal{O}(\alpha_s^2)},
\end{equation}
with
\begin{equation}
\sigma_{\textrm{PT}}^\textrm{QCD}\big|_{\mathcal{O}(\alpha_s^0)}=11.077\sigma_0^b,
\quad
\sigma_{\textrm{PT}}^\textrm{QCD}\big|_{\mathcal{O}(\alpha_s^1)}=-0.575\sigma_0^b,
\quad
\sigma_{\textrm{PT}}^\textrm{QCD}\big|_{\mathcal{O}(\alpha_s^2)}=-0.081\sigma_0^b,
\end{equation}
\end{subequations}
where $\sigma_0^b$ is given in \eq{born-crosssection}.
Next, the total integrals of the fixed-order singular cross section from $\tau_1^b=0$ to 1 defined in Eq.~\eqref{eq:hatsig_sing_fixed} are given by
\begin{subequations}
\begin{equation}
\sigma_{\textrm{PT}}^\textrm{s,fixed}
=
\sigma_{\textrm{PT}}^\textrm{s,fixed}\big|_{\mathcal{O}(\alpha_s^0)}
+
\sigma_{\textrm{PT}}^\textrm{s,fixed}\big|_{\mathcal{O}(\alpha_s^1)}
+
\sigma_{\textrm{PT}}^\textrm{s,fixed}\big|_{\mathcal{O}(\alpha_s^2)},
\end{equation}
with
\begin{equation}
\sigma_{\textrm{PT}}^\textrm{s,fixed}\big|_{\mathcal{O}(\alpha_s^0)}=11.077\sigma_0^b,
\quad
\sigma_{\textrm{PT}}^\textrm{s,fixed}\big|_{\mathcal{O}(\alpha_s^1)}=-1.984\sigma_0^b,
\quad
\sigma_{\textrm{PT}}^\textrm{s,fixed}\big|_{\mathcal{O}(\alpha_s^2)}=-0.194\sigma_0^b.
\end{equation}
\end{subequations}
At $\mathcal{O}(\alpha_s^0)$, the differential cross section consists entirely of the delta function contributions, so the total cross section at this order is accounted for by the singular contributions:
$\sigma_{\textrm{PT}}^\textrm{s,fixed}\big|_{\mathcal{O}(\alpha_s^0)}=\sigma_{\textrm{PT}}^\textrm{QCD}\big|_{\mathcal{O}(\alpha_s^0)}$.
The cumulant distributions of the residual nonsingular contributions are then computed using Eq.~\eqref{eq:rc1-from-analytic} as follows:
\begin{equation}
r_c
=
r_c\big|_{\mathcal{O}(\alpha_s^0)}
+
r_c\big|_{\mathcal{O}(\alpha_s^1)}
+
r_c\big|_{\mathcal{O}(\alpha_s^2)},
\end{equation}
with
\begin{equation}\label{eq:rc1_analytic_num}
r_c\big|_{\mathcal{O}(\alpha_s^0)}=0,
\quad
r_c\big|_{\mathcal{O}(\alpha_s^1)}=1.409\sigma_0^b,
\quad
r_c\big|_{\mathcal{O}(\alpha_s^2)}=0.112\sigma_0^b.
\end{equation}

\begin{figure}
    \centering
    \vspace{-1em}
    \begin{subfigure}
        \centering
        \includegraphics[width=0.49\linewidth]{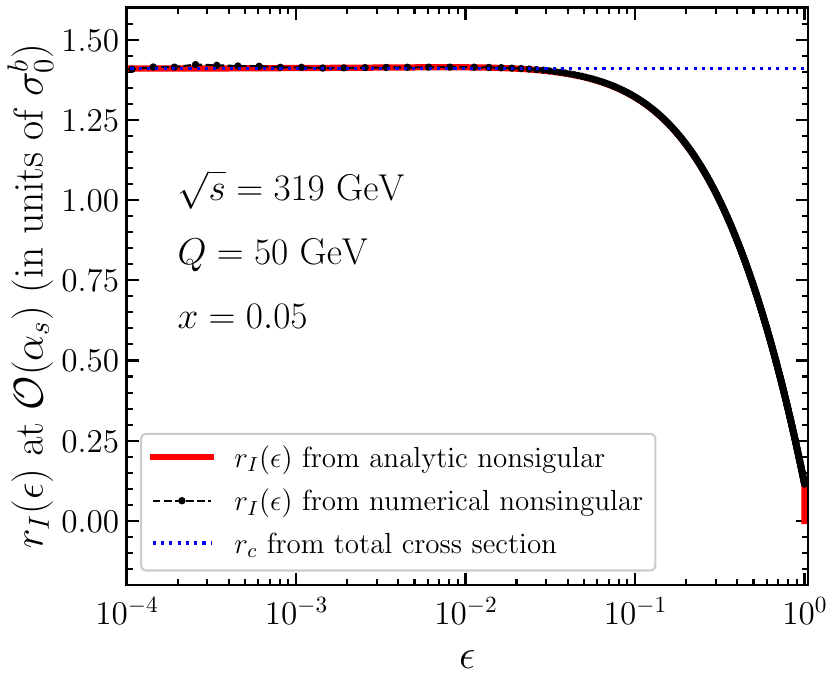}
    \end{subfigure}
    \begin{subfigure}
        \centering
        \includegraphics[width=0.49\linewidth]{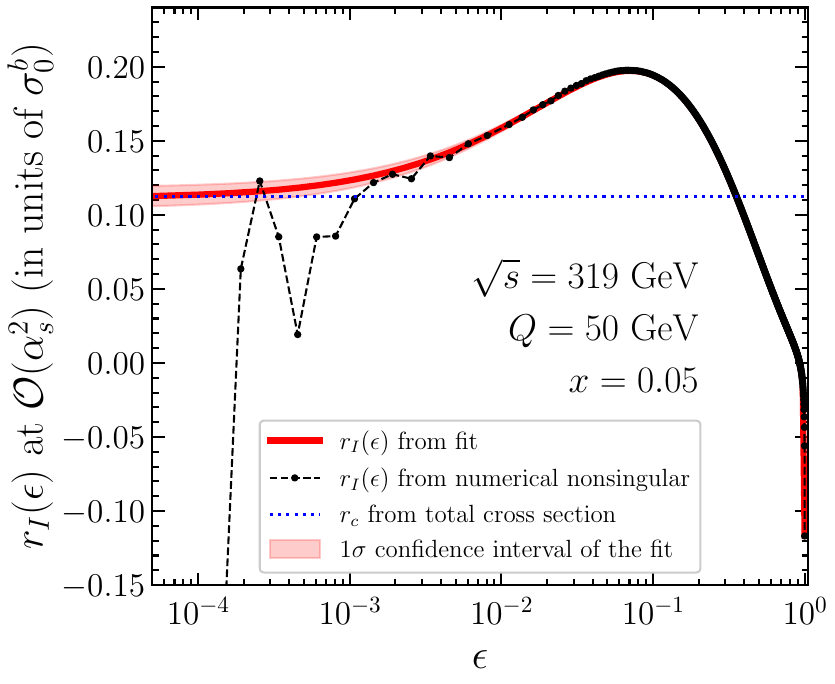}
    \end{subfigure}
    \vspace{-2em}
    \caption{
Determination of the non-logarithmic cross section $r_c\ = \lim_{\epsilon\to 0}
r_I(\epsilon)$ at $\mathcal{O}(\alpha_s)$ and at $\mathcal{O}(\alpha_s^2)$ for 
    at $\sqrt{s}=319~\textrm{GeV}$, $Q=50~\textrm{GeV}$, and $x=0.05$.}
    \label{fig:rc1}
\end{figure}

Finally, we validate the nonsingular distributions obtained from \texttt{NLOJet++} by checking whether $r_I(\epsilon)$ from Eq.~\eqref{eq:rcep} for sufficiently small $\epsilon$ agrees with the values of $r_c$ from Eq.~\eqref{eq:rc1_analytic_num}. This comparison confirms the accuracy of the fitted nonsingular distributions in Eq.~\eqref{eq:ns_interp_fit}.
In Fig.~\ref{fig:rc1}, we show the $r_I(\epsilon)$ predicted from Eq.~\eqref{eq:rcep} as a function of $\epsilon$ (black dots) and $r_c$ obtained from Eq.~\eqref{eq:rc1_analytic_num} (blue dotted lines) at $\mathcal{O}(\alpha_s)$ (left) and $\mathcal{O}(\alpha_s^2)$ (right). 
At $\mathcal{O}(\alpha_s)$, we also include $r_I(\epsilon)$ obtained from the analytic expression (red line) in Ref.~\cite{Kang:2014qba}. As expected, the numerical $r_I(\epsilon)$ (black dots) matches well with the analytic result (red line) before minor numerical instability sets at $\epsilon\lesssim 10^{-3}$. Additionally, the predicted plateau aligns precisely with the analytic $r_c$, confirming the validity of the nonsingular distributions computed using \texttt{NLOJet++} at $\mathcal{O}(\alpha_s)$. 

At $\mathcal{O}(\alpha_s^2)$, the $r_I(\epsilon)$ from \texttt{NLOJet++} (black dots) approaches the analytic $r_c$ (blue dotted line) as $\epsilon$ decreases. However, for $\epsilon\lesssim 10^{-2}$, numerical instabilities begin to affect the results before a clear plateau can be established. To resolve this issue, we rely on the fit function determined by the weighted fit in the region $\tau_1^b\in (0, \tau_\textrm{upper})$. The prediction of $r_I(\epsilon)$ from the fit/interpolation function in Eq.~\eqref{eq:ns_interp_fit} is shown as the red line in the right plot of Fig.~\ref{fig:rc1}, with $1\sigma$ confidence interval shaded. As shown, the fit/interpolation nonsingular cross section in Eq.~\eqref{eq:ns_interp_fit} accurately reproduces the trend in $r_I(\epsilon)$ for $\epsilon>10^{-2}$ and matches the analytically predicted $r_c$ from Eq.~\eqref{eq:rc1_analytic_num} as $\epsilon\to0$. 
In Appendix~\ref{app:rc_other_mu}, we present additional results for $r_I(\epsilon)$ at other scale settings. As shown in Fig.~\ref{fig:rc1_NLO_fit}, the nonsingular distributions from the fit/interpolation functions also reproduce the correct $r_c$ at the fixed scales, $\mu=Q$, $Q/2$, and $2Q$. 

At $\mathcal{O}(\alpha_s^2)$, the nonsingular (as $\tau_1^b\to 0$) distribution in Fig.~\ref{fig:ns_NLO_scale_variations} exhibits a characteristic singular or peaked behavior as $\tau_1^b\to 1$. At $\mathcal{O}(\alpha_s)$, this singular behavior manifests as a delta distribution at $\tau_1^b=1$ as shown in Ref.~\cite{Kang:2014qba}, corresponding to events where, in the Breit frame, the jet hemisphere is empty, and all final-state particles are contained within the beam hemisphere.\footnote{This behavior is also evident in Eq.~\eqref{eq:DIS-thrust}, where $\tau_1^b$ becomes its maximum value of $1$ in the absence of particles in the jet hemisphere $\mathcal{H}_J$.}
At $\mathcal{O}(\alpha_s^2)$, these empty jet hemisphere events appear to be slightly smeared, resulting in events with $\tau_1^b<1$ due to an additional final-state particle leaking a small amount of momentum into $\mathcal{H}_J$. In Sec.~\ref{sec:HERA-comparison}, we will discuss how this behavior has been observed experimentally in HERA data \cite{H1:2024aze, H1:2024nde}.

Even though we capture this behavior within the fixed-order nonsingular cross sections, convolving these cross sections with the shape function, as in Eq.~\eqref{eq:k-int-cumulant}, may obscure this behavior due to the universal shift in the distributions when $\tau_1^b\gg \Lambda_\textrm{QCD}/Q$. To preserve the singular behavior as $\tau_1^b\to 1$, we avoid using the convolved cross sections for the entire range of $\tau_1^b$. Instead, we revert to the fixed-order full QCD cross section for  $\tau_1^b>t_3$, where $t_3 = 0.8\tau_\textrm{max}^b$.\footnote{The definitions of the profile function parameters $t_i$ are provided in the subsequent section.}
To smoothly merge the two predictions at $\tau_1^b=t_3$, we apply the following formula for the final theory predictions, combining the convolved and fixed-order cross sections:
\begin{equation}
\label{eq:back-to-fixed}
\frac{d\sigma}{d\tau_1^b}
=
\left[1-f(\tau_1^b, t_3,\epsilon_0) \right]
\frac{d\sigma}{d\tau_1^b}\bigg|_\textrm{Eq.~\eqref{eq:k-int-cumulant}}
+
f\left(\tau_1^b, t_3,\epsilon_0 \right)
\frac{d\sigma}{d\tau_1^b}\bigg|_\textrm{fixed},
\end{equation}
where the transition function $f$ is defined in Eq.~\eqref{eq:smooth-function} with $\epsilon_0 = 0.03$.

Since the \texttt{NLOJet++} numerical results are provided as binned differential cross sections in $\tau_1^b$, it is natural to convolve the nonsingular cross sections with the shape function and perform renormalon subtraction directly in the differential distribution, rather than in the cumulative distribution as was done for the singular cross sections in Eq.~\eqref{eq:cumulant-to-differential}.
At small $\tau_1^b$ it is not strictly necessary to convolve the shape function with the nonsingular distribution as implied in \eq{k-int-cumulant},
since this corresponds to an estimate for the nonperturbative corrections that act on the already suppressed nonsingular distribution. 
For values of $\tau_1^b$ where the singular and nonsingular cross sections are of similar magnitude, it is important that whatever method is used to implement nonperturbative corrections on them separately, does not spoil perturbative cancellations between them.
We use a common shape function for both terms here, up to the split in \eq{back-to-fixed}, to deal with the subtraction of renormalons and the effect of the shift parameter $\Omega_1$ on the same footing across the whole range of $\tau_1^b$, as was done in \cite{Abbate:2010xh,Hoang:2014wka,Bell:2018gce}. As multi-jet nonperturbative power corrections are not a focus of our work here, we defer alternative treatments of the above choice of shape function convolution to future work.

\section{Profile functions and uncertainties}
\label{sec:profile}

We implement profile functions to characterize the $\tau_1^b$-dependence of the renormalization scales ($\mu_H$, $\mu_B$, $\mu_J$, $\mu_S$). 
The goals of this method are to smoothly connect nonperturbative (peak), perturbative resummation (tail), and fixed-order (far-tail) regions of the $\tau_1^b$ distributions; to robustly estimate theoretical uncertainties smoothly in each region; and to incorporate naturally the $R$-gap renormalon subtraction scheme in \sec{shape-renormalon}.
For comprehensive discussions of profile function methodology,  
see Refs.~\cite{Ligeti:2008ac,Abbate:2010xh,Kang:2013nha,Kang:2014qba,Bell:2018gce}.
In addition to the profile functions for the singular cross sections that determine the scales $\mu_{H,B,J,S}$, we introduce a dedicated profile function for the nonsingular cross sections governing the common scale $\mu_\textrm{ns}$.

The profile functions for the hard, beam, jet and soft functions in the factorization theorem are defined as follows:
%---------------
\begin{align}\label{eq:profile_function}
\begin{split}
%---------------
\mu_H 
&= 
\mu,
\\
\mu_{B,J}(\tau_1^b) 
&= 
\left[
1+e_{B,J}\,
g(\tau_1^b)
\right]
\sqrt{\mu\, \mu_\textrm{run}(\tau_1^b,\mu)},
\\
\mu_{S}(\tau_1^b) 
&= 
\left[
1+e_{S}\,
g(\tau_1^b)
\right]
\mu_\textrm{run}(\tau_1^b,\mu),
%---------------
\end{split}
\end{align}
%---------------
where $e_{B,J,S}$ are the parameters that vary respective scales $\mu_{B,J,S}$ enabling to estimate perturbative uncertainties, and the function $g$ is defined by
%---------------
\begin{equation}
%---------------
g(\tau_1^b, \{t_0, t_3\})
=
\begin{cases}
1,
&\textrm{for~$0\le \tau_1^b< t_0$},
\\
\left(t_3-\tau_1^b\right)^2
/\left(t_3-t_0\right)^2,
&\textrm{for~$t_0\le \tau_1^b\le t_3$},
\\
0,
&\textrm{for~$\tau_1^b>t_3$},
\end{cases}
%---------------
\end{equation}
%---------------
This implementation ensures the scale variations remain frozen  
in both the deep nonperturbative regime ($\tau_1^b < t_0$)  
and the fixed-order-dominated region ($\tau_1^b > t_3$).
The running scale $\mu_\textrm{run}$ is given by
%---------------
\begin{equation}
\label{eq:runnsing-scale}
%---------------
\mu_\textrm{run}
(\tau_1^b, \mu, \{\mu_0, r, t_0, t_1, t_2, t_3\})
=
\mu \times
\begin{cases}
\mu_0/\mu, 
&\textrm{for~$0\le \tau_1^b < t_0$},
\\
\zeta(\tau_1^b, \{t_0,\mu_0, 0\}, \{t_1, rt_1, r\}), 
&\textrm{for~$t_0\le \tau_1^b < t_1$},
\\
r \tau_1^b,
&\textrm{for~$t_1\le \tau_1^b < t_2$},
\\
\zeta(\tau_1^b, \{t_2,rt_2, r\}, \{t_3, 1, 0\}), 
&\textrm{for~$t_2\le \tau_1^b < t_3$},
\\
1,
&\textrm{for~$\tau_1^b \ge t_3$}.
\end{cases}
%---------------
\end{equation}
%---------------
The function $\zeta$ ensures a smooth transition between the canonical (tail) region ($t_1\le \tau_1^b\le t_2$) and the frozen scale (peak) regions at small and large $\tau_1^b$. 
It is defined as
%---------------
\begin{equation}
%---------------
\zeta(\tau,\{x_0,y_0,r_0\}, \{x_1,y_1,r_1\})
=
\begin{cases}
a\tau^2+b\tau+c,
&\textrm{for~$x_0\le \tau \le (x_0+x_1)/2$},
\\
d\tau^2+ e\tau+f,
&\textrm{for~$(x_0+x_1)/2\le \tau \le x_1$},
\end{cases}
%---------------
\end{equation}
%---------------
where the parameters $a,b,c,d,e,f$ ensure that $\zeta$ and its first derivative are continuous at its boundaries and across its domain: 
%---------------
\begin{align}
\begin{split}
%---------------
a
&=
\frac{(3r_0+r_1)(x_0-x_1) + 4(y_1 - y_0)}
{2(x_1-x_0)^2},
\\
b
&=
r_0 - 2a x_0,
\\
c
&=
y_0 - bx_0 -a x_0^2,
\\
d 
&=
\frac{(3r_1+r_0)(x_1-x_0) + 4(y_0 - y_1)}
{2(x_1-x_0)^2},
\\
e
&=
r_1 - 2d x_1,
\\
f
&=
y_1 - e x_1 -d x_1^2.
%---------------
\end{split}
\end{align}
%---------------

In Eq.~\eqref{eq:runnsing-scale}, $\mu_\textrm{run}$ is designed to be a monotonically increasing function in $\tau_1^b$. However, if the hard scale $\mu$ becomes too small (around $\mu\sim 10~\textrm{GeV}$), the monotonicity of Eq.~\eqref{eq:runnsing-scale} breaks down because the frozen scale $\mu_\textrm{run} = \mu_0$ could exceed the starting value of the scale in the canonical (tail) region, $\mu_\textrm{run} = rt_1$. In such cases where $\mu_0>rt_1$, to maintain the monotonic behavior of the running scale, we implement the following adjustment:
%---------------
\begin{equation}
%---------------
\mu_\textrm{run}
(\tau_1^b, \mu, \{\mu_0, r, t_0, t_1, t_2, t_3\})
=
\mu \times
\begin{cases}
\mu_0/\mu, 
&\textrm{for~$0\le \tau_1^b < t_0$},
\\
\zeta(\tau_1^b, \{t_0,\mu_0, 0\}, \{t_3, 1, 0\}), 
&\textrm{for~$t_0\le \tau_1^b < t_3$},
\\
1,
&\textrm{for~$\tau_1^b > t_3$}.
\end{cases}
%---------------
\end{equation}
%---------------

\begin{figure}
    \centering
    \vspace{-1em}
    \includegraphics[width=0.85\linewidth]{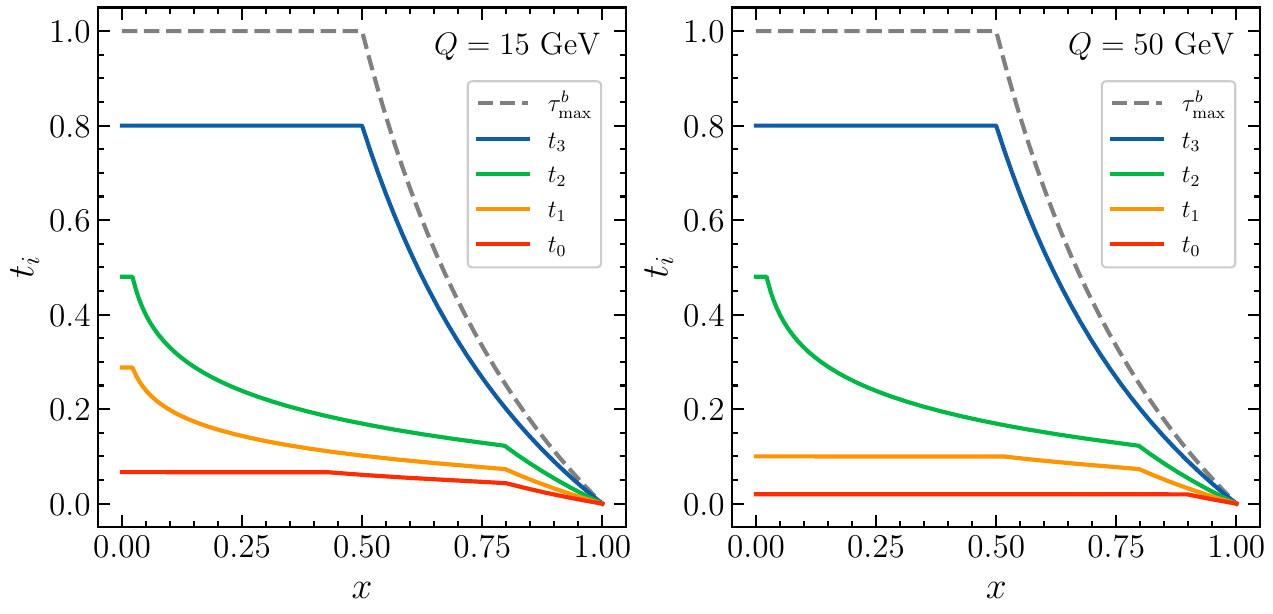}
    \vspace{-1em}
    \caption{The parameters $t_i$ for the profile functions at $Q=15$ GeV and~ $Q=50$ GeV as functions of $x$. The $t_i$ are continuous but undergo transitions at various thresholds, as discussed in the text.}
    \label{fig:t0123}
\end{figure}

\begin{figure}
    \centering    
    \vspace{-1em}
    \includegraphics[width=0.65\linewidth]{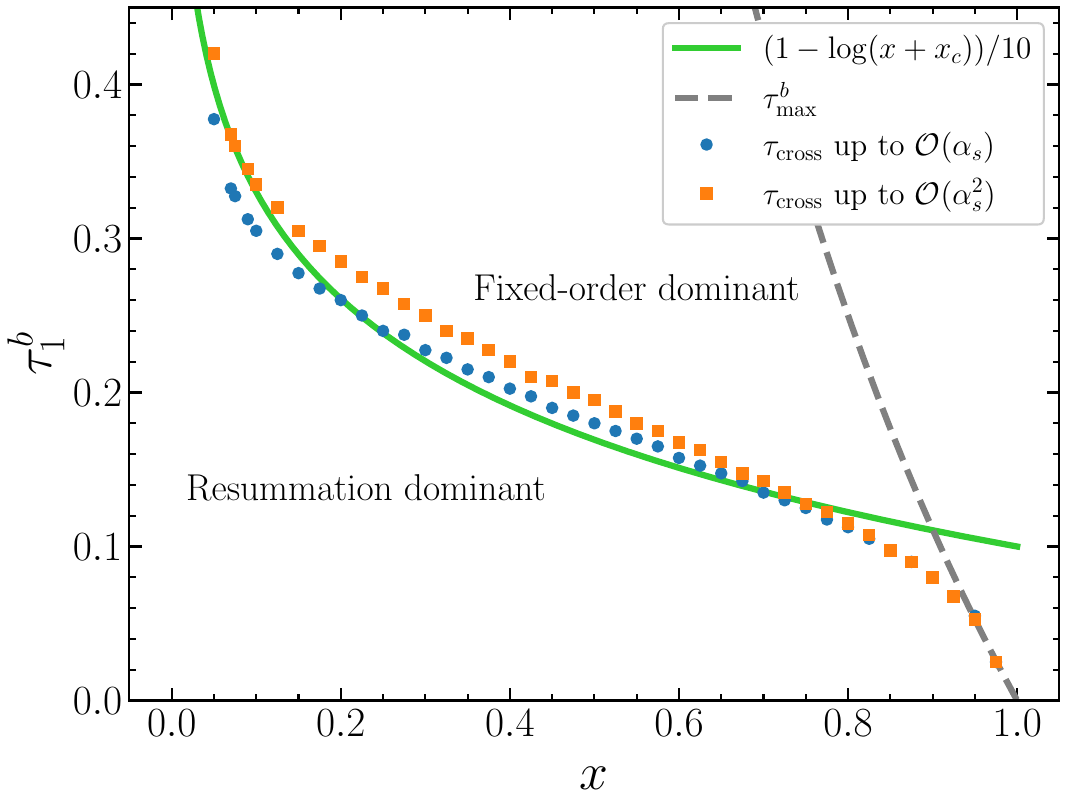}
    \vspace{-1em}
    \caption{The crossing points $\tau_\textrm{cross}$, where the nonsingular cross section becomes larger in magnitude than the singular cross section, are shown as a function of $x$ at  $Q=30~\textrm{GeV}$. In this $x$--$\tau_1^b$ plane, the  singular contribution dominates in the lower-left region (Resummation dominant), while the nonsingular contribution dominates in the upper-right region (Fixed-order dominant). The functional form of $t_2$ (green) captures the behavior of the crossing points at both $\mathcal{O}(\alpha_s)$ (blue) and $\mathcal{O}(\alpha_s^2)$ (orange). For $x>0.8$, the crossing points are constrained by $\tau_\textrm{max}^b$ (grey dashed line).}
    \label{fig:crossing}
\end{figure}

The profile function is uniquely specified by the parameters, $\mu, \mu_0, r, t_0, t_1, t_2, t_3$ and $e_{B,J,S}$. The default settings giving the central values of our prediction are given as
%---------------
\begin{align}\label{eq:profile_setting}
\begin{split}
%---------------
\mu &= Q,
\\
\mu_0&= 1.1~\textrm{GeV},
\\
r &= 1/\tau^b_\textrm{max},
\\
t_3 &= 0.8\tau_\textrm{max}^b,
\\
t_2 &= \textrm{min}
\left(\frac{1-\log(x+x_c)}{10},
0.6 t_3
\right),
\\
t_1 &= \textrm{min}
\left(\frac{5~\textrm{GeV}}{Q}, 0.6 t_2\right),
\\
t_0 &= \textrm{min}
\left(\frac{1~\textrm{GeV}}{Q}, 0.6 t_1\right),
\\
e_{B,J,S} &= 0,
%---------------
\end{split}
\end{align}
%---------------
where $x_c = 0.0001234$, a number determined empirically by comparing  fixed-order singular and nonsingular distributions at varying $x$, as explained below and in \fig{crossing}. As we discussed in Eq.~\eqref{eq:back-to-fixed}, to properly capture the singular behavior of $\tau_1^b$ cross section as $\tau_1^b\to 1$, we fix $t_3 = 0.8\tau_\textrm{max}^{b}$ and use the fixed-order cross sections directly for $\tau_1^b>t_3$. The minimization conditions in Eq.~\eqref{eq:profile_setting} ensure the hierarchy $t_0<t_1<t_2<t_3$ for all values of $x$ and $Q$ as illustrated in Fig.~\ref{fig:t0123}.

The $x$-dependent functional form for $t_2$ is designed to account for the fact that the boundaries between the fixed-order dominant and resummation dominant regions vary with $x$. In Fig.~\ref{fig:crossing}, the crossing point $\tau_1^b = \tau_\textrm{cross}$, where the nonsingular cross section overtakes the fixed-order singular cross section, is shown as a function of $x$ at $Q=30~\textrm{GeV}$.
The chosen functional form of $t_2$, $(1-\log(x+x_c))/10$, well mimics the behavior of the crossing points as a function of $x$ for $x<0.8$ even when the $\mathcal{O}(\alpha_s^2)$ contributions are included.
It should be noted that as $x$ decreases, $t_2$ increases,  
causing the resummation-dominant region to broaden  
relative to the fixed-order-dominant region.

\begin{figure}
    \centering
    \vspace{-1em}
    \includegraphics[width=1\linewidth]{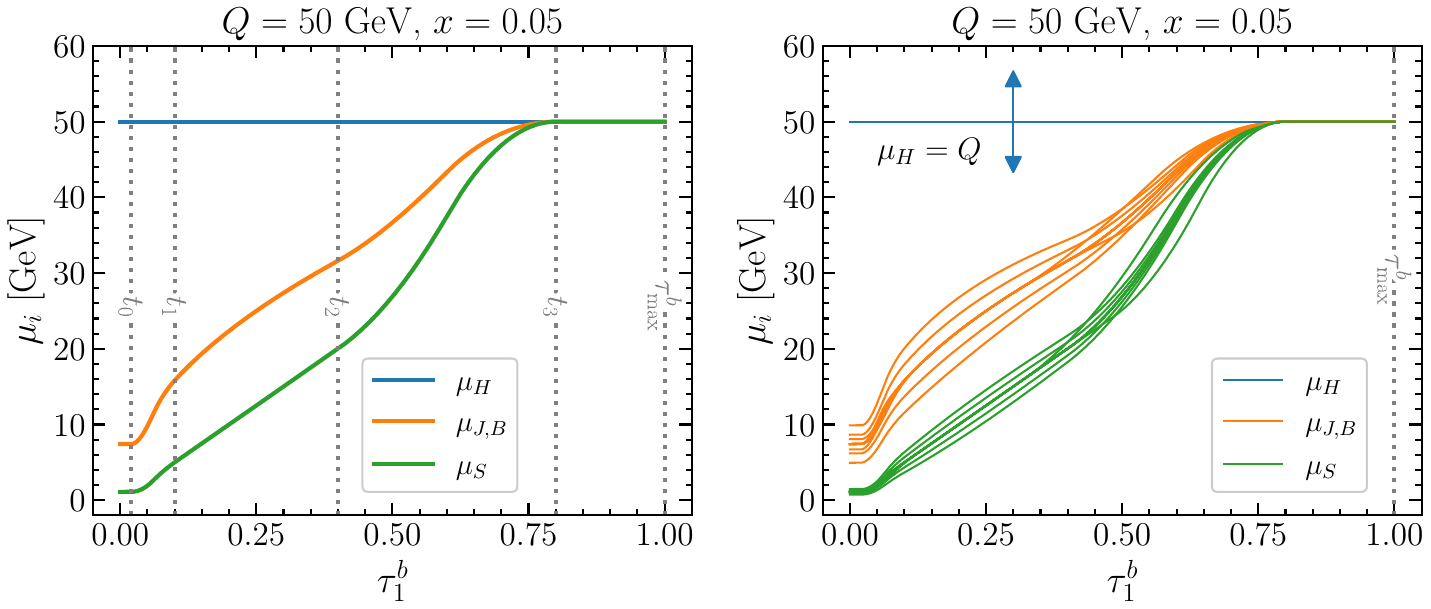}
    \vspace{-2em}
    \caption{In the left panel the profile function for $\mu_H$, $\mu_J$ and $\mu_S$ together with their transition points $t_i$ are shown. The right panel shows the profile scale variations at $Q=50~\textrm{GeV}$ and $x=0.05$. The double arrow illustrates the Variations 7--8 in Eq.~\eqref{eq:singular_scale_variations}.}
    \label{fig:profile_functions}
\end{figure}

To estimate the perturbative uncertainties in the theoretical predictions, we consider the following 16 scale variations in addition to the central setting in Eqs.~\eqref{eq:profile_function} and \eqref{eq:profile_setting}:
\begin{align}\label{eq:singular_scale_variations}
\begin{split}
\textrm{Variations 1--2}:~ & \mu_0 \to 1.1 \pm 0.2~\textrm{GeV},
\\
\textrm{Variations 3--4}:~ & t_1 \to (1\pm 0.2)t_1\big|_\textrm{central},
\\
\textrm{Variations 5--6}:~ & t_2 \to(1\pm 0.2)t_2\big|_\textrm{central},
\\
\textrm{Variations 7--8}:~ & \mu \to 2^{\pm 1} Q,
\\
\textrm{Variations 9--12}:~ & e_{B,J} \to \displaystyle\pm \frac{1}{3}, \pm\frac{1}{6},
\\
\textrm{Variations 13--16}:~ & e_{S} \to \displaystyle\pm \frac{1}{3}, \pm\frac{1}{6}.
\end{split}
\end{align}
These variations account for the uncertainties arising from the renormalization and factorization scales.
Variations 1--6 adjust the profile function parameters in Eq.~\eqref{eq:profile_setting}. 
Variations 7 and 8 uniformly shift all the scales in Eq.~\eqref{eq:profile_function} up or down by factors of 2 to estimate the fixed-order theoretical uncertainties.
Variations 9--12 and 13--16 provide additional uncertainty estimates at different orders of logarithmic accuracy in resummed perturbation theory, which cannot be achieved by varying a single scale $\mu$. 
Fig.~\ref{fig:profile_functions} shows the profile function at $Q=50~\textrm{GeV}$ and $x=0.05$, along with
its scale variations used to evaluate the perturbative uncertainties. 
The profile functions for various other values of $Q$ and $x$ are displayed in Appendix~\ref{app:profile_total}. 

As discussed in Sec.~\ref{sec:shape-renormalon}, we make the renormalon subtraction scale $R$ dependent on $\tau_1^b$ to avoid
large logarithms in the subtraction terms $\delta_i(R,\mu_S)$ for the soft function. An apparent choice would be $R(\tau_1^b)=\mu_S(\tau_1^b)$. 
However, in the peak region, it is advantageous to deviate from this choice so that the $O(\alpha_s)$ subtraction term $\delta_1(R,\mu_S)=-0.848826 \log(\mu_S/R)$ remains nonzero. Therefore, we adopt the form,
%---------------
\begin{equation}
\label{eq:R_profile_def}
%---------------
R(\tau_1^b)
=\mu_S(\tau_1^b)\big|_{\mu_0\to R_0},
%---------------
\end{equation}
%---------------
where the only free parameter is $R_0$, which sets the value of $R$ for $\tau_1^b<t_0$.
We choose $R_0=0.85\mu_0$ to give the one-loop subtraction $\delta_1(R,\mu_S)$
the appropriate sign to cancel the renormalon in the peak region. 
This setting makes $R(\tau_1^b)$ differ from $\mu_S$ only for $\tau_1^b<t_1$. 

The scale dependence of the nonsingular contributions is treated separately. The central value and the scale variations for the nonsingular contributions are defined as follows:
\begin{align}\label{eq:nonsingular_scale_variations}
\begin{split}
\textrm{Central}:~ & \mu_\textrm{ns} = \mu_J(\tau_1^b)\big|_\textrm{central},
\\
\textrm{Variation 1}:~ & \mu_\textrm{ns} = \frac{\mu_J(\tau_1^b) + \mu_S(\tau_1^b)}{2}\bigg|_\textrm{central},
\\
\textrm{Variation 2}:~ & \mu_\textrm{ns} = Q,
\\
\textrm{Variations 3--4}:~ & \mu_\textrm{ns} = 2^{\pm 1}Q.
\end{split}
\end{align}
To properly account for the correlation in uncertainties between the singular and nonsingular cross sections when varying of $\mu=2^{\pm 1}Q$, we take the Variations 7--8 for the singular contributions in Eq.~\eqref{eq:singular_scale_variations} and Variations 3--4 for the nonsingular contributions in Eq.~\eqref{eq:nonsingular_scale_variations} simultaneously.

\section{Results}
\label{sec:results}
We present our predictions for the $\tau_1^b$ distributions at fixed values of $x$ and $Q$, which are relevant for HERA and EIC experimental setups. 
For this analysis, we fix 
the $\overline{\textrm{MS}}$ coupling $\alpha_s(m_Z) = 0.118$, along with the nonperturbative shift parameter of $\Omega_1(R_\Delta, \mu_\Delta) =0.5~\textrm{GeV}$ and the gap parameter $\Delta(R_\Delta,\mu_\Delta)=0.05~\textrm{GeV}$, as defined in Sec.~\ref{sec:shape-renormalon}.
These parameters are referenced at the $R$-gap scheme scales $R_\Delta=\mu_\Delta=2~\textrm{GeV}$. 
Looking ahead, future work will involve performing global fits to experimental data in order to precisely determine the values of $\alpha_s$, $\Omega_1$, and other nonperturbative parameters. In this work, we focus on establishing the theoretical framework that will serve as the basis for such analyses, while also exploring the feasibility of this approach through several test cases. 

\subsection{Theoretical predictions}
\begin{figure}
    \centering    
    \vspace{-1em}
    \includegraphics[width=1\linewidth]{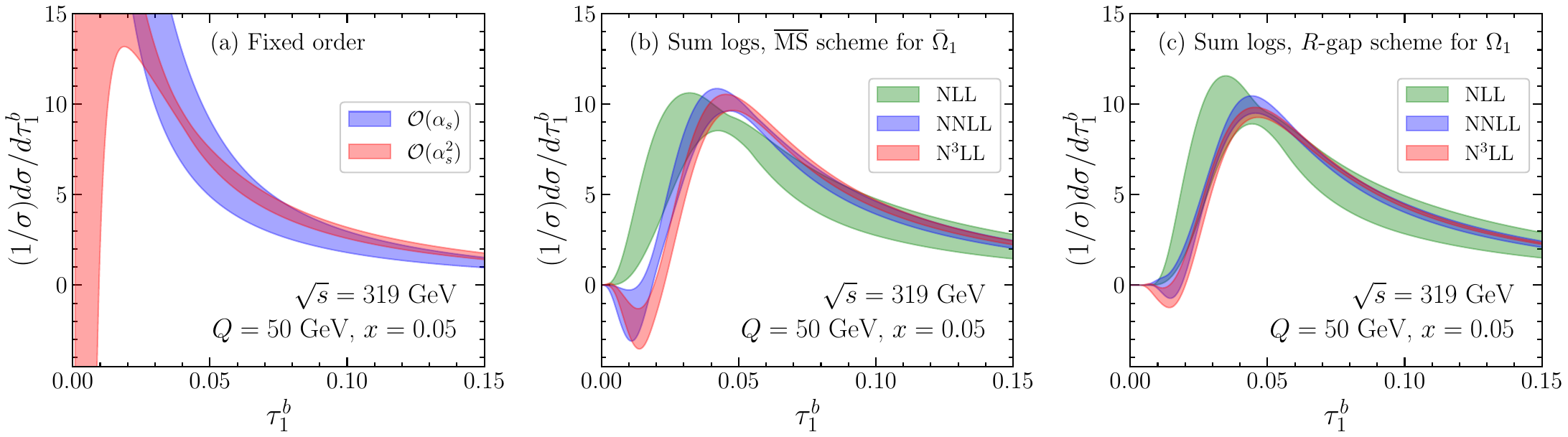}
    \vspace{-2em}
    \caption{The theory scan of perturbative QCD errors in the peak region, $\tau_1^b\in (0, 0.15)$. The panels show (a) fixed-order predictions at $\mathcal{O}(\alpha_s)$ (blue) and $\mathcal{O}(\alpha_s^2)$ (red), (b) resummation with a nonperturbative shape function using $\overline{\textrm{MS}}$ scheme for $\bar{\Omega}_1$, without renormalon subtraction, and (c) resummation with a nonperturbative shape function using $R$-gap scheme for ${\Omega}_1$, with renormalon subtraction.}
    \label{fig:PT_convergence_3}
\hspace{0.2cm}
    \centering
    \includegraphics[width=1\linewidth]{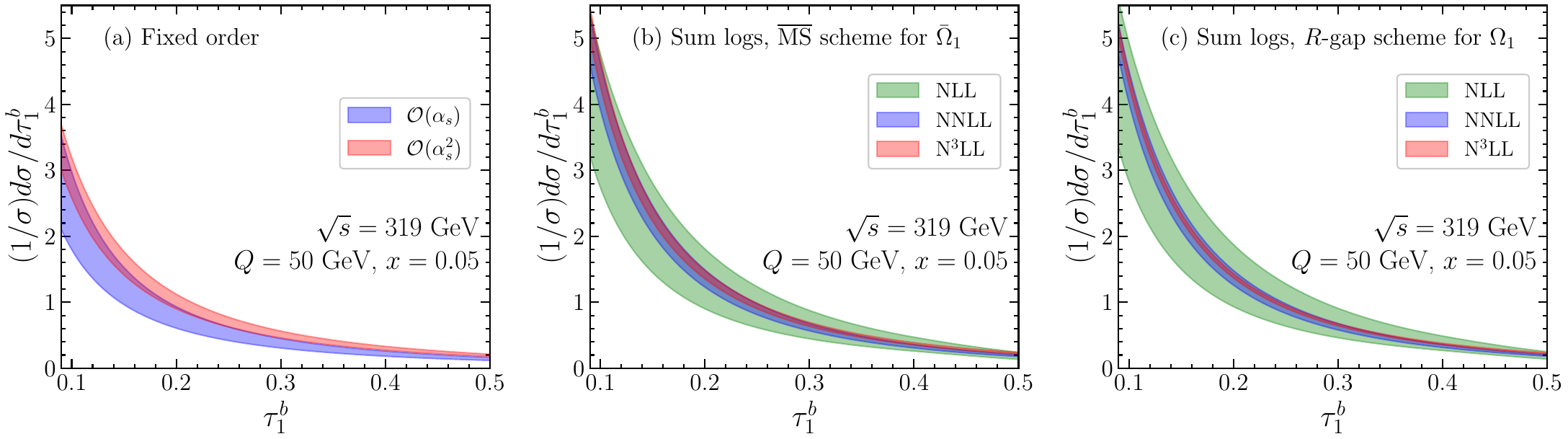}
    \caption{The same plots as Fig.~\ref{fig:PT_convergence_3}, focusing on the tail region, $\tau_1^b\in (0.1,0.5)$.}
    \label{fig:PT_convergence_2}
\end{figure}

First, in presenting the perturbative convergence of the theoretical predictions, we consider three different approaches at $\sqrt{s}=319~\textrm{GeV}$, $Q=50~\textrm{GeV}$, and $x=0.05$:
\begin{enumerate}[label=(\alph*)]
\item Fixed-order calculation: The cross section is calculated at fixed-order without any resummation, serving as a baseline for comparison. 
\item Resummation with $\overline{\textrm{MS}}$ scheme for $\bar{\Omega}_1$: This calculation includes resummation and convolution with the shape function, but without applying renormalon subtraction. 
\item Resummation with $R$-gap scheme with renormalon subtraction: This approach includes the subtraction of renormalon ambiguities using $R$-gap scheme, providing a more accurate treatment of both the perturbative and nonperturbative contributions.
\end{enumerate}
In Figs.~\ref{fig:PT_convergence_3} and \ref{fig:PT_convergence_2} we illustrate the results of using these three methods in the peak and tail regions, respectively. This provides a clearer picture of the convergence of the predictions in both the peak and tail regions. 
All distributions are normalized by their total cross section $\sigma$ to ensure that the area under each distribution is 1. 

Figs.~\ref{fig:PT_convergence_3}(a) and \ref{fig:PT_convergence_2}(a) show  the $\tau_1^b$ distributions at fixed-order, with blue curves representing  
$\mathcal{O}(\alpha_s)$ (LO) results and red curves showing  
$\mathcal{O}(\alpha_s^2)$ (NLO) predictions.
Each distribution is  normalized by the total cross section computed up to its respective fixed order. The $\mathcal{O}(\alpha_s)$ distribution is calculated analytically  
using the results from Ref.~\cite{Kang:2014qba}, while the  
$\mathcal{O}(\alpha_s^2)$ distribution is computed numerically  
via \texttt{NLOJet++} simulations. In both cases, the renormalization/factorization scales, along with their variations, are chosen according to the nonsingular scale $\mu_\textrm{ns}$ defined in Eq.~\eqref{eq:nonsingular_scale_variations}.
As $\tau_1^b$ approaches zero in the peak region, the fixed-order distributions diverge due to the presence of large Sudakov logarithms. These logarithms become increasingly significant at small $\tau_1^b$, necessitating their all-order resummation to obtain a physically meaningful prediction. In contrast, in the tail region (Fig.~\ref{fig:PT_convergence_2}), significant differences are observed between the fixed-order results [panel (a)] and the resummed results [panels (b) and (c)]. This demonstrates how resummation captures the correct cross-section behavior,
while fixed-order calculations alone underestimate theoretical uncertainties.
\begin{figure}
\centering
\vspace{-1em}
\includegraphics[width=0.6\linewidth]{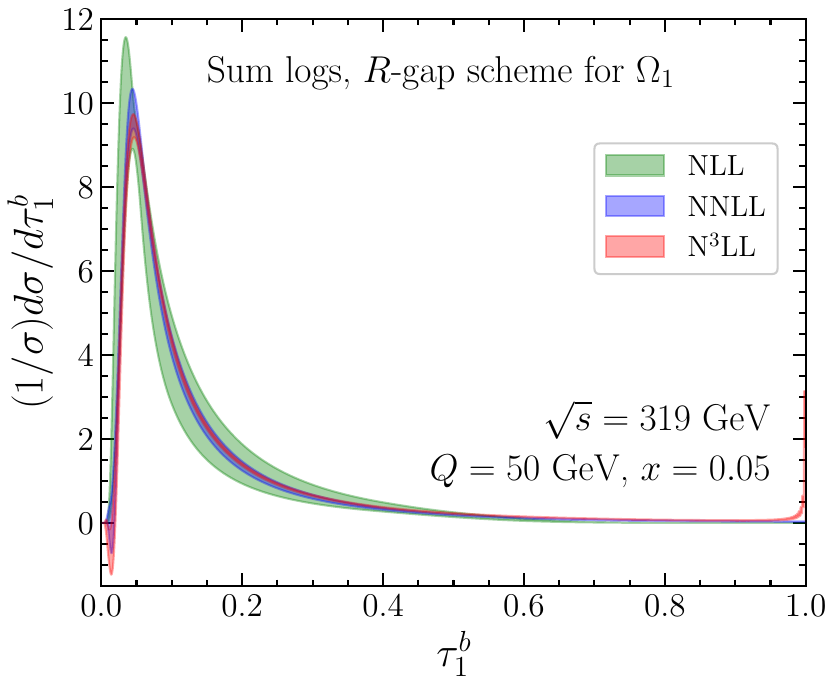}
\vspace{-1em}
    \caption{The differential cross section in $\tau_1^b$ across the full $\tau_1^b \in (0,1)$ range, incorporating resummation of logarithms and the shape function with $R$-gap renormalon subtraction.}
    \label{fig:PT_convergence_1}
\end{figure}

In Figs.~\ref{fig:PT_convergence_3}(b) and \ref{fig:PT_convergence_2}(b), we present the fully resummed distributions at NLL (green), NNLL (blue), and N$^3$LL (red) orders, incorporating the nonperturbative shape function $F$, as defined in Eq.~\eqref{eq:model-NP-soft}, without the renormalon subtraction scheme. 
Recall according to \tab{order} that NLL is matched only to tree-level fixed order, NNLL includes matching to 1-loop fixed order, and N$^3$LL is matched to 2-loop fixed order.
Here, the nonperturbative shift parameter $\bar{\Omega}_1$ is expressed in $\overline{\textrm{MS}}$ scheme via Eq.~\eqref{eq:no-gap}, without a gap. In the peak region, shown in Fig.~\ref{fig:PT_convergence_3}(b), the $\mathcal{O}(\Lambda_\textrm{QCD})$ renormalon becomes evident as $\tau_1^b$ approaches zero, leading to a negative cross section due to the renormalon ambiguity. Meanwhile, in the tail region shown in Fig.~\ref{fig:PT_convergence_2}(b), this implementation shifts the distribution by ${2\bar\Omega_1}/{Q}$ as expected, compared to the resummed distributions. In this case, the relative uncertainties in the tail region $\tau_1^b\in [0.1, 0.5]$ are about $\pm 30\%$ for NLL, $\pm 10\%$ for NNLL, and $\pm 5\%$ for N$^3$LL.

In Figs.~\ref{fig:PT_convergence_3}(c) and \ref{fig:PT_convergence_2}(c), we present the fully resummed distributions incorporating both the shape function and $R$-gap renormalon subtraction scheme, where $\Omega_1$ is now defined in the $R$-gap scheme. The adoption of the $R$-gap scheme significantly mitigates the renormalon ambiguities, as can be seen in Fig.~\ref{fig:PT_convergence_3}(c), where the behavior of the cross section stabilizes as $\tau_1^b$ approaches zero. (Subleading renormalons may still remain.) Moreover, the perturbative uncertainties in the tail region in Fig.~\ref{fig:PT_convergence_2}(c) are substantially reduced at both NNLL and N$^3$LL accuracy. After applying the renormalon subtraction beginning at NNLL accuracy, the relative uncertainties in the tail region decrease to about
$\pm 7.5\%$ for NNLL and $\pm 2.5\%$ for N$^3$LL, while remaining at $\pm 30\%$ for NLL. These results demonstrate the effectiveness of renormalon subtraction in improving theoretical precision. 
Based on these findings, we adopt the results obtained with $R$-gap renormalon subtraction scheme for further analysis in this work. Finally, in Fig.~\ref{fig:PT_convergence_1}, we present our final prediction for the $\tau_1^b$ distributions, incorporating all the key theoretical ingredients  discussed so far, covering the full $\tau_1^b\in (0,1)$ range.

\begin{figure}
    \centering
    \vspace{-1em}
    \begin{subfigure}
        \centering
        \includegraphics[width=0.32\linewidth]{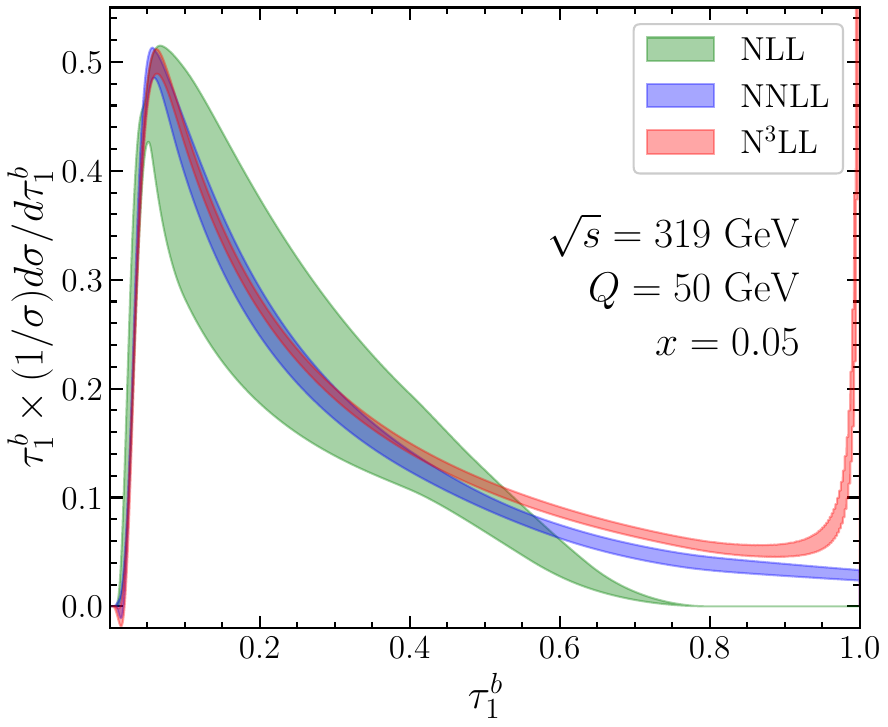}
    \end{subfigure}
    \begin{subfigure}
        \centering
        \includegraphics[width=0.32\linewidth]{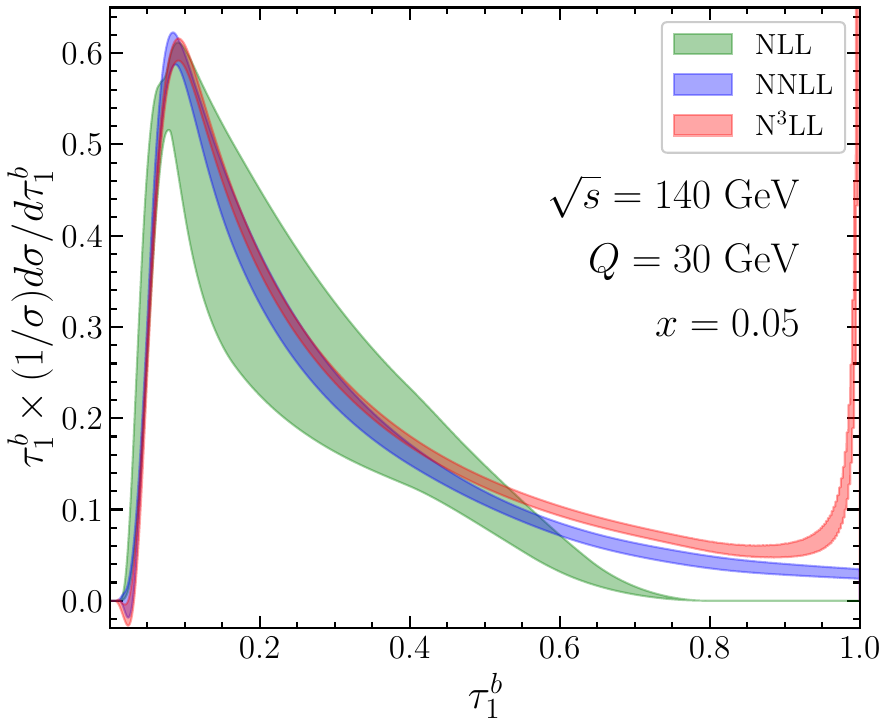}
    \end{subfigure}
    \begin{subfigure}
        \centering
        \includegraphics[width=0.32\linewidth]{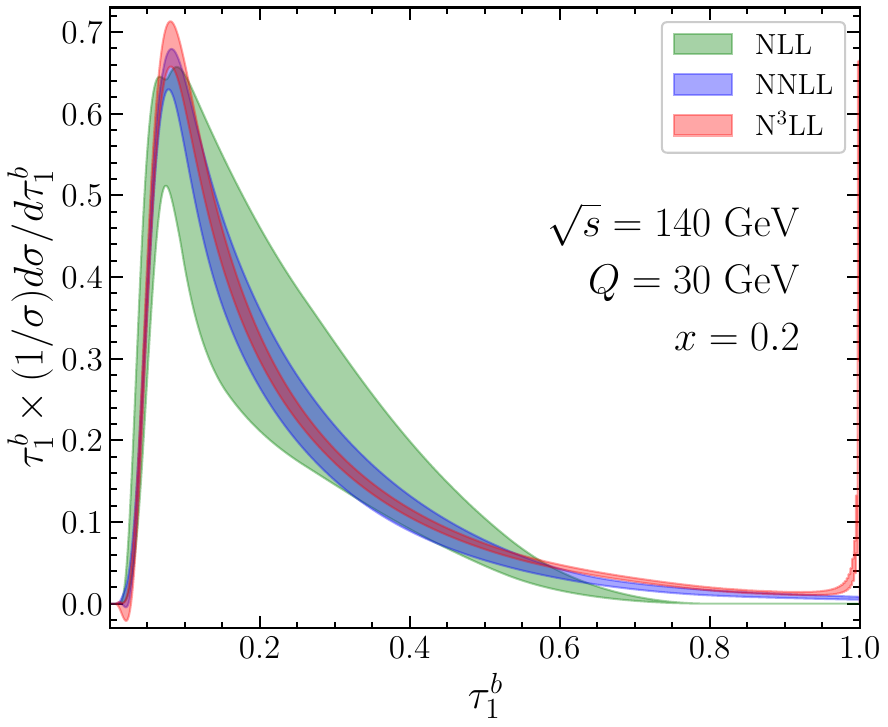}
    \end{subfigure}
    \vspace{-1em}
    \caption{The differential cross sections multiplied by $\tau_1^b$ at various $(\sqrt{s}, Q,x)$, normalized by each total cross section. 
    The first panel is the same as \fig{PT_convergence_1}, just multiplied by $\tau_1^b$.}
    \label{fig:weighted_distributions}
\end{figure}
In Fig.~\ref{fig:weighted_distributions}, we present the weighted distributions of the differential cross sections, expressed as  $\tau_1^b\times d\sigma/d\tau_1^b$, normalized by each total cross sections $\sigma$. The weighting by $\tau_1^b$ enhances the visibility of the distributions, as the differential cross sections decrease sharply with increasing $\tau_1^b$, as previously shown in Fig.~\ref{fig:PT_convergence_1}. 
The three parameter sets of $(\sqrt{s}, Q,x)$ for HERA and EIC kinematics are
\begin{enumerate}
    \item \makebox[5.5cm][l]{HERA (left panel):}        $(319~\mathrm{GeV}\,, 50~\mathrm{GeV}\,, 0.05)$
    \item \makebox[5.5cm][l]{EIC (center panel):}       $(140~\mathrm{GeV}\,, 30~\mathrm{GeV}\,, 0.05)$
    \item \makebox[5.5cm][l]{EIC at larger $x$ (right panel):} $(140~\mathrm{GeV}\,, 30~\mathrm{GeV}\,, 0.2)$
\end{enumerate}
The plots display the good perturbative convergence of the bands for the three resummation orders (NLL, NNLL, and N$^3$LL), extending up to $\tau_1^b\approx 0.4$. Beyond this point, the N$^3$LL results start to diverge from the NLL and NNLL, which can be attributed to significant contributions from nonsingular terms at $\mathcal{O}(\alpha_s^2)$. Notably, the characteristic singular behavior as $\tau_1^b\to1$ becomes evident, corresponding to events where the jet hemisphere is empty or near empty, a phenomenon also discussed in Ref.~\cite{Kang:2014qba} at $\mathcal{O}(\alpha_s)$. 

\subsection{Comparison with H1 measurements}\label{sec:HERA-comparison}
Recently, the H1 collaboration at HERA reported measurements of the 1-jettiness event shape observable \cite{H1:2024aze, H1:2024nde}. 
This allows a direct comparison of our N$^3$LL + $\mathcal{O}(\alpha_s^2)$ predictions with the experimental data.
The data, taken from the H1 detector from 2003 to 2007, results from electron/positron and proton collisions at the center-of-mass energy $\sqrt{s}=319~\textrm{GeV}$. The measurements provide triple binned differential cross sections in $Q^2$, $y$, and $\tau_1^b$ for 32 sets of $(Q^2, y)$ bins. 

As an illustration, we compare our theoretical predictions with two of the measurements from Ref.~\cite{H1:2024aze}. One corresponds to the range $1100<Q^2/\textrm{GeV}^2<1700$ and $0.4<y<0.7$, and the other to $700<Q^2/\textrm{GeV}^2<1100$ and $0.4<y<0.7$.
\begin{figure}
\vspace{-1em}
    \centering
    \begin{subfigure}
        \centering
        \includegraphics[width=0.49\linewidth]{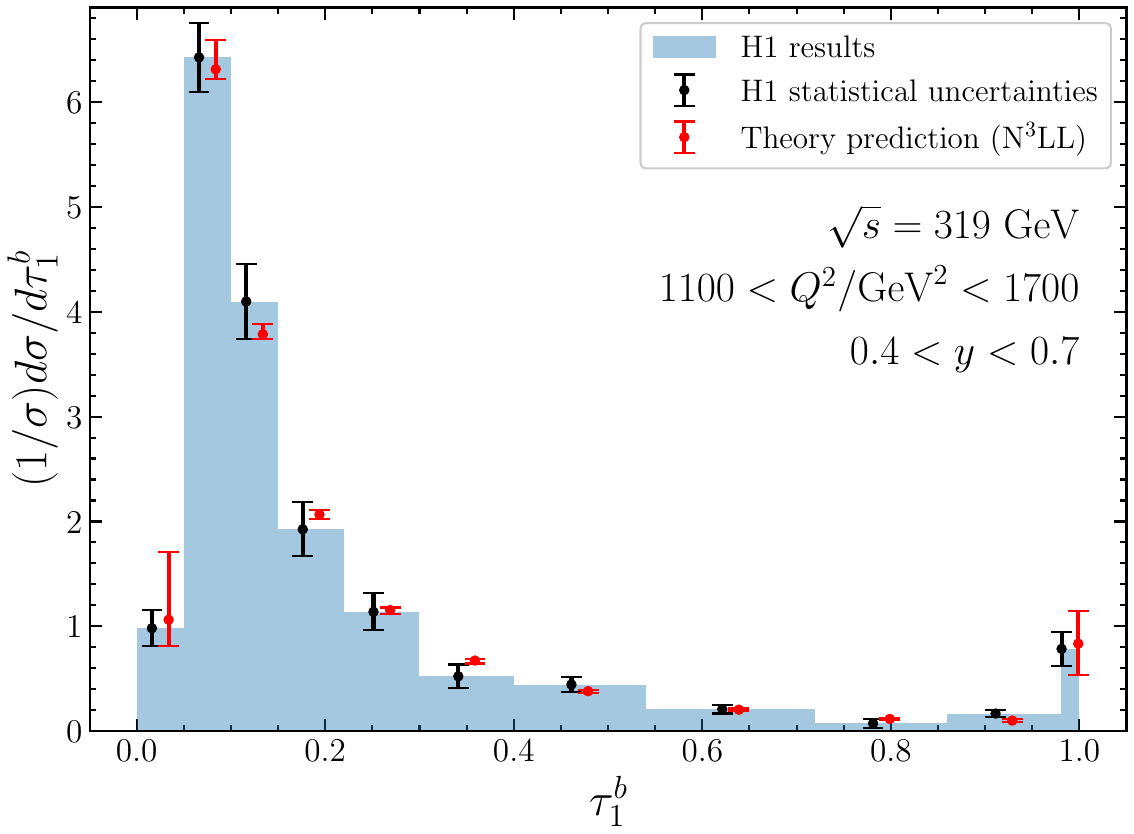}
    \end{subfigure}
    \begin{subfigure}
        \centering
        \includegraphics[width=0.49\linewidth]{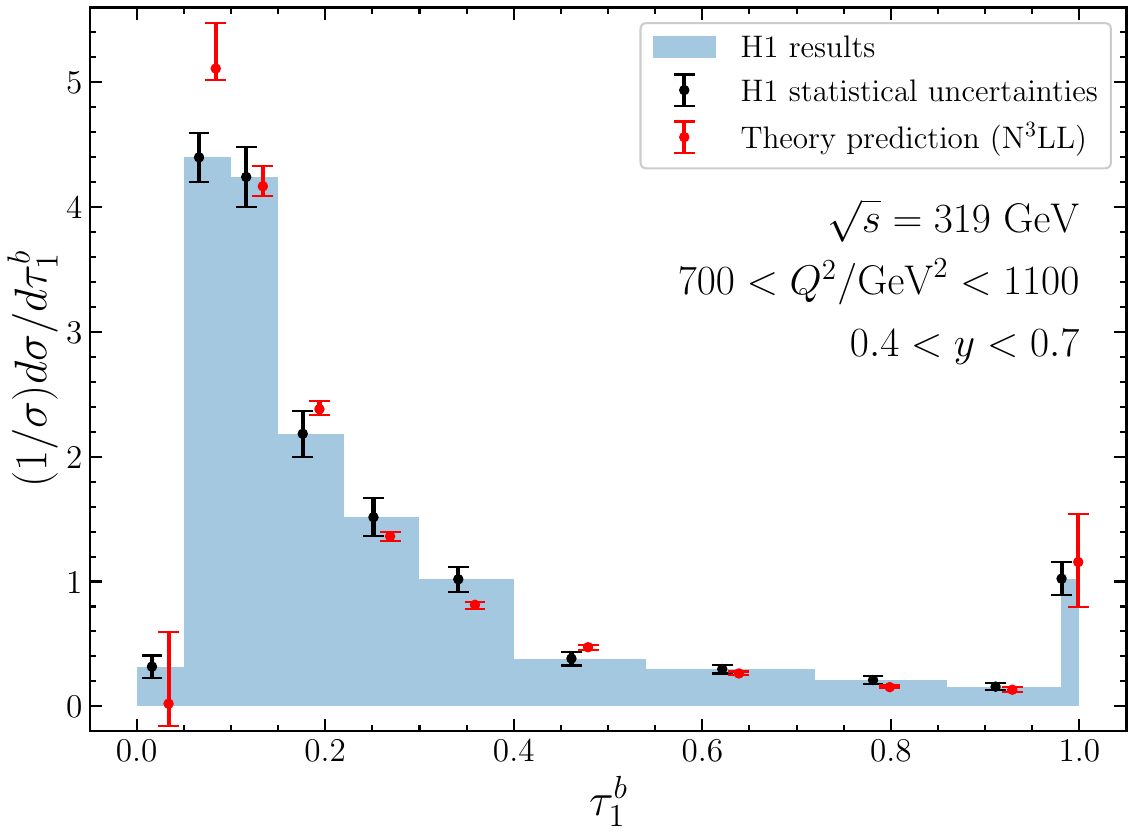}
    \end{subfigure}
    \vspace{-2em}
    \caption{Comparison of differential cross section measurements from the H1 collaboration at HERA \cite{H1:2024aze} (black points) with our theoretical predictions (red points). The left and right panels show results from two different bins in $Q^2$ with the same bin in $y$.}
    \label{fig:HERA_measurement}
\end{figure}
The histograms for these two measurements are displayed in Fig.~\ref{fig:HERA_measurement}. The central values of the cross sections are shown as blue histograms, with uncertainties for each bin represented by black error bars. Our theoretical predictions for the singular cross sections are computed as triply differential in $Q^2$, $x$, and $\tau_1^b$. To match the H1 data, we first  change variables from $x$ to $y$, using the kinematic constraint $Q^2=sxy$. Next, we generate a grid of $\tau_1^b$ distributions  for each of the H1 measurements (e.g., generating $7\times7=49$ distributions in $Q^2$ and $y$ within the range $1100<Q^2/\textrm{GeV}^2<1700$ and $0.4<y<0.7$). These distributions are then interpolated tricubically in $Q^2$, $y$, and $\tau_1^b$, and finally, integrated over $Q^2$, $y$ and $\tau_1^b$ to compute
$\Delta\sigma$ for finite $\Delta y$, $\Delta Q^2$, and $\Delta\tau_1^b$. For the nonsingular cross sections, we directly compute the fixed-order results using \texttt{NLOJet++}, applying the same kinematic cuts on $y$ and $Q^2$. To derive the nonsingular cross sections from the \texttt{NLOJet++} results, we should subtract them by the fixed-order singular cross sections. As we have done for the resummed singular cross sections, we compute the triple-binned fixed-order singular cross section using the same method.

In Fig.~\ref{fig:HERA_measurement}, we compare our theoretical predictions (shown with red error bars) to the H1 measurements, both normalized by their total cross sections.\footnote{For the first case in Fig.~\ref{fig:HERA_measurement}, the total cross section measured by H1 collaboration is $(23.0\pm 0.9)$ pb (with the statistical errors only), while our prediction is $21.6^{+0.8}_{-0.3}$ pb. In the second case, the H1 measurement is $(41.4\pm 1.2)$ pb (with the statistical errors only), while our prediction is  $39.8^{+1.5}_{-0.6}$ pb. The slight discrepancy in the total cross sections may be due to the missing $Z^0$-contribution in the nonsingular term. However, normalizing both results by their total cross sections mitigates the effect of this missing  contribution, enabling a more consistent comparison.} Our predictions show good agreement with the H1 measurements, particularly in the tail and far-tail region. 
The relatively larger deviation in the peak region, particularly for the lower $Q$ data (right panel of Fig.~\ref{fig:HERA_measurement}), may be attributed to the simplicity of the shape function chosen with $N=0$ in \eq{model-NP-soft}, which lacks free parameters to vary higher moments. Readers are referred to Appendix~\ref{app:with_c2} for a discussion on the effects of the higher moments.
This overall agreement underscores the potential of theoretical and experimental results for DIS event shapes to determine universal physical quantities in DIS, analogous to the event shape analyses in $e^+e^-$ collisions. Such studies are expected to play a crucial role in future hadronic physics research, particularly at the EIC.  

\begin{figure}[t!]
    \centering
    \begin{subfigure}
        \centering
        \includegraphics[width=0.49\linewidth]{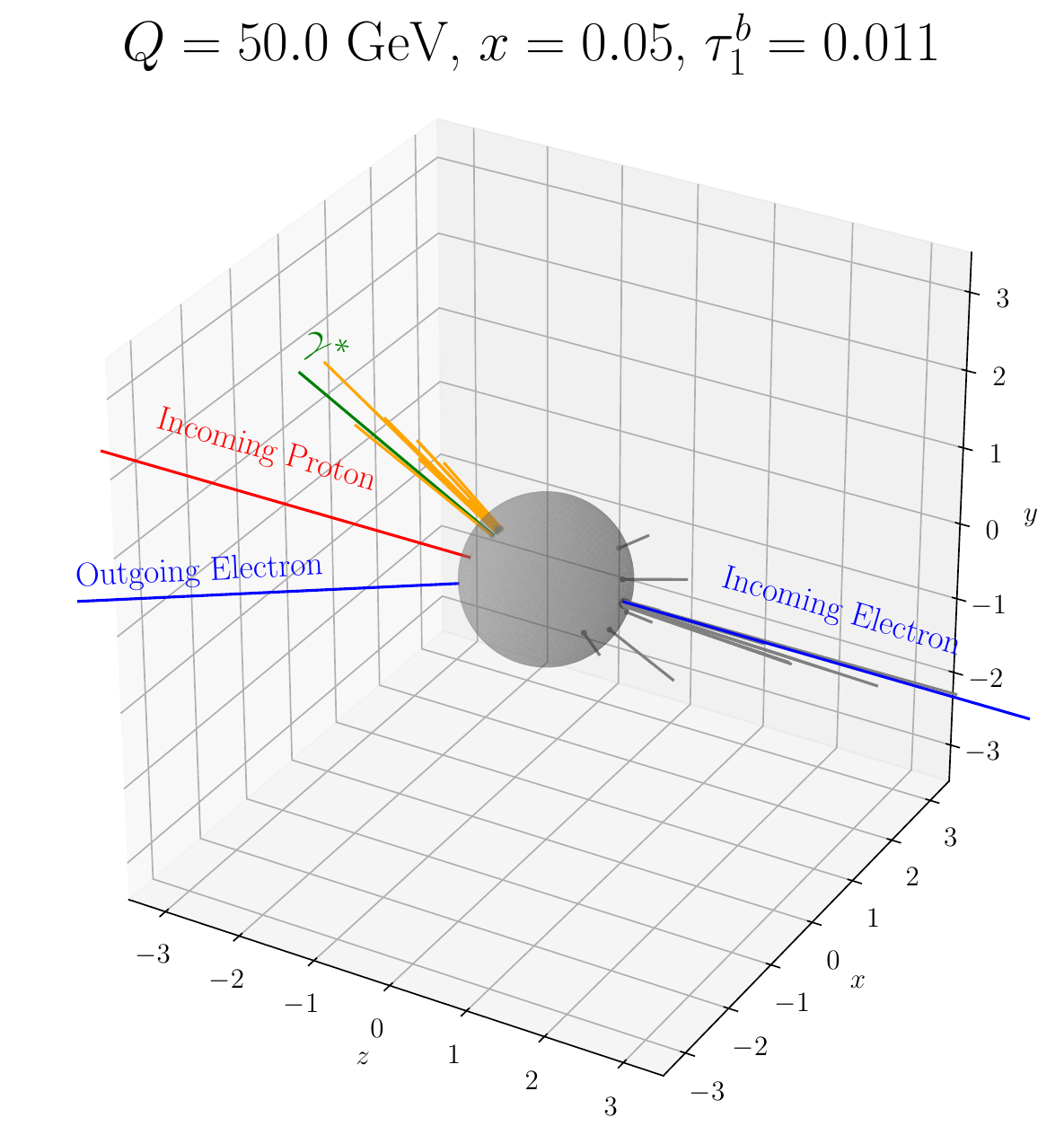}
    \end{subfigure}
    \begin{subfigure}
        \centering
        \includegraphics[width=0.49\linewidth]{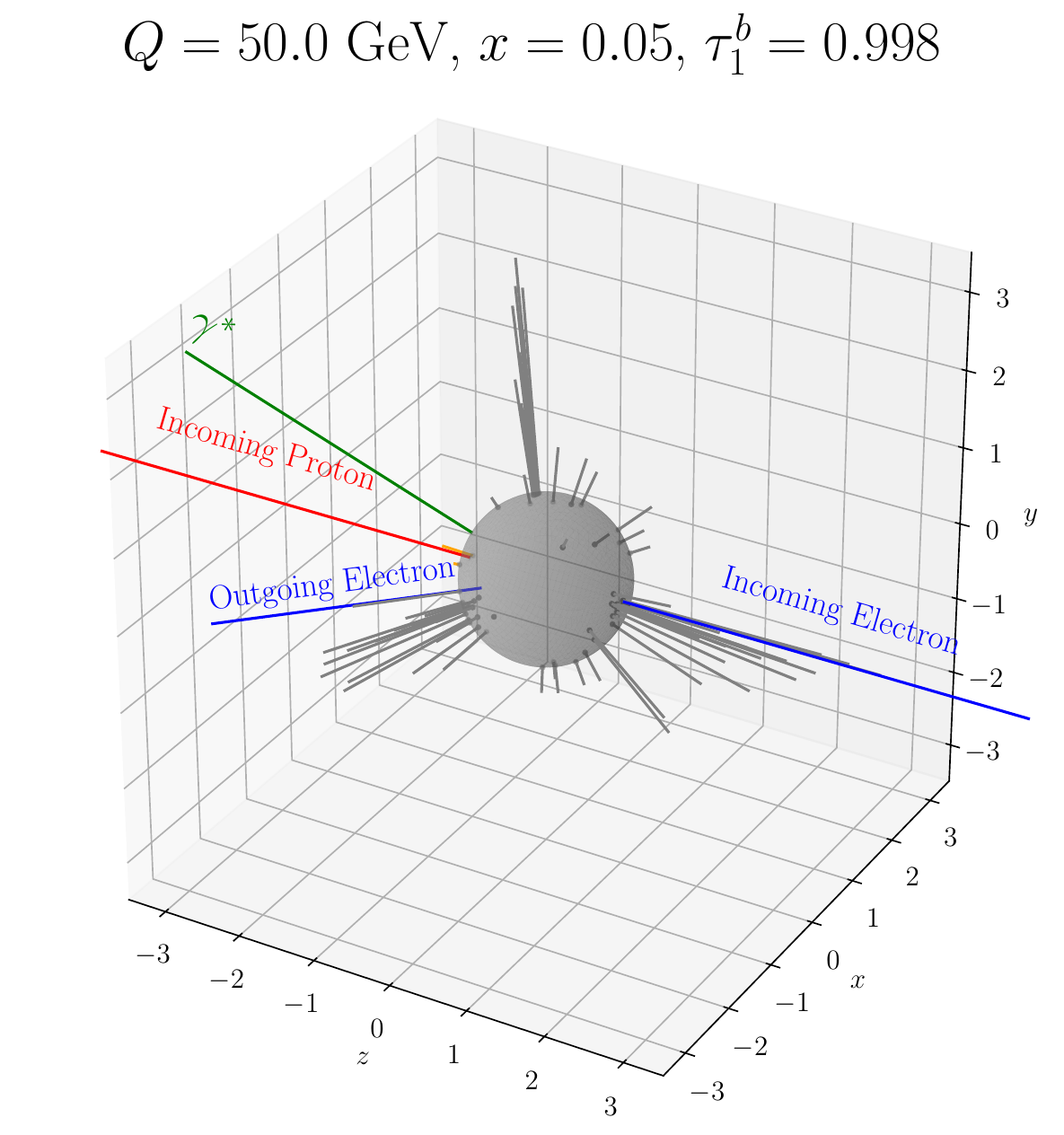}
    \end{subfigure}
    \vspace{-2em}
    \caption{Visualization of final-state momentum configurations in the CM frame at $\sqrt{s}=319~\textrm{GeV}$, generated using Pythia 8. The left panel shows an event with small $\tau_1^b$, while the right panel shows an event with large $\tau_1^b\approx 1$. The corresponding values of $Q$, $x$, and $\tau_1^b$ are indicated above each figure. The blue lines represent the incoming and outgoing electron momenta, the red line represents the incoming proton momentum, and the green line represents the outgoing off-shell photon momentum. The yellow lines correspond to the final state momenta in the jet hemisphere ($\mathcal{H}_J$), while the gray lines correspond to those in the beam hemisphere ($\mathcal{H}_B$). The lengths of the final-state momenta represent their energies (with $Q$ used for $\gamma^*$).}
    \label{fig:pythia}
\end{figure}
Interestingly, both our predictions and H1 results exhibit a peak structure as $\tau_1^b\to 1$. As noted in Ref.~\cite{H1:2024nde}, events with an empty current hemisphere ($\tau_1^b\approx 1$) predominantly feature two jets with $p_T>7~\textrm{GeV}$, indicating that these events are effectively 3-jet configurations, including contributions from initial state radiation.\footnote{Ref.~\cite{H1:2024nde} also highlights that, in jet multiplicity measurements, contributions from proton remnants and initial-state radiation are significantly suppressed due to limited detector acceptance. This implies the observed two jets in the empty jet hemisphere corresponds to mostly 3-jet events.} To further illustrate these findings, Fig.~\ref{fig:pythia} shows Pythia 8 \cite{Bierlich:2022pfr} simulations of events at $\sqrt{s}=319~\textrm{GeV}$, visualizing momentum configurations in the CM frame for small $\tau_1^b\approx 0$ (left) and large $\tau_1^b\approx 1$ (right).\footnote{The length of the $i$th final-state momentum represents its energy $E_i$, which is log-normalized as $\log(E_i/(1~\textrm{GeV}) + 1)$ for improved visualization. For the off-shell photon, we used $\log(Q/(1~\textrm{GeV}) + 1)$.}
In the small $\tau_1^b$ event (left), we observe a typical dijet configuration, with balanced momentum distributions in both hemispheres. In contrast, the large $\tau_1^b$ event (right) corresponds to a 3-jet configuration, with nearly all final-state momenta concentrated in the beam hemisphere. 
These configurations align with the findings of Ref.~\cite{H1:2024nde}, highlighting that cross sections at $\tau_1^b\approx 1$ effectively capture multi-jet events.

\subsection{Sensitivity to fundamental quantities of QCD}
\begin{figure}
    \centering
    \begin{subfigure}
        \centering
        \includegraphics[width=0.48\linewidth]{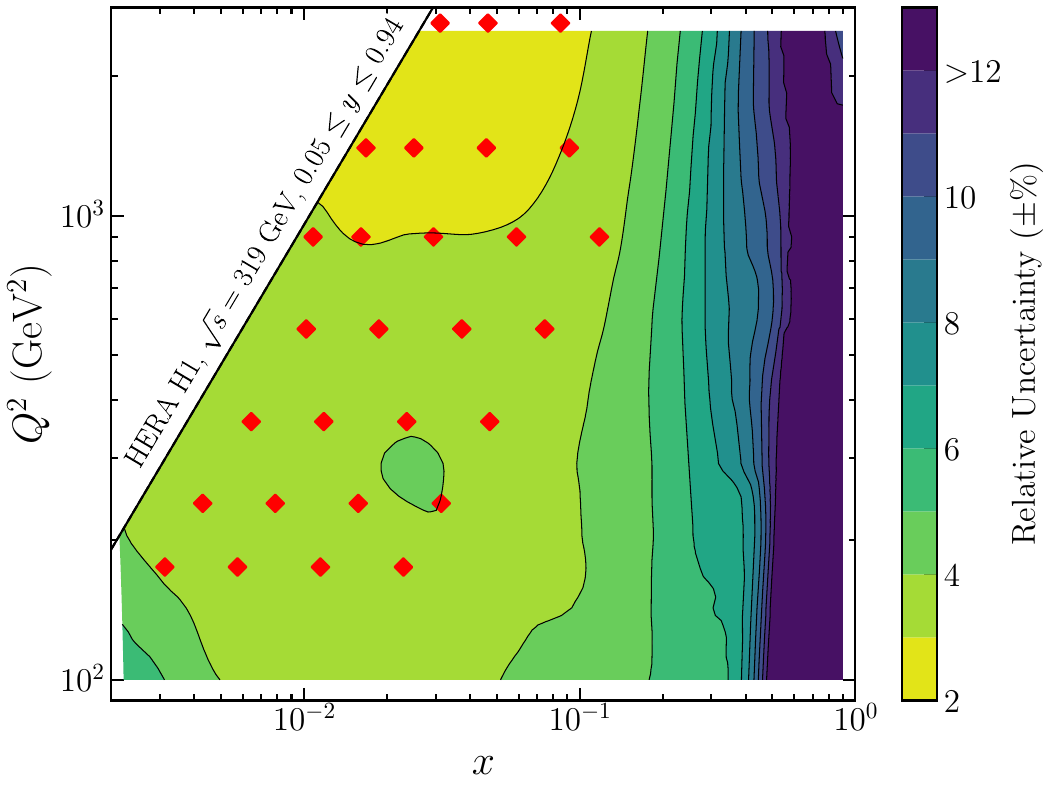}
    \end{subfigure}
    \begin{subfigure}
        \centering
        \includegraphics[width=0.48\linewidth]{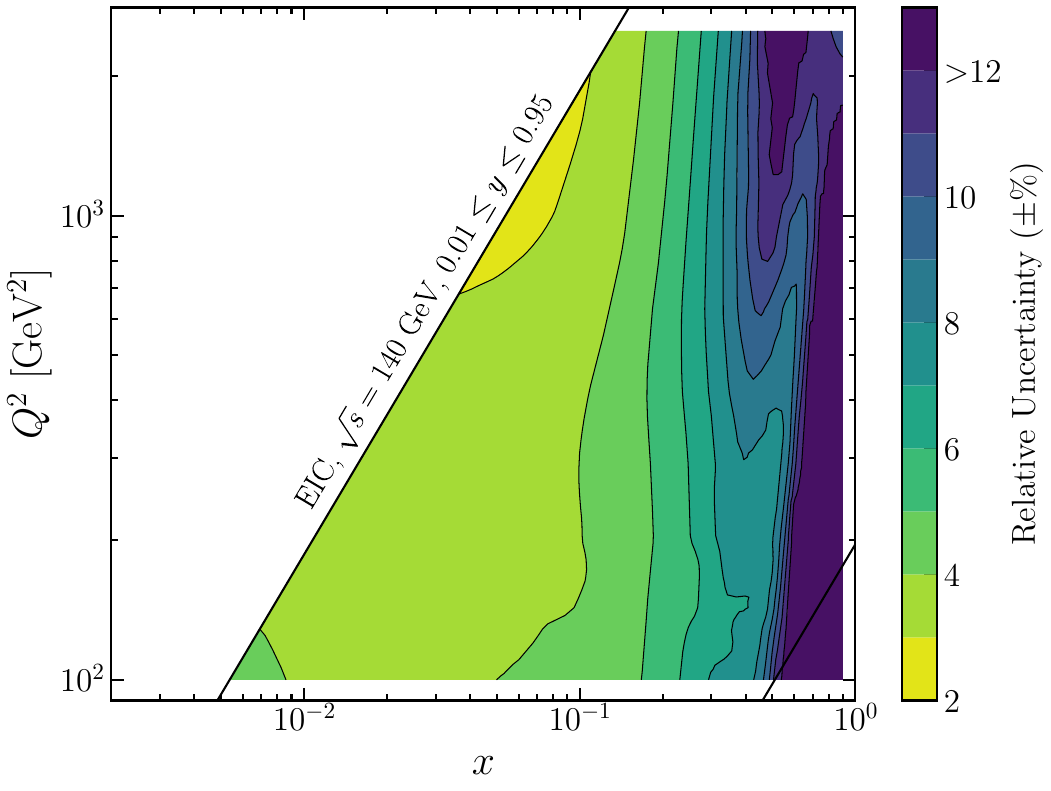}
    \end{subfigure}
    \vspace{-1em}
    \caption{Contours of  theoretical perturbative uncertainties for HERA (left) and EIC (right). In the left plot, the percent uncertainty is shown in $x$--$Q^2$ plane for HERA, where red diamonds indicating the values of $x$ and $Q$ corresponding to the H1 measurements in Ref.~\cite{H1:2024aze}. The right plot shows the contours of  theoretical perturbative uncertainties for the EIC.
}
\label{fig:HERA_EIC_sensitivity}
\end{figure}

In Fig.~\ref{fig:HERA_EIC_sensitivity}, we display $\text{N}^3\text{LL} + \mathcal{O}(\alpha_s^2)$ 
uncertainty contours in the $x$--$Q^2$ plane, overlaid on the kinematic coverage regions of HERA (left) 
and EIC (right).
Our theory predictions are shown for $Q\lesssim 50~\textrm{GeV}$ due to the absence of $Z^0$ boson contribution in the nonsingular cross section.\footnote{Both the analytic $\mathcal{O}(\alpha_s)$ results in Ref.~\cite{Kang:2014qba} and the \texttt{NLOJet++} calculations include only photon-channel contributions.} However, since we include the singular $Z^0$ boson contribution, our predictions remain largely valid even around $Q\sim m_Z$, particularly in the tail region of $\tau_1^b$ distributions, where singular contributions dominate. 
For the assessment of perturbative uncertainties, we focus on the tail region of $\tau_1^b$ [i.e., $\tau_1^b \in (t_1, t_2)$], where the profile functions follow the canonical form as described in Eq.~\eqref{eq:profile_setting}. 
The plots indicate optimal theoretical precision 
in the region of $x \sim 0.05$ at higher $Q$ values. While HERA's higher center-of-mass energy allows better precision in certain regions, the future EIC is expected to explore previously inaccessible regions of $x$ and $Q$ while offering significantly improved statistical precision due to its much higher luminosity. 

\begin{figure}
    \centering
    \begin{subfigure}
        \centering
        \includegraphics[width=0.49\linewidth]{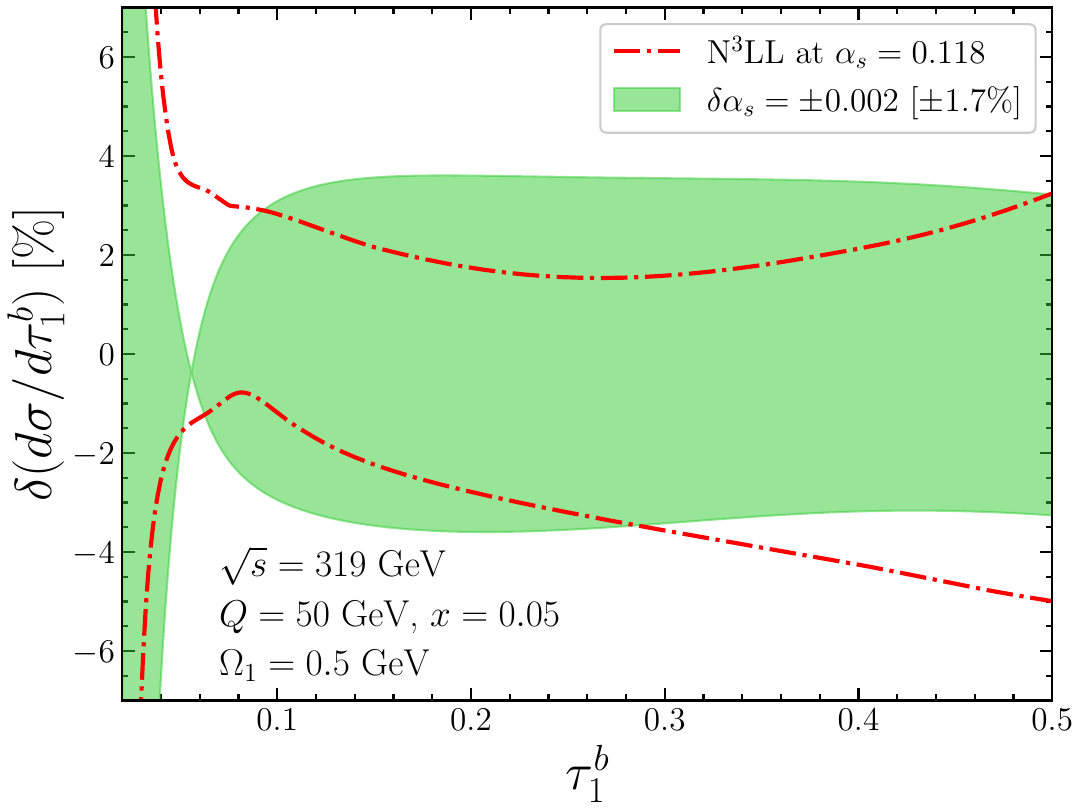}
    \end{subfigure}
    \begin{subfigure}
        \centering
        \includegraphics[width=0.49\linewidth]{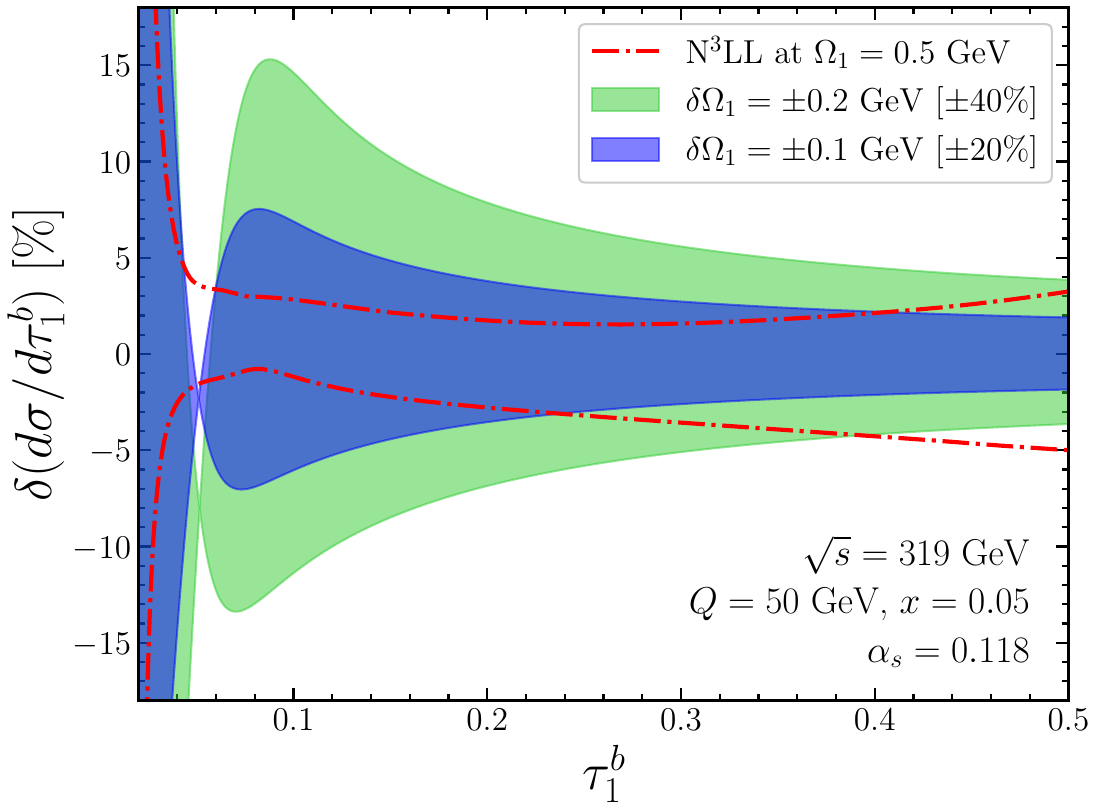}
    \end{subfigure}
    \vspace{-2em}
    \caption{Sensitivity of the differential cross section in $\tau_1^b$ to variations in $\alpha_s$ (left) and $\Omega_1$ (right) at $\sqrt{s}=319~\textrm{GeV}$, $Q=50~\textrm{GeV}$ and $x=0.05$. The red dot-dashed lines show the bounds for the theoretical perturbative uncertainty at this order.}
    \label{fig:sensitivity}
\end{figure}

To assess the sensitivity of our theoretical predictions to variations in $\alpha_s(M_Z)$ and $\Omega_1$, we present two sensitivity plots in Fig.~\ref{fig:sensitivity}. In the left panel, we vary $\alpha_s(M_Z)$ by $\pm 0.002$ from the central value $\alpha_s(M_Z)=0.118$ and compare the resulting variations to the relative uncertainty of the central prediction. Detecting a $\pm 0.002$ variation in $\alpha_s(M_Z)$ requires an accuracy of better than $\pm4\%$. The figure shows that in the tail region, our predictions yield uncertainties of approximately $\pm 1\%$ to $\pm 2\%$, indicating that they are sufficiently precise for determining $\alpha_s(M_Z)$ to this level. 
In the right panel of Fig.~\ref{fig:sensitivity}, we vary $\Omega_1$ by $\pm 0.1~\textrm{GeV}$ and $\pm 0.2~\textrm{GeV}$ from its central value $\Omega_1 = 0.5~\textrm{GeV}$. The results indicate that our predictions can determine $\Omega_1$ with an accuracy of within $\pm 0.1~\textrm{GeV}$ in the tail region. 
The sensitivity to $\alpha_s(M_Z)$ and $\Omega_1$ exhibit distinct dependencies on $\tau_1^b$. While $\Omega_1$ exhibits stronger constraints in the lower range of $\tau_1^b$ (especially around $\tau_1^b\sim 0.1$ for the kinematics in Fig.~\ref{fig:sensitivity}), $\alpha_s(M_Z)$ maintains relatively uniform sensitivity across a wider range of $\tau_1^b$.
These sensitivities will vary with different $x,Q$, and combining fits from many such kinematic points will presumably improve the sensitivity estimates given above for a single kinematic point in $x,Q$, especially as varying these can play a big role in breaking degeneracies.

\begin{figure}
    \centering
    \begin{subfigure}
        \centering
        \includegraphics[width=0.49\linewidth]{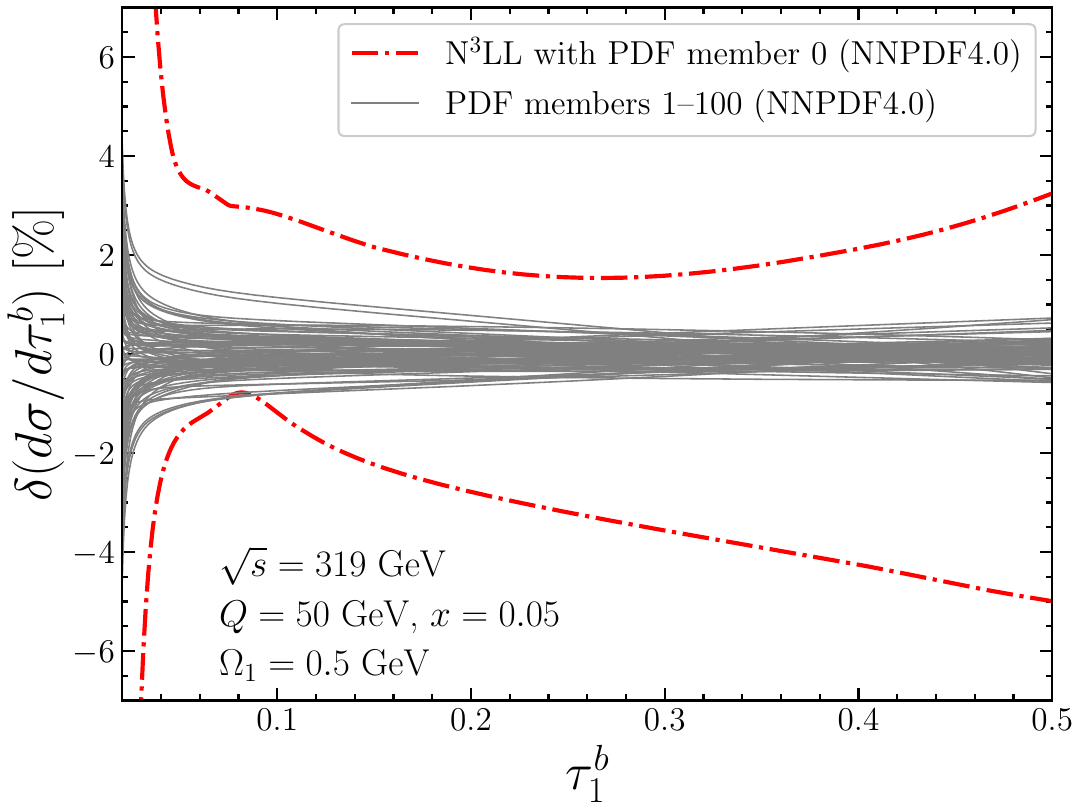}
    \end{subfigure}
    \begin{subfigure}
        \centering
        \includegraphics[width=0.49\linewidth]{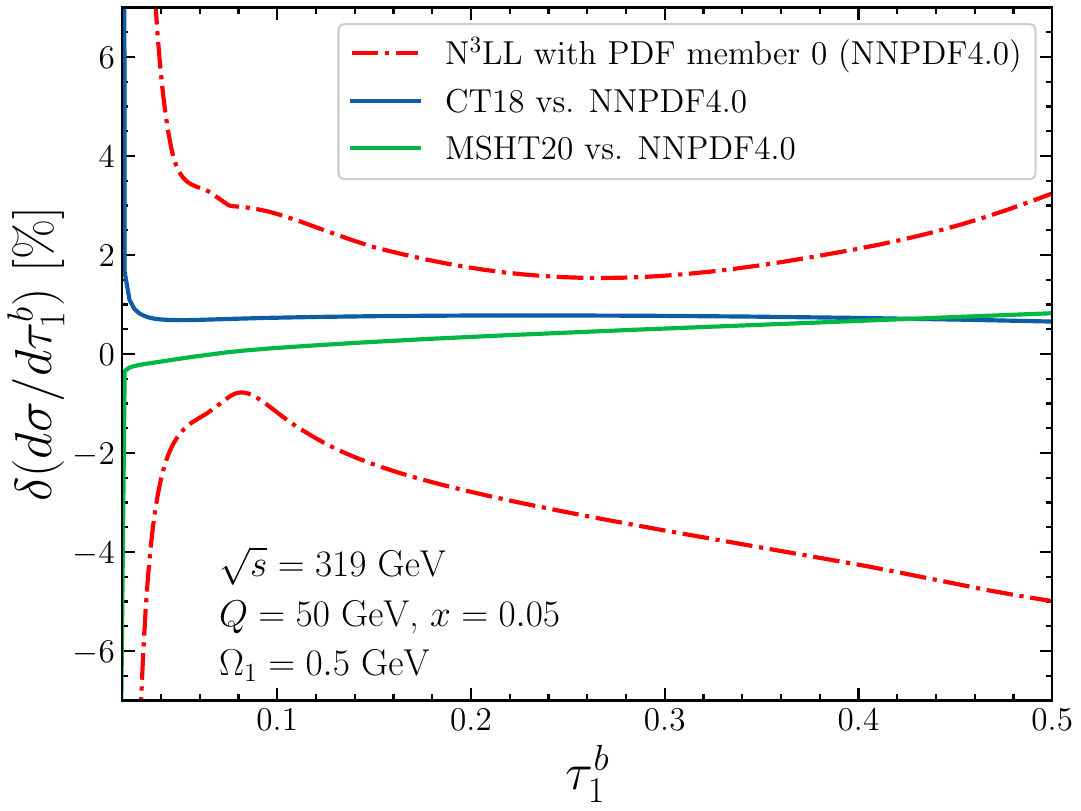}
    \end{subfigure}
    \vspace{-2em}
    \caption{Sensitivity of the differential cross section in $\tau_1^b$ to PDF uncertainties. The left panel shows the variations in $\tau_1^b$ distributions across different PDF members (1--100) of NNPDF4.0. The right panel shows the shifts in the $\tau_1^b$ distributions for different PDF sets (CT18 and MSHT20) relative to the central prediction from NNPDF4.0. The red dot-dashed lines indicate the bounds for the theoretical perturbative uncertainty at N$^3$LL.}
    \label{fig:pdf_sensitivity}
\end{figure}

Unlike $e^+e^-$ collisions, DIS is sensitive to PDF uncertainties. To estimate the impact of these uncertainties, we vary over PDFs provided by NNPDF4.0, as well as central PDF determinations from different collaborations. The NNPDF4.0 set \cite{NNPDF:2021njg} provides PDF member 0 as the central prediction and an additional 100 PDF members corresponding to 100 MC replicas, to assess PDF uncertainties. 
All predictions we have presented so far were obtained using the central PDF member 0. To quantify the impact of PDF variations, we compare the relative perturbative uncertainties at full N$^3$LL$+\mathcal{O}(\alpha_s^2)$ with the shifts in the central predictions across the 100 PDF members in Fig.~\ref{fig:pdf_sensitivity} (left), and with the shifts in the central predictions for different PDF sets, including CT18~\cite{Hou:2019efy} and MSHT20~\cite{Bailey:2020ooq} in Fig.~\ref{fig:pdf_sensitivity} (right). In both cases, the impact of the PDF uncertainties on the theoretical predictions of the $\tau_1^b$ distributions is estimated to be at the 1\% level, which remains well within our theoretical perturbative uncertainties. 

Overall, as shown in these sensitivity plots, our predictions depend on the two parameters $\alpha_s(M_Z)$ and  $\Omega_1$. Therefore, when fitting to experimental data, the best-fit values of these parameters will be correlated.
We expect that the broad kinematic coverage of $x$--$Q^2$ plane in DIS will provide an opportunity to disentangle these correlations, mitigate the impact of PDF uncertainties, and ultimately achieve precise determinations of $\alpha_s$ and $\Omega_1$. 

\subsection{Comparison with previous studies}
This work focuses on the DIS event shape $\tau_1^b$, building upon the factorization theorem and NNLL resummation established in Ref.~\cite{Kang:2013nha} using SCET, and the analytic $\mathcal{O}(\alpha_s)$ nonsingular calculation in Ref.~\cite{Kang:2014qba}. In this work, we extend these results by incorporating the missing $\mathcal{O}(\alpha_s^2)$ contributions using \texttt{NLOJet++} (photon channel only), adding nonperturbative corrections and renormalon subtractions. As a result, we achieve full N$^3$LL$+\mathcal{O}(\alpha_s^2)$ accuracy, including universal nonperturbative effects.

For other 1-jettiness variables, such as $\tau_1$ (similar to $\tau_1^a$ in the notation of Ref.~\cite{Kang:2013nha}), theoretical results are also available at N$^3$LL + $\mathcal{O}(\alpha_s^2)$ \cite{Cao:2024ota}. These event shapes use a reference vector $q_J$, determined by a jet algorithm, which aligns with the jet momentum $p_J$ and simplifies the beam function (from the transverse momentum dependent beam function to the ordinary beam function). 
One of the differences from our analysis lies in the treatment of profile functions.
The profile function in Ref.~\cite{Cao:2024ota} is applied uniformly across $x$ and $Q$, whereas our approach dynamically adjusts the profile function based on the dominance of singular or nonsingular contributions, depending on the value of $x$ and $Q$. This flexibility enables a  more extensive study of DIS over a broad kinematic range. 
Ref.~\cite{Cao:2024ota} also employed \texttt{NLOJet++} to compute the $\mathcal{O}(\alpha_s)$ and $\mathcal{O}(\alpha_s^2)$ fixed-order contributions, demonstrating asymptotic agreement between fixed-order singular and full QCD results. 
It would interesting to use these nonsingular results to perform the analysis for $\tau_{1}$ that we performed for $\tau_{1}^b$ in Sec.~\ref{sec:nonsing}.
%Furthermore, while their hard function accounts for $Z^0$-boson contributions, it omits the flavor-singlet terms at $\mathcal{O}(\alpha_s^2)$, as detailed in Appendix~\ref{app:flavor-singlet}. 

Another key improvement in our work is the inclusion of $\mathcal{O}(\Lambda_\textrm{QCD})$ renormalon subtraction, which reduces renormalon ambiguities in the peak region and enhances perturbative convergence and precision in the tail region, implemented through a gapped shape function model that makes universality of the first moment parameter $\Omega_1$ manifest. We anticipate that applying a similar renormalon subtraction procedure to the $\tau_1$ analysis in Ref.~\cite{Cao:2024ota} could further improve the theoretical predictions.

\section{Conclusions}
\label{sec:conclusion}
We present high-precision predictions for the DIS 1-jettiness event shape (thrust), $\tau_1^b$, at N$^3$LL resummed accuracy matched to $\mathcal{O}(\alpha_s^2)$ fixed-order accuracy in QCD, including universal nonperturbative corrections and $\mathcal{O}(\Lambda_\textrm{QCD})$ renormalon subtractions. The resummation is implemented through RG evolution of the factorization formula within SCET. The fixed-order $\mathcal{O}(\alpha_s^2)$ nonsingular contributions are included using \texttt{NLOJet++} results, and the contributions are rigorously verified and validated. Nonperturbative soft effects are modeled using a gapped shape function respecting the universality of the first moment parameter $\Omega_1$, with leading-power renormalon ambiguities removed through the $R$-gap scheme, stabilizing predictions in both the peak region ($\tau_1^b\to 0$) and the tail region ($\Lambda_\text{QCD}/Q\ll\tau_1^b\ll 1$). 
Perturbative uncertainties are estimated using profile scale functions designed to be valid across a wide range of values of $x$ and $Q$, ensuring proper treatment of the interplay between singular and nonsingular contributions.

Our predictions are compared with the recent HERA H1 measurements of the same DIS event shape $\tau_1^b$, showing excellent agreement in the tail region. In our analysis we demonstrated the characteristic peaked behavior of the event shape distributions as $\tau_1^b\to 1$ and
compared to the experimentally observed similar behavior as $\tau_1^b\to 1$, where the jet hemisphere is nearly empty. The level of agreement observed in this behavior is encouraging.
The inclusion of $\mathcal{O}(\alpha_s^2)$ nonsingular contributions and the renormalon subtraction proved essential for achieving quantitative agreement and the required sensitivity for determining $\alpha_s(M_Z)$ and $\Omega_1$.
Furthermore, we expect that our results will help break the degeneracy between $\alpha_s(M_Z)$ and $\Omega_1$ by enabling their precise determination in measurements at many different values of $x$ and $Q$. 
The goodness of agreement so far between our results and the HERA H1 measurements bodes well for a future program e.g. at EIC to use both highly precise theory and plentiful experimental data to determine fundamental quantities in QCD such as the strong coupling, PDFs, and hadronization effects. This method complements existing determinations, such as those from $e^+e^-$ collisions, offering a new avenue to resolve open questions and tensions. 

The next objective is to perform a global fit of our predictions to the full set of HERA H1 data and explore the potential for a more accurate determination of the fundamental QCD parameters with these and future experimental measurements. These efforts will lay the groundwork for high-precision analyses in the upcoming Electron-Ion Collider, establishing an essential benchmark for precision QCD in this new era.

\acknowledgments
JHE, DK, and CL would like to thank Korea University for their generous hospitality hosting us for an extended collaboration visit in 2022, during which a significant amount of progress on this project occurred.
DK is supported by the National Natural Science Foundation of China (NSFC) through National Key Research and Development Program under the contract No. 2024YFA1610503.
IS was supported in part by the U.S.\ Department of Energy  Office of Nuclear Physics under contract DE-SC0011090 and by the Simons Foundation through the Investigator grant 327942.
JHE and CL were supported by the U.S. Department of Energy, Office of Science, Office of Nuclear Physics and the Laboratory Directed Research and Development program of Los Alamos National Laboratory (LANL) under project numbers 20200022DR and 20230857PRD2. Early stages of this work were also supported by LDRD project 20140671PRD2. JHE and CL used resources provided by the LANL Institutional Computing Program. LANL is operated by Triad National Security, LLC, for the National Nuclear Security Administration of U.S. Department of Energy (Contract No. 89233218CNA000001).

\appendix

\section{Coefficients of the fixed-order functions}
\label{app:fixed-order-coeff}
In this appendix we collect the fixed-order formula for the hard function in Eq.~\eqref{eq:CVA}, the soft function in eq.~\eqref{eq:s-hemi-orig}, the jet function in Eq.~\eqref{eq:exp-jet-function}, and the integrated beam function in Eq.~\eqref{eq:integrated-beam-2}.

\subsection{Hard function}

The final form of the hard function in Eq.~\eqref{eq:hard-before-simp} can be written as a power series in $\alpha_s$,
\begin{equation}\label{eq:hard-after-simp}
H_{q,\bar q}(y,Q^2,\mu) 
=
\sum_{n=0}^\infty
\left[\frac{\alpha_s(\mu)}{4\pi}\right]^n
H^{(n)}_{q,\bar{q}}(y,Q^2,\mu),
\end{equation}
where $H^{(n)}_{q,\bar{q}}$ are the $n$th order coefficient functions.
\subsubsection{Flavor-diagonal}

The explicit form of the flavor-diagonal coefficient in Eq.~\eqref{eq:CVA} is given by
\begin{equation}\label{eq:C}
C(q^2,\mu^2) = 1 + \frac{\alpha_s C_F}{4\pi} \Bigl( -L^2 + 3L - 8 + \frac{\pi^2}{6}\Bigr) +\Bigl( \frac{\alpha_s}{4\pi}\Bigr)^2 ( C_F^2 H_F + C_F C_A H_A + C_F T_F n_f H_T),
\end{equation}
where 
\begin{equation}
L=\ln \frac{-q^2}{\mu^2} = \ln \frac{Q^2}{\mu^2}.
\end{equation}
The 1-loop coefficients are given in \cite{Manohar:2003vb,Bauer:2003di}, and the 2-loop coefficients are given by \cite{Idilbi:2006dg,Becher:2006mr}:
\begin{align}
H_F &= \frac{L^4}{2} - 3L^3 + \Bigl(\frac{25}{2} - \frac{\pi^2}{6}\Bigr) L^2 + \Bigl(-\frac{45}{2}-\frac{3\pi^2}{2}+24\zeta_3\Bigr)L + \frac{255}{8} + \frac{7\pi^2}{2} - \frac{83\pi^4}{360} - 30\zeta_3, 
\nonumber
\\
H_A &= \frac{11}{9}L^3 + \Bigl(-\frac{233}{18} + \frac{\pi^2}{3}\Bigr) L^2 + \Bigl( \frac{2545}{54} + \frac{11\pi^2}{9} - 26\zeta_3\Bigr)L - \frac{51157}{648} - \frac{337\pi^2}{108} + \frac{11\pi^4}{45} + \frac{313}{9}\zeta_3, 
\nonumber
\\
H_T &= -\frac{4}{9}L^3 + \frac{38}{9} L^2 + \Bigl(-\frac{418}{27} - \frac{4\pi^2}{9}\Bigr)L + \frac{4085}{162} + \frac{23\pi^2}{27} + \frac{4}{9}\zeta_3.
\end{align}
The hard coefficient functions in Eq.~\eqref{eq:hard-after-simp} are given by
\begin{align}\label{eq:hard-coefficient-012}
\begin{split}
H^{(0)}_{q,\bar{q}}(y,Q^2,\mu)
&= H^{(0)}(Q^2,\mu)L_{q,\bar{q}}(y,Q^2),
\\
H^{(1)}_{q,\bar{q}}(y,Q^2,\mu)
&= 
H^{(1)}(Q^2,\mu)
L_{q,\bar{q}}(y,Q^2),
\\
H^{(2)}_{q,\bar{q}}(y,Q^2,\mu)
&=
H^{(2)}(Q^2,\mu) L_{q,\bar{q}}(y,Q^2)
+
\frac{16}{3} I_2\left(-\frac{Q^2}{4m_t^2}\right)
L_{q,\bar q}^{\textrm{sing}}(y,Q^2),
\end{split}
\end{align}
where
\begin{align}
\begin{split}
H^{(0)}(Q^2,\mu) &= 1,
\\
H^{(1)}(Q^2,\mu) &= 2C_F\Bigl( -L^2 + 3L - 8 + \frac{\pi^2}{6}\Bigr),
\\
H^{(2)}(Q^2,\mu) &= 
C_F^2 H_F + C_F C_A H_A + C_F T_F n_f H_T
+
C_F^2
\Bigl( -L^2 + 3L - 8 + \frac{\pi^2}{6}\Bigr)^2,
\end{split}
\end{align}
and
\begin{align}
\begin{split}
L_{q,\bar{q}}(y,Q^2) 
&= 
L_{gqq}^{VV} + L_{gqq}^{AA} \mp 2 r(y) L_{ \epsilon qq}^{VA},
\\
L_{q,\bar q}^{\text{sing}} 
&=
L^{AA}_{gbq} \mp r(y) L^{AV}_{\epsilon bq}.
\end{split}
\end{align}
Here the upper sign is for $L_q$ and $L_q^\textrm{sing}$, and the lower sign is for $L_{\bar{q}}$ and $L_{\bar{q}}^\textrm{sing}$.
The last term of the two-loop hard coefficient,
$H^{(2)}_{q,\bar{q}}(y,Q^2,\mu)$, in Eq.~\eqref{eq:hard-coefficient-012} originates from the flavor-singlet contributions, $C^\textrm{sing}_{fq}$ in \eq{CVA}, which are depicted in Fig.~\ref{fig:anomaly}. The detailed derivation of this term is provided in the following subsection. 

\subsubsection{Flavor-singlet}
\label{app:flavor-singlet}
The contribution $C^\textrm{sing}_{fq}$ in \eq{CVA} comes from the triangle anomaly graphs, e.g., in \fig{anomaly}. Since it begins at $\mathcal{O}(\alpha_s^2)$, it multiplies only a tree-level Wilson coefficient when plugged into \eq{CVA}, yielding the formula in \eq{Hsing}.
The sum over flavors $f$ in \eq{Hsing} goes over the flavors in the triangle loop in \fig{anomaly}. For massless flavors, the result of the loop integral is the same, with each flavor's contribution proportional to the axial charge $a_f$ appearing in Eq.~\eqref{eq:leptonic}. Since $a_u = a_c = -a_d = -a_s$, the sum over the four lightest flavors vanishes. The sum is then only over $f=b,t$, for which the large mass splitting breaks the degeneracy and gives a non-cancelling result. Using $a_b = -a_t$, we obtain
\begin{equation}
\label{eq:Hsingb}
H^\textrm{sing}_{q,\bar q} = ({C^\textrm{sing}_{bq}}^* + C^\textrm{sing}_{bq} - {C^\textrm{sing}_{tq}}^* - C^\textrm{sing}_{tq}) \bigl[ L^{AA}_{gbq} \mp r(y) L^{AV}_{ \epsilon bq}\bigr].
\end{equation}
Now, $C^\textrm{sing}_{bq,tq}$ is independent of the flavor $q$ of the external (light) quark in the diagram---it does not affect the value of the loop integrals. So it comes out of the sum over $q$ in the cross sections \eq{tau1b-FT-2}.

Let us briefly recall what happens in the case of $e^+e^-$ thrust \cite{Abbate:2010xh}. There, the factorization theorem like \eq{tau1b-FT-2} would contain a $J_q$ in place of ${B}_q$, and $J_q$ is independent of quark flavor. Thus in $e^+e^-$, there is just a sum over the index $q$ in the leptonic factors in \eq{Hsingb}. Then, two simplifications occur. First, in the first term, $L^{AA}_{bq}$ is proportional just to $a_q$, and so the sum over $q=\{u,d,s,c\}$ cancels, leaving only $q=b$ ($q=t$ is not produced in the final state at the relevant collision energies). Second, in the second leptonic term in \eq{Hsingb}, 
\begin{equation}
L_{\epsilon bq}^{AV} = \frac{a_b a_e}{1+m_Z^2/Q^2}\Bigl( Q_q - \frac{2v_q v_e}{1+m_Z^2/Q^2}\Bigr),
\end{equation}
the sum over $q$ does not cancel for the light flavors, but there is still a sum over $H^\textrm{sing}_q$ and $H^\textrm{sing}_{\bar q}$ in the cross section. Since the quark and antiquark jet functions are the same, we just sum $H^\textrm{sing}_{q,\bar q}$ given in \eq{Hsingb}, and the $AV$ term cancels due to the opposite signs in front of $r(y)$. This is a manifestation of Furry's theorem \cite{Furry:1937zz}, since in $e^+e^-$ thrust, one has another (cut) fermion triangle at the right-hand vertex (see, e.g., Fig.~2 in \cite{Abbate:2010xh}) which vanishes for the vector current. 

However, in DIS, the same simplifications do not occur, because the ${B}_q$ in \eq{tau1b-FT-2} does depend on flavor, so the whole sum over $q$ over both terms of \eq{Hsingb} remains. Also, the cancellation between $q$ and $\bar q$ for the second term of \eq{Hsingb} also does not occur for DIS due to the differing quark and antiquark beam functions ${B}_{q,\bar q}$ which multiply $H_{q,\bar q}$ in \eq{tau1b-FT-2}. Furry's theorem does not apply for the $q$ leg in the diagram, because it does not close into a triangle, but instead enters the proton.

From \eq{Hsingb}, we now have the expression for the flavor singlet contribution to the hard function as
\begin{equation}
\label{eq:Hsingq}
H^\textrm{sing}_{q,\bar q} =2\textrm{Re}( C^\textrm{sing}_b - C^\textrm{sing}_t) L_{q,\bar q}^{\textrm{sing}} \equiv H_Q^{\textrm{singlet}}(q^2) L_{q,\bar q}^{\textrm{sing}},
\end{equation}
where
\begin{equation}
\label{eq:Lsingq}
L_{q,\bar q}^{\text{sing}} \equiv
 L^{AA}_{gbq} \mp r(y) L^{AV}_{\epsilon bq}
=
\frac{a_b}{1+m_Z^2/Q^2}\Bigl[ \frac{a_q(v_e^2+a_e^2)}{1+m_Z^2/Q^2} \mp r(y) a_e\Bigl(Q_q - \frac{2v_qv_e}{1+m_Z^2/Q^2}\Bigr) \Bigr]\,.
\end{equation}
In \eq{Hsingq}, we have dropped the $q$ index on $C^\textrm{sing}$ to indicate it is in fact independent of the light flavor $q$ in the final state. We can read off from the formula in \eq{Hsingq} that the function $H_Q^\text{singlet}$ that was given in \cite{Abbate:2010xh} is simply equal to the combination $2\textrm{Re}(C^\textrm{sing}_b - C^\textrm{sing}_t)$, that is, the coefficient of the leptonic factor. However, there it was given for $q^2 = Q^2 >0$. We need the form for $q^2 = -Q^2 <0$, appropriate for DIS. Luckily, both of these forms were given in \cite{Kniehl:1989qu}. Using the notation of \cite{Abbate:2010xh}, we find
\begin{equation}
H_Q^{\text{singlet}}(q^2=-Q^2) = 2\textrm{Re}(C^\textrm{sing}_b-C^\textrm{sing}_t)(-Q^2)= \frac{1}{3}\left[\frac{\alpha_s(\mu)}{\pi}\right]^2 I_2(r_t)\,,
\end{equation}
where $r_t = -Q^2/(4m_t^2)$, and 
\begin{figure}
\vspace{-1em}
    \centering
    \includegraphics[width=0.6\linewidth]{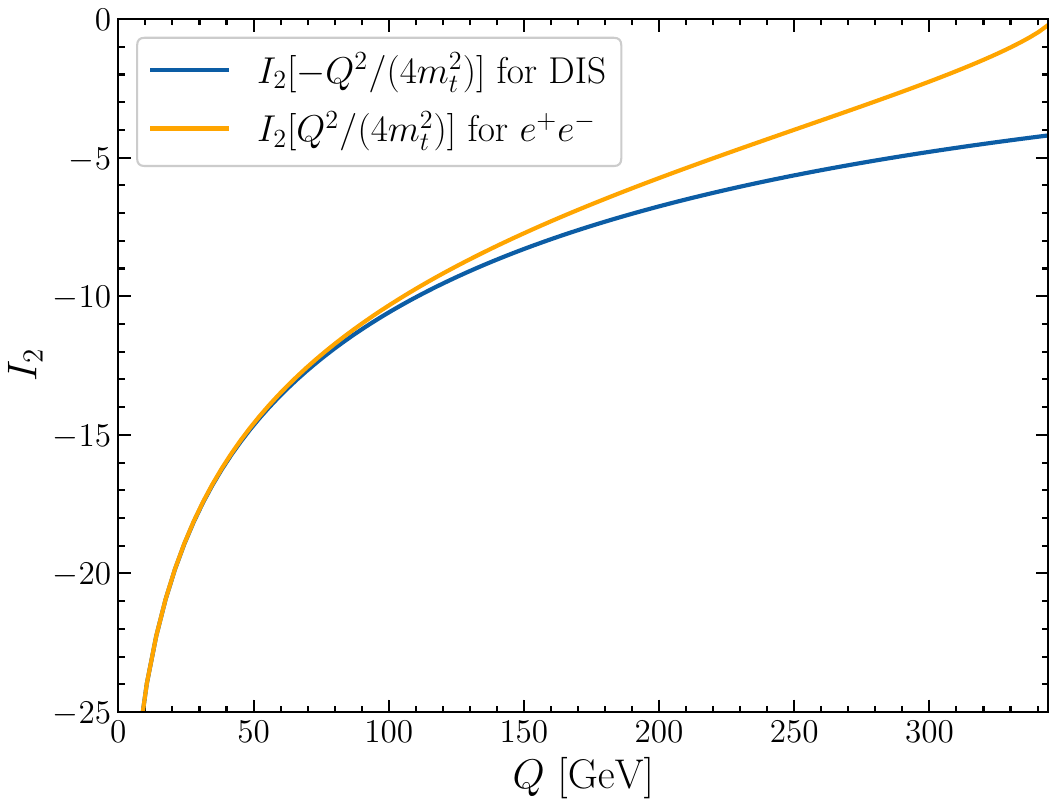}
    \vspace{-1em}
    \caption{The numerical differences between $I_2[-Q^2/(4m_t^2)]$ for DIS (blue) and $I_2[Q^2/(4m_t^2)]$ for $e^+e^-$ (orange).}
    \label{fig:Isinglet}
\end{figure}
\begin{align}
\label{eq:I2DIS}
I_2(r) &= \frac{\pi^2}{3} + 6g(r) - 10\left[f(r)\right]^2 \\
&\quad + \frac{1}{r^2} \Bigl\{-\Li_3(r_-^2) + \zeta_3 - f(r) \Bigl[ \Li_2(r_-^2) + \frac{\pi^2}{6}\Bigr] - \left[f(r)\right]^2 \Bigl[ \frac{f(r)}{3} + 1\Bigr] \Bigr\} \nn\\
&\quad + \frac{2}{r} \Bigl\{ 2\Li_3 (r_-^2) - \Li_3(r_-^4) - \zeta_3 + 2 f(r) \Bigl[ -\Li_2(r_-^4) + \frac{\pi^2}{6} \Bigr] + \left[f(r)\right]^2 [-4f(r) + 4g(r) + 3]\Bigr\} \nn\\
&\quad + \sqrt{1-\frac{1}{r}} \, \biggl\{ \frac{1}{r} \Bigl[ \Li_2 (r_-^2) - \frac{\pi^2}{6} + f(r)[f(r) - 2g(r)] \Bigr] \nn \\
&\qquad\qquad\qquad  + 6\Li_2(r_-^2) - 4\Li_2(r_-^4) - \frac{\pi^2}{3} + 2f(r) [ -5f(r) + 2g(r) + 8h(r) - 3] \biggr\}\,, \nn
\end{align}
where 
\begin{equation}
r_\pm \equiv \sqrt{1-r} \, \pm \sqrt{-r}\,,\quad f(r) \equiv \ln r_+\,,\quad  g(r) \equiv \ln(r_+ - r_-)\,, \quad h(r) \equiv \ln(r_+ + r_-)\,.
\end{equation}
In \fig{Isinglet}, we compare the function $I_2$ in Eq.~\eqref{eq:I2DIS} for DIS versus the same function for $0<q^2<4m_t^2$ for $e^+e^-$ given in \cite{Kniehl:1989qu,Abbate:2010xh}. They are numerically of similar size. We thus expect the flavor singlet anomaly contribution to be of similar importance in DIS as in $e^+e^-$, although, in DIS, the $AV$ terms also contribute to \eq{Hsingq}, with one term in \eq{Lsingq} suppressed by one less power of $(1+m_Z^2/Q^2)$, and all flavors $q$ in the final state contribute, weighted by the PDFs.
The singlet contribution should be non-negligible when we perform precise numerical analysis to extract $\alpha_s$ to percent-level precision, but are not particularly noticeable at the level we are currently working.

\subsection{Soft function}

The soft function as well as jet and beam functions is expressed as a series in the plus distribution $\mathcal{L}_m$ defined in Eq.~\eqref{eq:plus_equation}. 
We follow the convention for plus distributions from Ref.~\cite{Ligeti:2008ac}, where for a generic function $q(x)$, the distribution $[\quad]_+$ acts as:
\[
[q(x)]_+ = 
[\theta(x)q(x)]_+
=
\lim_{\epsilon\to 0}
\frac{d}{dx}
\left[
\theta(x-\epsilon)Q(x)
\right],
\]
where $Q(x)=\int_1^x dx'\, q(x').$
This choice ensures the boundary condition 
$\int_0^1 dx\left[q(x)\right]_+=0$, which will be used 
in Eq.~\eqref{eq:plus-int-identity}.

The coefficient $S_m (\alpha_s)$ defined in Eq.~\eqref{eq:s-hemi-orig}. can be expanded as a power series in $\alpha_s$:
%---------------
\begin{align}\label{eq:s-hemi-orig-Smn}
%---------------
S_m(\alpha_s)
&=
\sum_{n=0}^\infty
\left[\frac{\alpha_s(\mu)}{4\pi}\right]^n
S_m^{(n)},
\end{align}
%---------------
where the superscript $n$ denotes the order in $\alpha_s$.

The nonvanishing coefficients of the soft function, starting from tree level, are provided below
\cite{Fleming:2007xt}:
%---------------
\begin{equation}
%---------------
S_{-1}^{(0)}
=
1,
\quad
S_1^{(1)}
= 
-16C_F,
\quad
S_{-1}^{(1)}
= 
\frac{C_F\pi^2}{3},
%---------------
\end{equation}
%---------------
followed by the two-loop corrections given by
\cite{Kelley:2011ng, Monni:2011gb, Hornig:2011iu}
%---------------
\begin{align}
\begin{split}
%---------------
S_{3}^{(2)}
&=
128C_F^2,
\\
S_{2}^{(2)}
&=
C_A C_F \left(\frac{176}{3}\right)
+
C_FT_F n_f\left(-\frac{64}{3}\right),
\\
S_{1}^{(2)}
&=
C_F^2\left(-48\pi^2\right)
+C_A C_F
\left(
-\frac{1072}{9}+\frac{16\pi^2}{3}
\right)
+
C_F T_F n_f
\left(\frac{320}{9}\right),
\\
S_{0}^{(2)}
&=
C_F^2
\left(256\zeta_3\right)
+C_A C_F
\left(
\frac{1616}{27}-\frac{44\pi^2}{9}
-56\zeta_3
\right)
+
C_FT_Fn_f
\left(
-\frac{448}{27}+\frac{16\pi^2}{9}
\right),
\\
S_{-1}^{(2)}
&=
C_F^2\left(- \frac{3\pi^4}{10}\right)
+C_AC_F
\left(
\frac{268\pi^2}{27}
-\frac{4\pi^4}{9}
+\frac{352\zeta_3}{9}
\right)
+
C_FT_Fn_f
\left(
-\frac{80\pi^2}{27}
-\frac{128\zeta_3}{9}
\right)
+2s_2,
%---------------
\end{split}
\end{align}
%---------------
where the analytic expression for the two-loop constant term $2s_2$ is
%---------------
\begin{align}
\label{eq:s2-const}
\begin{split}
%---------------
2s_2 
&=
C_F C_A
\left(
-\frac{2140}{81}
-\frac{871\pi^2}{54}
+\frac{14\pi^4}{15}
+\frac{286\zeta_3}{9}
\right)
+
C_F T_F n_f
\left(
\frac{80}{81}
+\frac{154\pi^2}{27}
-\frac{104\zeta_3}{9}
\right).
%---------------
\end{split}
\end{align}
%---------------

\subsection{Jet function}
The coefficient $J_m (\alpha_s)$ as a power series in $\alpha_s$ is
%---------------
\begin{align}\label{eq:exp-jet-function-Jmn}
%---------------
J_m(\alpha_s)
=
\sum_{n=0}^\infty
\left[\frac{\alpha_s(\mu)}{4\pi}\right]^n
J_m^{(n)}.
%---------------
\end{align}
%---------------
The nonvanishing coefficients of the jet function, starting from tree level, are provided below:
%---------------
\begin{equation}\label{eq:J-series-01}
%---------------
J_{-1}^{(0)}
=1,
\quad
J^{(1)}_{1}
=
4C_F,
\quad
J^{(1)}_{0}
=
-3C_F,
\quad
J^{(1)}_{-1}
=
C_F(7-\pi^2),
%---------------
\end{equation}
%---------------
followed by the two-loop contributions given by
\cite{Becher:2006qw, Hoang:2008fs}
%---------------
\begin{align}\label{eq:J-series-2}
%---------------
J^{(2)}_3
={}&
8C_F^2,
\nonumber
\\
J^{(2)}_2
={}&
C_F^2(-18)
+
C_AC_F
\left(
-\frac{22}{3}
\right)
+
C_F T_F n_f
\left(
\frac{8}{3}
\right),
\nonumber
\\
J^{(2)}_1
={}&
C_F^2
\left(
37-\frac{20\pi^2}{3}
\right)
+
C_A C_F
\left(
\frac{367}{9}-\frac{4\pi^2}{3}
\right)
+
C_F T_F n_f
\left(
-\frac{116}{9}
\right),
\nonumber
\\
J^{(2)}_0
={}&
C_F^2
\left(
-\frac{45}{2}+7\pi^2-8\zeta_3
\right)
+
C_A C_F
\left(
-\frac{3155}{54}+\frac{22\pi^2}{9}+40\zeta_3
\right)
+
C_FT_F n_f
\left(
\frac{494}{27}-\frac{8\pi^2}{9}
\right),
\nonumber
\\
J^{(2)}_{-1}
=&
C_F^2
\left(
\frac{205}{8}-\frac{67\pi^2}{6}
+\frac{14\pi^4}{15}
-18\zeta_3
\right)
+
C_A C_F
\left(
\frac{53129}{648}
-\frac{208\pi^2}{27}
-\frac{17\pi^4}{180}
-\frac{206\zeta_3}{9}
\right)
\nonumber
\\
&
+
C_FT_Fn_f
\left(
-\frac{4057}{162}
+\frac{68\pi^2}{27}
+\frac{16\zeta_3}{9}
\right).
%---------------
\end{align}
%---------------

\subsection{Integrated Beam Function}

Next, let us consider the coefficient functions for the kernel of the integrated beam function in Eq.~\eqref{eq:integrated-beam}.
The radiative kernel $\mathcal{I}_{ij}$ in Eq.~\eqref{eq:transverse-dep-beam} is expressed as a series of plus distributions:
%---------------
\begin{align}
\label{eq:Iij-exp-dimless-2}
\mathcal{I}_{ij}(t, z,\mathbf{k}_\perp^2,\mu)
=
\frac{1}{\pi t}
\frac{1}{\mu^2}
\sum_{m=-1} \mathcal{I}_{ij,m}(z, r)
\mathcal{L}_m(t/\mu^2)\,,
%---------------
\end{align}
%---------------
where $r \equiv \mathbf{k}_\perp^2/t$. Writing $\mathbf{k}_\perp^2 = tr$
and factoring out $1/t$ dependence in Eq.~\eqref{eq:Iij-exp-dimless-2}, we make the coefficients $\mathcal{I}_{ij,m}(z, r)$ dimensionless. 
%Here, $\sum_{m=-1} \mathcal{I}_{ij,m}(z, r)
%\mathcal{L}_m(t/\mu^2)$ is equal to $t\times  \mathcal{I}_{ij}(t,z,\mathbf{k}_\perp^2,\mu)$ in the notation of Ref.~\cite{Gaunt:2014xxa}.
Inserting Eq.\eqref{eq:Iij-exp-dimless-2} into Eq.~\eqref{eq:Jij-Iij-relation}, 
we have 
%---------------
\begin{align} \label{eq:Jij-n-Iijm}
%---------------
\mathcal{J}_{ij}(t, z,\mu)
=
\frac{1}{\mu^2}
\sum_{m=-1}
\int_{1-\frac{1-z}{z}}^1 \frac{dy}{y}\,
\mathcal{L}_m(ty/\mu^2)
\mathcal{I}_{ij,m}\left(z, \frac{1-y}{y}\right),
%---------------
\end{align}
%---------------
where the integration over $y$ encapsulates the transverse momentum contribution. 
%That is, the coefficients functions $\mathcal{J}_{ij,m}(z)$ are determined by performing the integration over $y$, using the kernel $\mathcal{I}$ given for the transverse-momentum-dependent beam function in Refs.~\cite{Jain:2011iu, Gaunt:2014xxa}. 

One can identify the coefficients $\mathcal{J}_{ij,m}(z;\alpha_s)$ of $\mathcal{L}_m(t/\mu^2)$ in Eq.~\eqref{eq:Jij-n-Iijm} using the rescaling identity in \eq{rescale}, and then expand them as a power series in $\alpha_s$ as
%---------------
\begin{align}\label{eq:Jij-plus-Jijmn}
\mathcal{J}_{ij,m}(z;\alpha_s) 
=
\sum_{n=0}^\infty
\left[\frac{\alpha_s(\mu)}{4\pi}\right]^n
\mathcal{J}^{(n)}_{ij,m}(z)\,.
\end{align}
%---------------
The corresponding beam function coefficient ${B}_{i,m}(x,\mu;\alpha_s)$ can be obtained using the convolution in Eq.~\eqref{eq:Bim-Jijm}.
The coefficients $\mathcal{J}^{(n)}_{ij,m}$ decompose into two parts: 
(i)~a contribution matching the ordinary beam function, and (ii)~an additional term absent in the conventional  beam function, given by
\begin{align}
\cJ^{(n)}_{ij,m}(z) = \mathcal{I}^{(n)}_{ij,m}(z) + \Delta\mathcal{I}^{(n)}_{ij,m}(z)\,,
\end{align}
The first term on the right-hand side can be obtained from  \cite{Stewart:2010qs,Gaunt:2014xga}, while the second term at one-loop is computed in \cite{Kang:2013nha}. At one-loop, the only non-vanishing contribution arises for $m=-1$ and is given by
%Performing the $y$ integration in Eq.~\eqref{eq:Jij-n-Iijm} with respect to the tree-level and one-loop kernel in Ref.~\cite{Jain:2011iu}, we obtain the coefficient functions $\mathcal{J}_{ij,m}^{(n)}(z)$ at tree-level and one-loop as 
%---------------
\begin{align}
\Delta\mathcal{I}_{ij,-1}^{(1)}(z)
&=
2P_{ij}^{(0)}(z)
\ln z
\,.
\end{align}
%---------------
$P_{ij}^{(n)}(z)$ is the $(n+1)$-loop terms of the PDF anomalous dimensions in the 
$\overline{\textrm{MS}}$, which is defined by
\begin{equation}
P_{ij}(z,\alpha_s)
=
\sum_{n=0}^\infty \left(\frac{\alpha_s}{2\pi}\right)^{n+1} P_{ij}^{(n)}(z).
\end{equation}
Note that $P_{ij}^{(n)}(z)$ are written in terms of the quark and gluon splitting functions. The explicit forms for $P_{ij}^{(n)}(z)$ at one- and two-loop are listed in Appendix~A.3 of Ref.~\cite{Gaunt:2014xga}.
%$I_{ij}^{(1)}(z)$ is the $\mu$-independent part of the one-loop matching coefficient of the ordinary beam function and it is listed in Appendix~A.2 of Ref.~\cite{Gaunt:2014xga}.  

Next, performing the $y$ integration in Eq.~\eqref{eq:Jij-n-Iijm} with respect to the two-loop kernel $\mathcal{I}_{ij}^{(2)}(t,z,\mathbf{k}_\perp^2,\mu)$ in Ref.~\cite{Gaunt:2014xxa}, we obtain the coefficient functions $\Delta\mathcal{I}_{ij,m}^{(2)}(z)$. These coefficients are nonzero for $m=-1,0,$ and $1$, and are given by 
%---------------
\begin{align}\label{eq:two-loop-genBF-total}
%---------------
\Delta\mathcal{I}_{ij,1}^{(2)}(z)
={}&
2 \Gamma_0^q P^{(0)}_{ij}(z)\, \ln z 
\,,\nonumber\\
\Delta\mathcal{I}_{ij,0}^{(2)}(z)
={}&
-
\left(\gamma_{B0}^q +2\beta_0 \right)
P_{ij}^{(0)}(z)\ln z
+
4
\sum_k
\hat{P}_{ikj}^{(1)}(z) 
\nonumber
\,,\\
\Delta\mathcal{I}_{ij,-1}^{(2)}(z)
={}&
\Gamma_0^q P_{ij}^{(0)}(z)
\left[
4\zeta_3
+\frac{\pi^2}{3}\ln z 
+2\ln z \Li_2(z)
-4\Li_3(z)
\right]
\nonumber
\\
&
-
\frac{1}{2}\left(\gamma_{B0}^q +2\beta_0\right)
P_{ij}^{(0)}(z)\ln^2 z
+
2
\sum_k
\hat{P}_{ikj}^{(2)}(z) +4 \hat{J}^{(1)}_{ij}(z) 
\,,
%---------------
\end{align}
%---------------
where $\beta_0 = (11C_A - 4T_F n_f)/3$ and $\Gamma_n^q$ and $\gamma_{Bn}^q$ are the $(n+1)$-loop terms of the $\overline{\textrm{MS}}$ cusp and beam function anomalous dimensions defined by
\begin{equation}\label{eq:anomalous-for-beam}
\Gamma_\textrm{cusp}^q(\alpha_s)
=
\sum_{n=0}^\infty
\Gamma_n^q 
\left(
\frac{\alpha_s}{4\pi}
\right)^{n+1},
\quad
\gamma_B^q(\alpha_s)
=
\sum_{n=0}^\infty
\gamma_{Bn}^q
\left(
\frac{\alpha_s}{4\pi}
\right)^{n+1}.
\end{equation}
The explicit forms of the coefficients $\Gamma_n^q$ (up to three loops) and $\gamma_{Bn}^q$ (up to three loops) are listed in Appendix~A.1 of Ref.~\cite{Gaunt:2014xga}.
The four-loop coefficient of the cusp anomalous dimension $\Gamma_3^q$ is given in Ref.~\cite{Henn:2019swt}.
% and $I_{ij}^{(2)}(z)$ is the $\mu$-independent part of the two-loop matching coefficient of the ordinary beam function given in section 4 of Ref.~\cite{Gaunt:2014xga}.  

We define the convolutions of the splitting functions as 
\begin{equation}
\hat{P}_{ikj}^{(n)}(z)
\equiv
\int_z^1 dy \frac{\ln^n y}{y^2}
\left[
\delta\left(\frac{1}{z}-\frac{1}{y}\right)
P_{ik}^{(0)}(z)
\right]
\otimes_z
P_{kj}^{(0)}(z)
=
[\ln^n z P_{ik}^{(0)}](z)
\otimes_z P_{kj}^{(0)}(z),
\end{equation}
with $\otimes_z$ defined as
\begin{equation}
A(z)\otimes_z B(z) = 
\int_z^1 \frac{d\omega}{\omega} A(\omega) B(z/\omega).
\end{equation}
The explicit forms of $\hat{P}_{ikj}^{(n)}(z)$ are collected in Appendix~\ref{app:some-identies}.

The functions $ \hat{J}^{(n)}_{ij}(z) $ are derived from the integral of $ J_{ij}^{(2)}(t, z, \mathbf{k}_\perp^2) $, as defined in Eq.~(2.7) of Ref.~\cite{Gaunt:2014xxa}, and are given by:
%%%
\begin{align} \label{eq:Jn}
\hat{J}^{(n)}_{ij}(z)
&\equiv \int_z^1 dy\,\frac{\ln^n y}{y}\, t\,J^{(2)}_{ij}\big(ty,z,t(1-y)\big)
%=\int_z^1 dy\,\frac{\ln^n y}{y^2}\,\cJ_{ij}\big(z,(1-y)/y\big)
\,,\nn\\ 
\hat{J}^{(n)}_{q_i q_j}(z)&= \hat{J}^{(n)}_{\bar q_i \bar q_j}(z) 
= C_F\big(\delta_{ij} \hat{J}^{(n)}_{qqV}+\hat{J}^{(n)}_{qqS} \big)
\,,\nn\\ 
\hat{J}^{(n)}_{q_i \bar q_j}(z)&= \hat{J}^{(n)}_{\bar q_i q_j}(z) 
= C_F \big(\delta_{ij} \hat{J}^{(n)}_{q\bar qV}+\hat{J}^{(n)}_{qqS}  \big)
\,,\nn\\ 
\hat{J}^{(n)}_{q_i g}(z)&= \hat{J}^{(n)}_{\bar q_i g}(z) 
= T_F \hat{J}^{(n)}_{qg}
\,,\end{align}
%%%
where individual functions $\hat{J}^{(n)}_{X}(z)$ for $X \ni  \{qqV,\,q\bar q V,\,qqS,\, qg\}$ are defined by $J^{(2)}_{X}(t,z,\mathbf{k}_\perp^2/t)$ in a similar way to \eq{Jn}. Note that for $n=0$, $\hat{J}^{(n)}_{ij}(z)$ corresponds to  the coefficient of $(1/\mu^2)\mathcal{L}_0(t/\mu^2)$ in the ordinary beam function at two loop \cite{Gaunt:2014xga}.
The explicit forms of $\hat{J}^{(n)}_{X}(z)$ for $n=1$ are collected in Appendix~\ref{app:JnX}.

\subsubsection{Explicit forms of $\hat{P}_{ikj}^{(n)}$ convolutions}
\label{app:some-identies}
\begin{subequations}
At $n=0$, $\hat{P}^{(0)}_{ikj}(z)=P_{ik}^{(0)}(z)\otimes_z P_{kj}^{(0)}(z)$, which is given in \cite{Gaunt:2014xga}.
At $n=1$, $\hat{P}^{(1)}_{ikj}(z)
=\left[\ln z\, P_{ik}^{(0)}(z)\right]\otimes_z P_{kj}^{(0)}(z)$, and
the relevant results are
%---------------
\begin{align}\label{eq:P1}
%---------------
\hat{P}^{(1)}_{qqq}(z)
&= 
C_F^2
\frac{ \ln z}{2 (1-z)}\left[1+4 z+z^2-(1+3 z^2) \ln z+4 (1+z^2) \ln (1-z)\right],
\nonumber
\\
\hat{P}^{(1)}_{qgq'}(z)
&=
C_F T_F  
\Bigg[
-\frac{13 \left(1-z^3\right)}{9 z} -\frac{3+6 z+4 z^2}{3} \ln z+(1+z) \ln^2 z
\Bigg],
\nonumber
\\
\hat{P}^{(1)}_{qgg}(z)
&=
C_A T_F \Bigg\{
-\frac{(1-z) \left(26+17 z+71 z^2\right)}{18 z}
-\frac{3-6 z+22 z^2}{3} \ln z
+(1+4 z) \ln^2 z
\nonumber
\\
&
\quad\quad\quad\quad
+\Big[ \frac{\pi^2}{3}-2\Li_2(z)\Big] P_{qg}(z)
\Bigg\}
+\beta_0 T_F \frac{\ln z}{2}P_{qg}(z),
\nonumber
\\
\hat{P}^{(1)}_{qqg}(z)
&=
C_F T_F  
\Bigg\{
-\frac{1}{2} (1-z) (1-5 z) -(1-3 z) \ln z+\frac{1-2z}{2} \ln^2 z  
\nonumber 
\\
&
\quad\quad\quad\quad
-[\ln^2 z+2 \Li_2(1-z)] P_{qg}(z)
\Bigg\}.
%---------------
\end{align}
%---------------
At $n=2$, $\hat{P}^{(2)}_{ikj}(z)
=\left[\ln^2 z\, P_{ik}^{(0)}(z)\right]\otimes_z P_{kj}^{(0)}(z)$, and
the relevant results are
%---------------
\begin{align}\label{eq:P2}
%---------------
\hat{P}^{(2)}_{qqq}(z)
&= 
\frac{C_F^2}{1-z}
\Bigg\{
-4 (1-z)^2
+\frac{2 [-3 (1-z^2)+\pi ^2 (1+z^2)] }{3}\ln z
+\frac{1+4 z+z^2}{2} \ln^2 z
\nonumber
\\ 
& \qquad\qquad
-\frac{1+ 3 z^2 }{3} \ln^3 z
+2 \left(1+z^2 \right) \big[\ln (1-z) \ln^2 z+2\ln z \Li_2(z) -4 \Li_3(z)+4 \zeta_3\big]
\Bigg\},
\nonumber
\\
\hat{P}^{(2)}_{qgq'}(z)
&=
C_F T_F  \Bigg[
\frac{(1-z) (89+224 z+89 z^2)}{27 z}
+\frac{2 (27+27 z+13 z^2)}{9}  \ln z
-\frac{3+6 z+4 z^2}{3}  \ln^2 z
\nonumber
\\
&
\quad\quad\quad\quad
+\frac{2(1+z)}{3}  \ln^3 z
\Bigg],
\nonumber
\\
\hat{P}^{(2)}_{qgg}(z)
&=
C_A T_F 
\Bigg\{
\frac{(1-z) \left(178+529 z+1231 z^2\right)}{54 z}
+\frac{63+126 z+134 z^2 +6\pi ^2 P_{qg}(z)}{9} \ln z
\nonumber
\\
&
\quad\quad\quad\quad
-\frac{3-6 z+22 z^2}{3} \ln^2 z
+\frac{2(1+4z)}{3}  \ln^3 z
-4 P_{qg}(z) \left[\Li_3(z)- \zeta_3\right]
\Bigg\}
\nonumber
\\
&
\quad
+\beta_0 T_F \frac{\ln^2z}{2}P_{qg}(z),
\nonumber 
\\
\hat{P}^{(2)}_{qqg}(z)
&=
C_F T_F  \Bigg\{
-\frac{(1-z)(11-9z)}{2}
-(3-2 z) \ln z
-(1-3 z) \ln^2 z
+\frac{1-2z-2 P_{qg}(z)}{3} \ln^3 z
\nonumber
\\ 
&
\qquad\qquad
+4 P_{qg}(z)  
\left[\frac{\ln (1-z)\ln^2 z}{2} + \ln z \Li_2(z)- \Li_3(z)+ \zeta_3\right]
\Bigg\}.
%---------------
\end{align}
%---------------
\end{subequations}

\subsubsection{Explicit forms of $\hat{J}^{(1)}_{X}(z)$}\label{app:JnX}
\begin{subequations}
Here, we provide the expressions for $ \hat{J}^{(1)}_{X}(z) $ for $ X \in \{qqV,\,q\bar{q}V,\,qqS,\,qg\} $:
%%%
\begin{align} \label{eq:JnX}
%%%%%%%% J1qqV
\hat{J}^{(1)}_{qqV}(z) &=
C_F \Bigg\{%
-6+5z -\frac{\pi^2(7-4z+9z^2)}{6(1-z)}-\frac{1+6z-4z^2}{1-z}\ln z
-2\ln ^2 z
\nn\\ &\qquad \qquad
+4\frac{1-z+z^2}{1-z}\ln z\ln (1-z)-2(1-z)\ln(1-z) +\frac{3-4z+5z^2}{1-z}\Li_2(z)
\nn\\ &\qquad \qquad
 +\frac{1+z^2 }{1-z}
  \bigg[ \frac{\pi^2}{6} \ln z +\ln^3 z-6\ln^2 z \ln(1-z)+\ln z \ln^2 (1-z)
  \nn\\ &\qquad\qquad
  -2\ln z \Li_2(z) -8\Li_3(1-z)-4\Li_3\Big( 1-\frac1z\Big) \bigg]
\Bigg\}
\nn\\ &\quad
+C_A\Bigg[ %
3-2 z
+\frac{5-3z +2z^2 }{3 (1-z)}\ln z
+\frac{1+z^2}{1-z}
\Bigg(\frac{\pi^2}{6}(1-2\ln z)+\frac12 \ln^2 z -2\ln z \ln(1-z) 
\nn\\ &\qquad \qquad
 +2\ln^2 z \ln(1-z) -2(1-\ln z) \Li_2(z) +2 \Li_3(1-z)
\Bigg)
\Bigg]
\nn\\ &\quad
+\beta_0\Bigg[%
\frac{1-z}{2}
-\frac{1+z^2 }{2 (1-z)}\Big[\frac{\pi^2}{6}-\frac{8}{3}\ln z-\ln^2 z - \Li_2(z)\Big]
 \Bigg]
\,,%\nn\\ 
\end{align}

%%%%%%%% J1qqbV
\begin{align} \label{eq:JqqbV}
\hat{J}^{(1)}_{q\bar qV} (z) &=\frac{2 C_F-C_A}{ 1+z}\Bigg[
(1+z)\Big[-4+4z -(1+3z)\ln z +z \ln^2 z \Big]
\nn\\ &\qquad \qquad\qquad
+(1+z^2)\Big[ \frac{\ln^3 z}{6}-\frac{\ln^2 z \ln (1+z)}{2} -\ln z \Li_2 \left(\frac{1}{1+z} \right)\Big]
\nn\\ &\qquad \qquad\qquad
+(1+z^2)\left[ \Li_3\left(\frac{z}{1+z}\right)- \Li_3\left(\frac{1}{1+z}\right) \right]
\Bigg]
\,,%\nn\\ 
\end{align}

%%%%%%%% J1qqS
\begin{align} \label{eq:JqqS}
\hat{J}^{(1)}_{qqS} (z) &=
T_F\Bigg[
-\frac{(1-z) (169+154 z+241 z^2)}{27 z}
-\frac{\pi ^2}{18}   (3+6 z+4 z^2)
-\frac{81+63 z+83 z^2}{9}  \ln z
\nn \\ &\qquad\qquad 
+\frac{21+21 z+20 z^2}{6}  \ln^2 z-\frac{13 (1-z^3)}{9 z} \ln (1-z)
+\frac{3+6 z+4 z^2}{3} \Li_2(z)
\nn \\ &\qquad \qquad
+(1+z)\bigg[\frac{ \pi ^2  }{3}  \ln z -\ln^3 z -2 \Li_3(z)+2 \zeta_3\bigg]
\Bigg]
\,,
%\nn\\ 
\end{align}

%%%%%%%% J1qg
\begin{align} \label{eq:Jqg}
 \hat{J}^{(1)}_{qg} (z) &=
C_F\Bigg[ 
  \frac{3}{2}(1-z)(5-7z)
  +\Big[\frac{8-15 z+5 z^2}{2} -\frac{\pi ^2}{6}  \left(15 - 30 z + 32 z^2\right)\Big] \ln z 
+\frac{3(1-4z)}{4}  \ln^2 z
\nn\\ &\qquad \quad
-\frac{ 1-2 z }{2}  \ln^3 z
-\frac{(1-z) (1-5z)}{2}  \ln (1-z)
-(1-3z)\bigg[ \ln z\ln(1-z) +	\Li_2(1-z) 	\bigg]
\nn\\ &\qquad \quad
+(1-2 z+2 z^2) \bigg[ 7 \ln^2 z \ln (1-z) +8 \ln z \Li_2(1-z) +4\Li_3(1-z)+2 
\bigg.
\nn\\ &\qquad \qquad
-\Li_3\left(1-\frac{1}{z}\right) \bigg]
+(13-26 z+28z^2)  \big[ \Li_3(z) - \zeta_3 \big]
\Bigg]
\nn\\ &
+C_A\Bigg[ 
-\frac{(1-z) (338+443 z+ 2210 z^2)}{54 z}
+\bigg[\frac{ \pi ^2 }{3} (1+4z) -\frac{180+324 z+607 z^2}{18}  \bigg] \ln z
\nn\\ &\qquad \quad
+\frac{21-12 z+110 z^2}{6}   \ln^2 z
-\frac{5+22 z -2 z^2}{6}       \ln^3 z
-\frac{  (1-z) (26+17 z+ 71 z^2) }{18 z} \ln (1-z)
\nn\\ &\qquad \quad
 -\frac{3-6 z+22 z^2}{3}\bigg[ \ln z\ln(1-z)+\Li_2 (1-z) \bigg]
  \nn\\ &\qquad \quad
 -\frac{1-2 z+2 z^2}{2}\bigg[  \ln^2 z\ln(1-z)-2\Li_3\left(1-\frac{1}{z}\right)	\bigg]
 \nn\\ &\qquad \quad
 -\frac{1+2 z+2 z^2}{2}\bigg[\ln^2 z \ln(1+z)+2\ln z \Li_2\left(\frac{1}{1+z}\right)+2 \Li_3\left(\frac{1}{1+z}\right)
  \nn\\ &\qquad \qquad
 -2 \Li_3\left(\frac{z}{1+z}\right)			\bigg]
-  (1+10 z-2 z^2) \big[ \Li_3(z) - \zeta_3 \big]
\Bigg]
\,.\end{align}
%%%
\end{subequations}

\section{Evolution factors at N$^\mathbf{3}$LL}
\label{app:N3LL-evolution}
The evolution factors in Eq.~\eqref{eq:RG-simple} are written as 
\cite{Kang:2013nha}
%---------------
\begin{eqnarray}
\label{eq:evol-summary}
%---------------
U_H(Q,\mu_0,\mu)
&=&
e^{K_H}
\left(\frac{Q}{\mu_0}\right)^{\eta_H},
\nonumber \\
U_{B_q}(t,\mu_0,\mu)
&=&
\frac{e^{K_{B}-\gamma_\textrm{E}\eta_{B}}}{\Gamma(1+\eta_{B})}
\left[
\frac{\eta_{B}}{\mu_0^2}
\mathcal{L}^{\eta_{B}}
\left(\frac{t}{\mu_0^2}\right)
+\delta(t)
\right],
\nonumber \\
U_{J}(t,\mu_0,\mu)
&=&
\frac{e^{K_{J}-\gamma_\textrm{E}\eta_{J}}}{\Gamma(1+\eta_{J})}
\left[
\frac{\eta_{J}}{\mu_0^2}
\mathcal{L}^{\eta_{J}}
\left(\frac{t}{\mu_0^2}\right)
+\delta(t)
\right],
\nonumber \\
U_S^2(k,\mu_0,\mu)
&=&
\frac{e^{2K_S-2\gamma_\textrm{E}\eta_{S}}}{\Gamma(1+2\eta_S)}
\left[
\frac{2\eta_S}{\mu_0}
\mathcal{L}^{2\eta_S}\left(\frac{k}{\mu_0}\right)
+\delta(k)
\right],
%---------------
\end{eqnarray}
%---------------
where the plus distribution $\mathcal{L}^{a}(x)$ is defined as follows:
\begin{equation}
\mathcal{L}^a(x)
\equiv
\left[
\frac{\theta(x)}{x^{1-a}}
\right]_+
=
\lim_{\epsilon\to0}
\frac{d}{dx}
\left[
\theta(x-\epsilon)
\frac{x^a-1}{a}
\right].
\end{equation}
$K_{H,B,J,S}$ and $\eta_{H,B,J,S}$ are the functions 
of $\mu_0$ and $\mu$, and they are written in terms of the three functions, $K_{\Gamma^q}(\mu_0,\mu)$, $\eta_{\Gamma^q}(\mu_0,\mu)$ and $K_\gamma(\mu_0,\mu)$ defined in Ref.~\cite{Kang:2013nha}.
These expressions are given up to NNLL accuracy in Eq.~(D26) of Ref.~\cite{Kang:2013nha}. In this appendix, let us add the N$^3$LL contributions to those functions \cite{Abbate:2010xh} (suppressing the superscript $q$ on $\Gamma^q$):
\begin{align}\label{eq:N3LL-extention-func}
\begin{split}
K_{\Gamma}(\mu_0,\mu)\big|_{\textrm{N$^3$LL}}
={}&
K_{\Gamma}(\mu_0,\mu)\big|_{\textrm{NNLL}}
+
\frac{\Gamma_0}{4\beta_0^2}
\frac{\alpha_s^2(\mu_0)}{(4\pi)^2}
\bigg\{
\left[
\left(\frac{\Gamma_1}{\Gamma_0} - \frac{\beta_1}{\beta_0}\right)B_2 + \frac{B_3}{2}
\right]
\frac{(r^2-1)}{2}
\\
&
+
\left(
\frac{\Gamma_3}{\Gamma_0}
-
\frac{\Gamma_2\beta_1}{\Gamma_0\beta_0}
+
\frac{B_2 \Gamma_1}{\Gamma_0}
+
B_3
\right)
\left(
\frac{r^3-1}{3} 
-
\frac{r^2 - 1}{2}
\right)
\\
&
-
\frac{\beta_1}{2\beta_0}
\left(
\frac{\Gamma_2}{\Gamma_0}
-
\frac{\Gamma_1\beta_1}{\Gamma_0\beta_0}
+
B_2
\right)
\left(
r^2\ln r - \frac{r^2-1}{2}
\right)
\\
&
-\frac{B_3}{2}\ln r
-
B_2\left(\frac{\Gamma_1}{\Gamma_0}-\frac{\beta_1}{\beta_0}\right)
(r-1)
\bigg\}\,,
\\
\eta_{\Gamma}(\mu_0,\mu)\big|_{\textrm{N$^3$LL}}
={}&
\eta_{\Gamma}(\mu_0,\mu)\big|_{\textrm{NNLL}}
-
\frac{\Gamma_0}{2\beta_0}
\frac{1}{3}\frac{\alpha_s^3(\mu_0)}{(4\pi)^3}
\bigg[
\frac{\Gamma_3}{\Gamma_0}
-\frac{\beta_3}{\beta_0}
+
\frac{\Gamma_1}{\Gamma_0}
\left(
\frac{\beta_1^2}{\beta_0^2}
-
\frac{\beta_2}{\beta_0}
\right)
\\
&
-\frac{\beta_1}{\beta_0}
\left(
\frac{\beta_1^2}{\beta_0^2}
-
2\frac{\beta_2}{\beta_0}
+
\frac{\Gamma_2}{\Gamma_0}
\right)
\bigg]
(r^3-1)\,,
\\
K_{\gamma}(\mu_0,\mu)\big|_{\textrm{N$^3$LL}}
={}&
K_{\gamma}(\mu_0,\mu)\big|_{\textrm{NNLL}}
-\frac{\gamma_0}{2\beta_0}
\frac{\alpha_s^2(\mu_0)}{(4\pi)^2}
\left(
\frac{\gamma_2}{\gamma_0}
-
\frac{\beta_1\gamma_1}{\beta_0\gamma_0}
+
\frac{\beta_1^2}{\beta_0^2}
-
\frac{\beta_2}{\beta_0}
\right)
\frac{r^2-1}{2},
\end{split}
\end{align}
where $r = \alpha_s(\mu)/\alpha_s(\mu_0)$ and the coefficients are $B_2=\beta_1^2/\beta_0^2 - \beta_2/\beta_0$ and $B_3 = - \beta_1^3/\beta_0^3 + 2\beta
_1\beta_2/\beta_0^2 - \beta_3/\beta_0$. 
These results are expressed in terms of series expansion of the cusp anomalous dimension in Eq.~\eqref{eq:anomalous-for-beam}, and series expansion of the QCD beta function and the $\overline{\textrm{MS}}$ non-cusp anomalous dimensions in powers of $\alpha_s$ as 
\begin{equation}
\beta(\alpha_s) = -2\alpha_s\sum_{n=0}^\infty \beta_n
\left(\frac{\alpha_s}{4\pi}\right)^{n+1},
\quad
\gamma(\alpha_s) = \sum_{n=0}^\infty \gamma_n
\left(\frac{\alpha_s}{4\pi}\right)^{n+1}.
\end{equation}
The coefficients $\beta_n$ up to three loops are listed in Appendix~D of Ref.~\cite{Kang:2013nha}, while the four-loop coefficient of the beta function $\beta_3$ is provided in Ref.~\cite{vanRitbergen:1997va}.
The $\overline{\textrm{MS}}$ anomalous dimensions for the hard, jet, beam, and soft functions are listed in Appendix~D of Ref.~\cite{Kang:2013nha} up to three loops.

Here, to be consistent with the NNPDF4.0 NNLO PDF sets, we use the three-loop QCD running coupling given in Ref.~\cite{DelDebbio:2007ee}:
\begin{align}
\begin{split}
\alpha_s(\mu)
={}&
\alpha_s(\mu)\big|_\textrm{LO}
\bigg[
1+\frac{\alpha_s(\mu)\big|_\textrm{LO}}{4\pi}
\frac{
\alpha_s(\mu)\big|_\textrm{LO}
-
\alpha_s(m_Z)
}{4\pi}
\left(
\frac{\beta_2}{\beta_0}
-
\frac{\beta_1^2}{\beta_0^2}
\right)
\\
&
\quad\quad\quad\quad
+
\frac{\alpha_s(\mu)\big|_\textrm{NLO}}{4\pi}
\frac{\beta_1}{\beta_0}
\ln\frac{\alpha_s(\mu)\big|_\textrm{NLO}}{\alpha_s(m_Z)}
\bigg],
\end{split}
\end{align}
with
\begin{align}
\begin{split}
\alpha_s(\mu)\big|_\textrm{LO}
&=
\frac{\alpha_s(m_Z)}{1+\beta
_0 \frac{\alpha_s(m_Z)}{4\pi}\ln(\mu^2/m_Z^2)},
\\
\alpha_s(\mu)\big|_\textrm{NLO}
&=
\alpha_s(\mu)\big|_\textrm{LO}
\left[
1-
\frac{\beta_1}{\beta_0}
\frac{\alpha_s(\mu)\big|_\textrm{LO}}{4\pi}
\ln\left(
1+\beta_0 \frac{\alpha_s(m_Z)}{4\pi} \ln(\mu^2/m_Z^2)
\right)
\right].
\end{split}
\end{align}
In the current work, we use $\alpha_s(m_Z) = 0.118$
and $m_Z = 91.1876~\textrm{GeV}$.

\section{Formulae for renormalon subtractions}
\label{app:renormalon-sub}

The anomalous dimension coefficients for the $R$ evolution of $\Delta(R,\mu_S)$ in Eq.~\eqref{eq:RG-for-del} up to three loops are given by
\cite{Hoang:2008fs, Abbate:2010xh}%\footnote{Note that in Eq.~(60) of \cite{Hoang:2008fs}, the factor $e^{\gamma_\textrm{E}}$ is missing. This omission arises from the absence of $Re^{\gamma_\textrm{E}}$ factor in Eq.~(56) of \cite{Hoang:2008fs}.}
%---------------
\begin{align}
\label{eq:gammaiR}
\begin{split}
%---------------
\gamma_0^R
={}&0,
 \\
\gamma_1^R
={}&
e^{\gamma_\textrm{E}}
\left[
C_A C_F
\left(
-\frac{808}{27}
-\frac{22}{9}\pi^2
+28\zeta_3
\right)
+
C_F T_F
n_f
\left(
\frac{224}{27}+\frac{8}{9}\pi^2
\right)
\right],
\\
\gamma_2^R
={}&
e^{\gamma_\textrm{E}}
\bigg[
C_A^2 C_F
\left(
\frac{35552}{81}
+\frac{662}{27}\pi^2
-\frac{1232}{3}\zeta_3
\right)
+C_A C_F T_F n_f
\left(
-\frac{22784}{81}
-\frac{524}{27}\pi^2
+\frac{448}{3}\zeta_3
\right)
\\
&
\quad
+
C_F(T_Fn_f)^2
\left(
\frac{3584}{81}
+\frac{128}{27}\pi^2
\right)
+4C_F^2 T_F n_f \pi^2
+4s_2
\left(
\frac{11}{3}C_A - \frac{4}{3} T_F n_F
\right)
+
\gamma_{S2}
\bigg],
%---------------
\end{split}
\end{align}
%---------------
where $s_2$ is given in Eq.~\eqref{eq:s2-const}, and $\gamma_{S\,2}$ is the 3-loop non-cusp anomalous dimension for the soft function which can be obtained from $\gamma_S = -\gamma_C^q - \gamma_B^q$ from the RG consistency. Readers refer to Appendix~D of Ref.~\cite{Kang:2013nha} for the 3-loop results for the anomalous dimensions. The $R$ evolution factor from the $R$-evolution equation in Eq.~\eqref{eq:RG-for-del} is given by $D[\alpha_s(R),\alpha_s(R_\Delta)]$ and it is 0 for NLL because $\gamma_0^R=0$. The nonvanishing result of $D[\alpha_s(R),\alpha_s(R_\Delta)]$ up to $\textrm{N}^3\textrm{LL}$ is given by \cite{Bell:2018gce, Bell:2023dqs}
\begin{align}\label{eq:R-evol-factor}
\begin{split}
D[\alpha_s(R),\alpha_s(R_\Delta)]
&=
\frac{R_\Delta}{2\beta_0}
e^{-G[\alpha_s(R_\Delta)]}
\left(
\frac{2\pi}{\beta_0}
e^{i\pi}
\right)^\frac{\beta_1}{2\beta_0^2}
\\
&
\quad
\times
\bigg\{
-\frac{\gamma_1^R}{2\beta_0}
G_1(R, R_\Delta)
+\frac{1}{4\beta_0^2}
\left[
\gamma_R^2 - \frac{\gamma_R^1}{\beta_0}
\left(
\beta_1 + \frac{B_2}{2}
\right)
\right]
G_2(R, R_\Delta)
\bigg\},
\end{split}
\end{align}
where we define $G_i(R, R_\Delta)$ as
\begin{equation}
G_i(R, R_\Delta)
\equiv 
\Gamma\left(-\frac{\beta_1}{2\beta_0^2}-i,
-\frac{2\pi}{\beta_0\alpha_s(R)}\right)
-
\Gamma\left(-\frac{\beta_1}{2\beta_0^2}-i,
-\frac{2\pi}{\beta_0\alpha_s(R_\Delta)}\right),
\end{equation}
and $G[\alpha]$ is the anti-derivative of $1/\beta[\alpha]$
%---------------
\begin{equation}
\label{eq:G_alpha}
%---------------
G[\alpha_s]
=
\frac{2\pi}{\beta_0}
\left[
\frac{1}{\alpha_s}
+\frac{\beta_1}{4\pi\beta_0}
\log \alpha_s
-
\frac{B_2}{(4\pi)^2}
\alpha_s+
\frac{B_3}{(4\pi)^3}
\frac{\alpha^2_s}{2}
\right],
%---------------
\end{equation}
%---------------
with
\begin{equation}
B_2 =- \frac{\beta_2}{\beta_0} + \frac{\beta_1^2}{\beta_0^2},
\quad
B_3 = -\frac{\beta_3}{\beta_0} + \frac{2\beta_1\beta_2}{\beta_0^2}
-\frac{\beta_1^3}{\beta_0^3}.
\end{equation}
Note that $D[\alpha_s(R),\alpha_s(R_\Delta)]$ is real because the complex phase $e^{i\pi(\beta_1/(2\beta_0^2))}$ cancels the imaginary part coming from the upper incomplete gamma functions defined as
%---------------
\begin{eqnarray}
%---------------
\Gamma(c,t)
=
\int_t^\infty dx\, x^{c-1}e^{-x}.
%---------------
\end{eqnarray}
%---------------
As shown in Table~\ref{tab:order}, we maintain the first term in the curly braces of Eq.~\eqref{eq:R-evol-factor} and up to $\mathcal{O}(\alpha_s)$ term in Eq.~\eqref{eq:G_alpha} at NNLL accuracy.
At N$^3$LL accuracy, we retain the first and second terms in the curly braces in Eq.~\eqref{eq:R-evol-factor} and up to $\mathcal{O}(\alpha_s^2)$ term in Eq.~\eqref{eq:G_alpha}.
This choices ensure precise subtraction of renormalon ambiguities at the appropriate order of perturbation theory. 

\section{Shape functions with different higher moments}
\label{app:with_c2}
\begin{figure}
    \centering
    \includegraphics[width=0.6\linewidth]{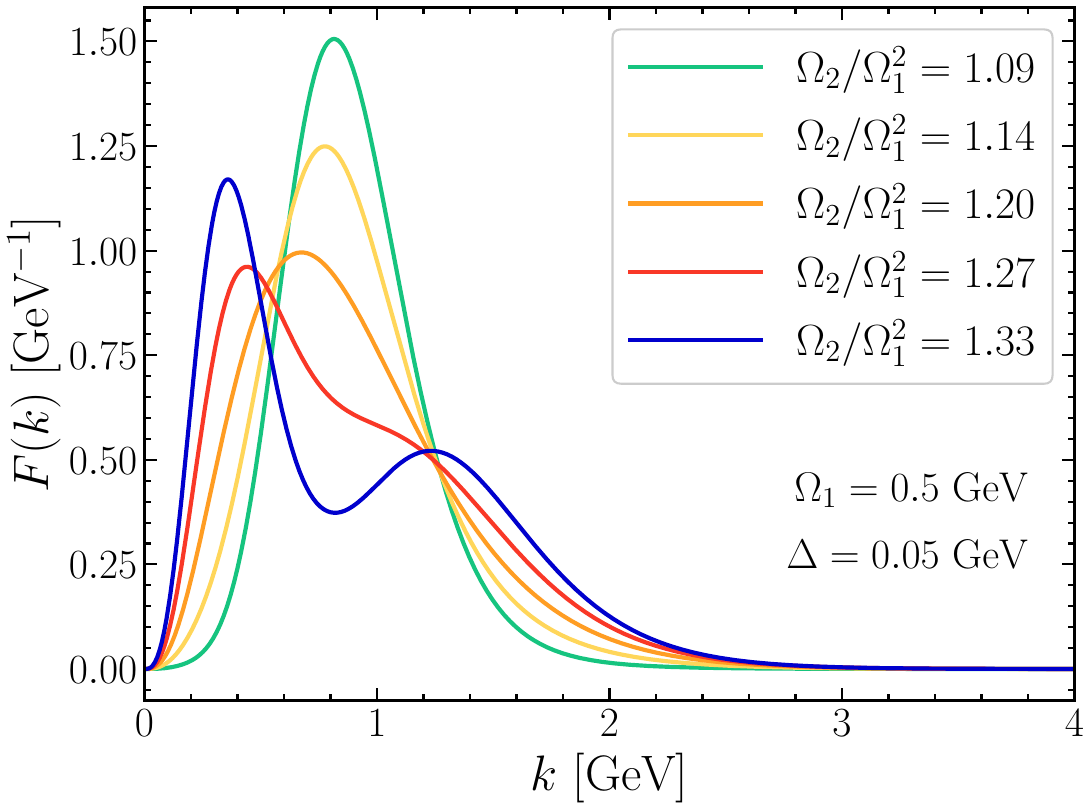}
    \vspace{-1em}
    \caption{Shape functions for different values of the second moment $\Omega_2$, with the first moment $\Omega_1$ and the gap parameter $\Delta$ held fixed.}
    \label{fig:shape_function_c2}
\end{figure}
A nonzero $c_2$ introduces the first generalization of the shape function presented in the main text (Fig.~\ref{fig:fk}):\footnote{Variations in the parameter $c_1$ are strongly correlated with changes in $\lambda$ and are therefore omitted for simplicity. We set $c_1=0$, as discussed in \cite{Abbate:2010xh}.}
%---------------
\begin{equation}
\label{eq:shape_with_c2}
%---------------
F(k) = \frac{1}{\lambda}
\left[
c_0 f_0\left(\frac{k}{\lambda}\right)
+c_2 f_2\left(\frac{k}{\lambda}\right)
\right]^2.
%---------------
\end{equation}
%---------------
The generalized shape function satisfies the same normalization conditions as the original shape function, and we assume the first moment $\Omega_1$ remains unchanged [Eq.~\eqref{eq:renormalon-free-omega}], as $\Omega_1$ primarily determines the shift in the $\tau_1^b$ distribution in the tail region:
%---------------
\begin{align}
\label{eq:shape-function-requirements}
\begin{split}
%---------------
\int dk\, F(k-2\Delta)&=1,
\\
\int dk\, kF(k-2\Delta)&=2\Omega_1.
%---------------
\end{split}
\end{align}
%---------------
The first condition imposes $c_0^2 + c_2^2=1$. Without loss of generality, we take $c_0>0$, allowing $c_0$ to be expressed in terms of $c_2$ as $c_0=\sqrt{1-c_2^2}$. This explicit dependence on $c_2$ is retained in Eq.~\eqref{eq:shape_with_c2}. The second condition determines $\lambda$ as a function of $c_2$, ensuring that $\Omega_1$ remains constant for any $c_2$:
\begin{equation}
\lambda(c_2) = \frac{2(\Omega_1-\Delta)}
{c_0^2+c_0c_2\cdot 0.201354
+c_2^2\cdot 1.10031}.
\end{equation}
This ensures consistent shifting of the $\tau_1^b$ distributions in the tail region, regardless of the contributions from higher moments. 

Effectively, $c_2$ serves as a proxy for the second moment $\Omega_2$, defined as
\begin{align}\label{eq:2nd-moment}
\begin{split}
\Omega_2(c_2)
&=
\Delta^2
+
\Delta\lambda(c_2)
\left(
c_0^2+c_0c_2\cdot 0.201354
+c_2^2\cdot 1.10031
\right)
\nonumber \\
{}&
\quad
+
\frac{[\lambda(c_2)]^2}{4}
\left(
1.25
c_0^2+c_0c_2\cdot 1.03621
+c_2^2\cdot 1.78859
\right).
\end{split}
\end{align}
As an example, we calculate $\Omega_2$ for $c_2=0$ and variations of $c_2$ by $\pm 0.15$ and $\pm 0.30$:
\begin{equation}
\Omega_2(c_2) = 
\begin{cases}
0.271646~\textrm{GeV}^2,
&\textrm{for $c_2=-0.30$,}
\\
0.284885~\textrm{GeV}^2,
&\textrm{for $c_2=-0.15$,}
\\
0.300625~\textrm{GeV}^2,
&\textrm{for $c_2=0$,}
\\
0.316631~\textrm{GeV}^2,
&\textrm{for $c_2=0.15$,}
\\
0.331386~\textrm{GeV}^2,
&\textrm{for $c_2=0.30$.}
\end{cases}
\end{equation}

These results demonstrate approximately 10\% variations of the $\Omega_2$ for $c_2=\pm 0.3$. Importantly, the parameters $\Omega_1 = 0.5~\textrm{GeV}$ and $\Delta = 0.05~\textrm{GeV}$ remain fixed across these variations.
\begin{figure}
    \centering
    \begin{subfigure}
        \centering
        \includegraphics[width=0.49\linewidth]{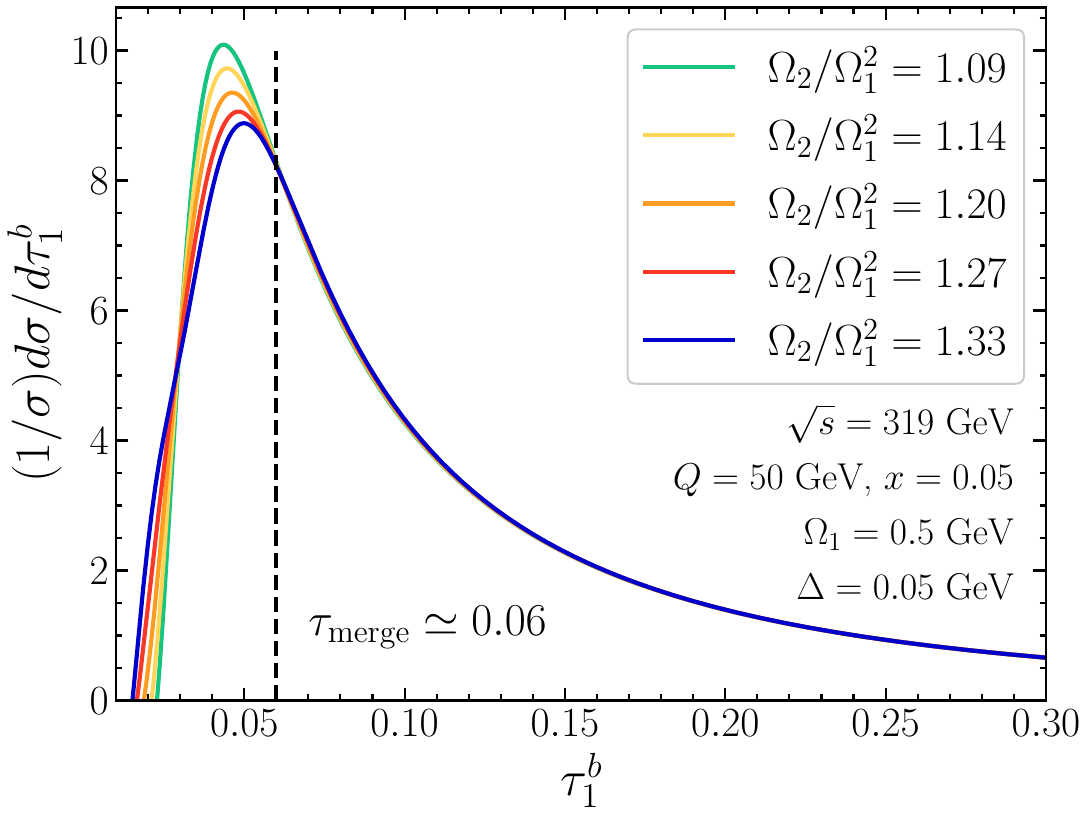}
    \end{subfigure}
    \begin{subfigure}
        \centering
        \includegraphics[width=0.49\linewidth]{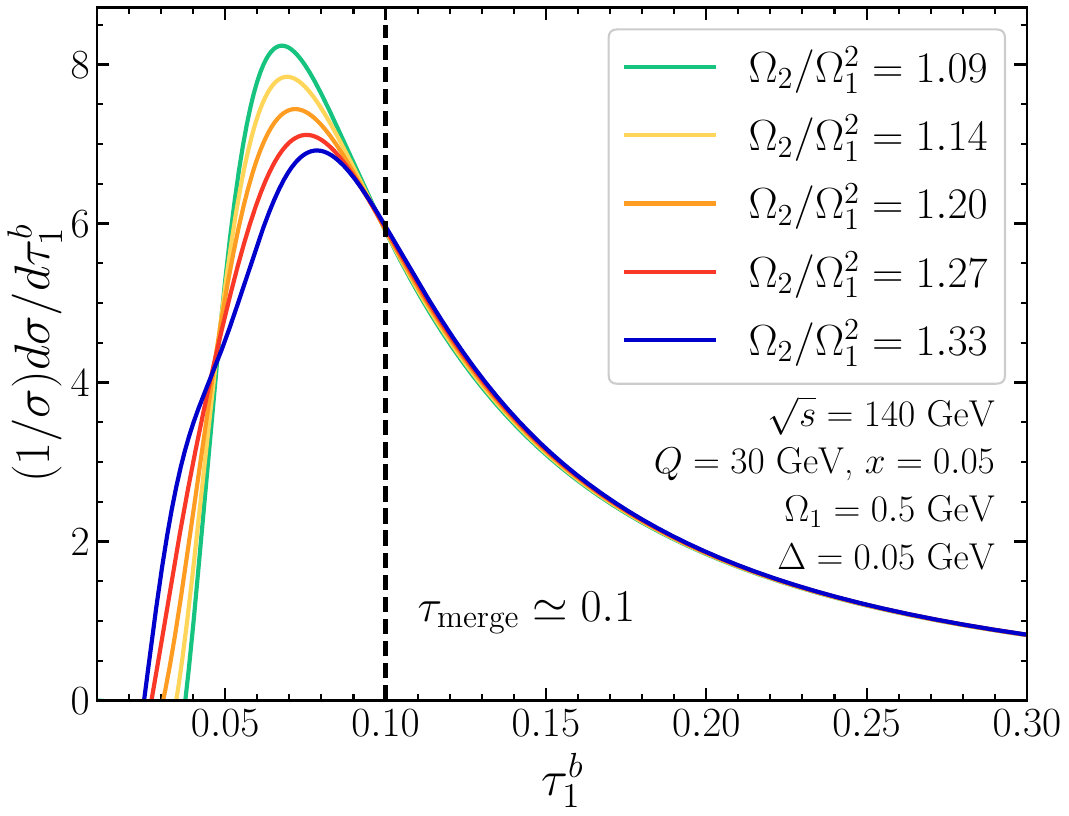}
    \end{subfigure}
    \vspace{-2em}
    \caption{The $\tau_1^b$ distributions for various $\Omega_2$ at $Q=50~\textrm{GeV}$ (left) and $Q=30~\textrm{GeV}$ (right). The black dashed line marks $\tau_1^b\simeq\tau_\textrm{merge}$, beyond which higher-order power corrections contribute at most $\mathcal{O}(1\%)$. For $\tau_1^b\ge \tau_\textrm{merge}$, the impact of $\Omega_2$ becomes negligible, and the leading contribution from $\Omega_1$ dominates. }
    \label{fig:tau1b_with_c2}
\end{figure}
To better illustrate the impact of $c_2$, we also compute the dimensionless ratios $\Omega_2/\Omega_1^2$:
\begin{equation}
\frac{\Omega_2(c_2)}{\Omega_1^2} = 
\begin{cases}
1.08659,
&\textrm{for $c_2=-0.30$,}
\\
1.13954,
&\textrm{for $c_2=-0.15$,}
\\
1.2025,
&\textrm{for $c_2=0$,}
\\
1.26653,
&\textrm{for $c_2=0.15$,}
\\
1.32555,
&\textrm{for $c_2=0.30$.}
\end{cases}
\end{equation}
Fig.~\ref{fig:shape_function_c2} shows how the shape function changes with respect to $\Omega_2$, while keeping $\Omega_1$ fixed.

According to the OPE of the cross section in Eq.~\eqref{eq:OPE_shape}, the relative contributions of the higher-order moments of the shape function scale as ${\alpha_s \Lambda_\textrm{QCD}}/(Q\tau_1^b)$ and ${\Lambda_\textrm{QCD}^2}/(Q^2{\tau_1^b}^2)$.
To ensure that these higher-order power corrections remain below 1\% compared to the leading nonperturbative contributions due to $\Omega_1$, we derive a lower bound on $\tau_1^b$ for safely neglecting these terms:
\begin{equation}
\tau_1^b\gtrsim 100\alpha_s \frac{\Lambda_\textrm{QCD}}{Q}
\quad
\textrm{and}
\quad
\tau_1^b\gtrsim 10 \frac{\Lambda_\textrm{QCD}}{Q}.
\end{equation}
Taking typical values $\alpha_s\sim 0.1$ and $\Lambda_\textrm{QCD}\sim 0.3~\textrm{GeV}$, we estimate the transition point $\tau_\textrm{merge}$ beyond which higher-order power corrections contribute at most $\mathcal{O}(1\%)$:
\begin{equation}\label{eq:tau_merge_def}
\tau_\textrm{merge} \simeq \frac{3~\textrm{GeV}}{Q}.
\end{equation}
Fig.~\ref{fig:tau1b_with_c2} illustrates the variation of the $\tau_1^b$ distribution for different $\Omega_2$ values at $Q=50~\textrm{GeV}$ (left) and $Q=30~\textrm{GeV}$ (right).
While the shape function exhibits a significant dependence on $\Omega_2$, this effect is primarily confined to the peak region ($\tau_1^b<\tau_\textrm{merge}$). Beyond $\tau_\textrm{merge}$, the distributions converge, where the OPE becomes valid and the leading contribution from $\Omega_1$ dominates. 

As expected from the definition of $\tau_\textrm{merge}$ in Eq.~\eqref{eq:tau_merge_def}, we observe that for larger $Q$, the merging point shifts to smaller values of $\tau_1^b$, as shown in Fig.~\ref{fig:tau1b_with_c2}.
Notably, $\tau_\textrm{merge}$ is significantly smaller than the profile function parameter $t_1$ ($t_1=0.1$ for the left panel and $t_1=0.167$ for the right panel), which justifies the study of $\alpha_s$ and $\Omega_1$ in the tail region, typically for $\tau_1^b\gtrsim t_1$.

\section{Profile functions}
\label{app:profile_total}
\begin{figure}
    \centering
    \includegraphics[width=1\linewidth]{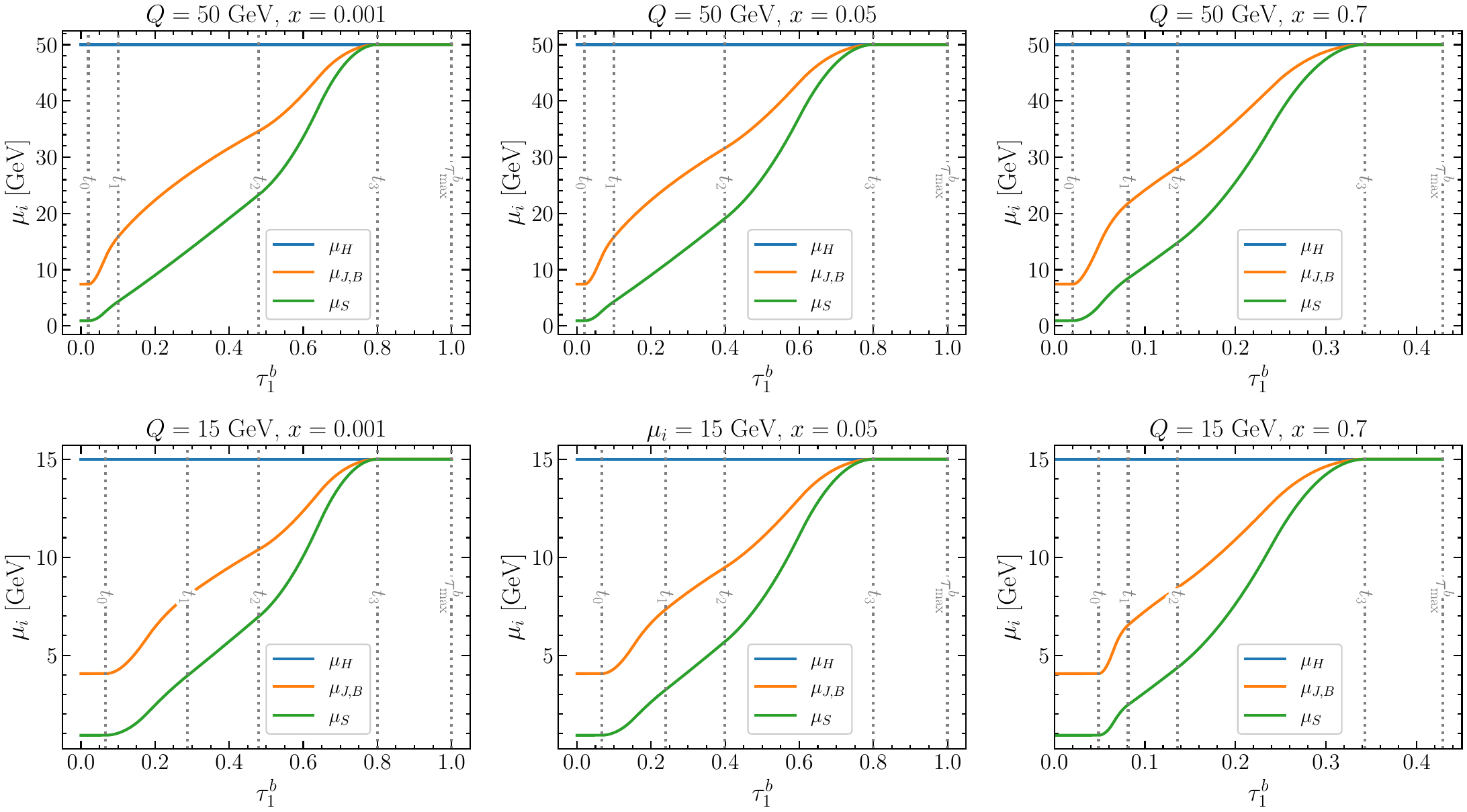}
    \vspace{-1em}
    \caption{The profile functions at various values of $x$ and $Q$. The $t_i$ denote transition points where the functional form of $\mu_B$, $\mu_J$, and $\mu_S$ change.}
    \label{fig:profile_cent}
\end{figure}

\begin{figure}
    \centering
    \includegraphics[width=1\linewidth]{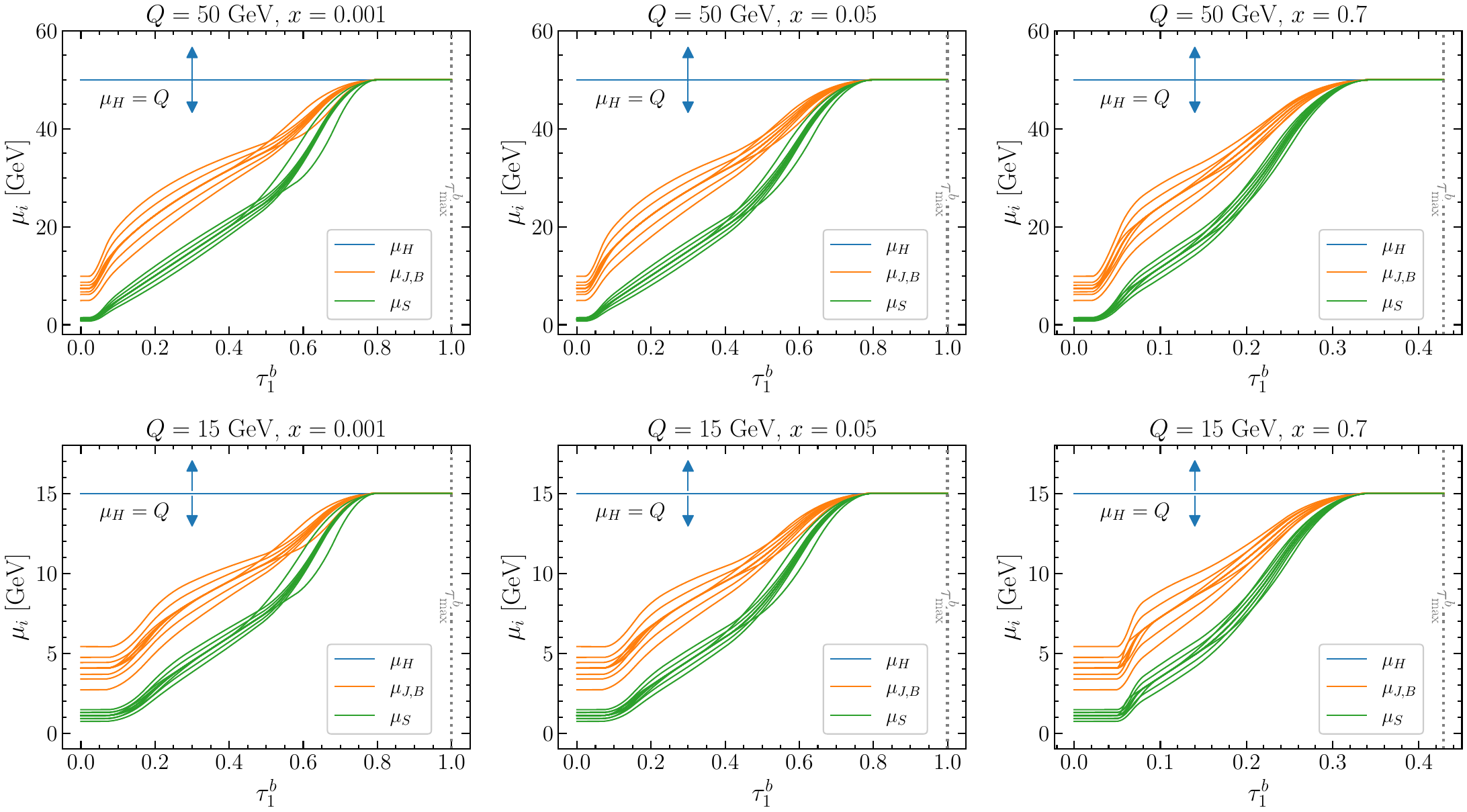}
    \vspace{-1em}
    \caption{The scale variations (16 variations) which modify the central profile functions in Fig.~\ref{fig:profile_cent}. The double arrow illustrates the up/down variation by a factor of two in Eq.~\eqref{eq:singular_scale_variations}.}
    \label{fig:profile_scale_var}
\end{figure}

In this appendix, we provide examples of the profile functions for various values of $x$ and $Q$. As outlined in Eq.~\eqref{eq:profile_setting}, the parameters of the profile function vary depending on $x$ and $Q$. 
Specifically, $t_0$ and $t_1$ are set based on $Q$, while $t_2$ and $t_3$ are set based on $x$. In particular, we select $t_2$ to reflect the relative dominance of singular and nonsingular contributions. For $\tau_1^b<t_2$, the singular contribution is dominant, whereas for $\tau_1^b>t_2$ the nonsingular contribution is dominant.
Fig.~\ref{fig:profile_cent} illustrate how the profile function changes with $x$ in the horizontal comparison (e.g., $x = 0.001, 0.05, 0.7$) and with $Q$ in the vertical comparison (e.g., $Q = 50~\textrm{GeV}$, $15~\textrm{GeV}$). Fig.~\ref{fig:profile_scale_var} displays the scale variations applied to the profile functions shown in Fig.~\ref{fig:profile_cent}.

\section{$r_I(\epsilon)$ for the other $\mu$ variations}
In this appendix, we present the $r_c$ analysis in Sec.~\ref{sec:nonsing} for different choices of the scales $\mu$ other than $\mu=Q$. We can compute the $r_c$ for the fixed scales $\mu=Q$, $Q/2$, and $2Q$, as the coefficients $A$ and $B$ in Eq.~\eqref{eq:full_QCD_fixed} remain independent of $\tau_1^b$. 
Using \texttt{QCDNUM}, we compute the inclusive QCD cross sections at each order in $\alpha_s$, and from the fixed-order SCET, we compute the cumulant singular contributions. By subtracting these two contributions, we obtain the values for $r_c$. The numerical results at $\mathcal{O}(\alpha_s^2)$ for $\sqrt{s}=319$ GeV, $Q=50$ GeV, and $x=0.05$ are summarized in Table~\ref{tab:rc_fixed_scale}. 
\begin{table}
\centering
\begin{tabular}{c|c|c|c}
\hline
\hline
\rule{0pt}{2.8ex}
Scale choices & $\sigma_\textrm{PT}^\textrm{QCD}\big|_{\mathcal{O}(\alpha_s^2)}$ & $\sigma_\textrm{PT}^\textrm{s,fixed}\big|_{\mathcal{O}(\alpha_s^2)}$ & $r_c\big|_{\mathcal{O}(\alpha_s^2)}$ 
\\[0.6ex]
\hline
\hline
\rule{0pt}{3.4ex}
$\mu=Q$ & $-0.081\sigma_0^b$ & $-0.193\sigma_0^b$ & $\phantom{+}0.112\sigma_0^b$
\\[0.6ex]
\rule{0pt}{2.8ex}
$\mu=Q/2$ & $-0.040\sigma_0^b$ & $\phantom{+}0.009\sigma_0^b$ & $-0.049\sigma_0^b$
\\[0.6ex]
$\mu=2Q$ & $-0.150\sigma_0^b$
 & $-0.369\sigma_0^b$
 & $\phantom{+}0.219\sigma_0^b$
\\[0.6ex]
\hline
\end{tabular}
\caption{Values of the inclusive QCD cross section, the fixed-order cumulant singular cross section, and $r_c$ at $\mathcal{O}(\alpha_s^2)$ for fixed scales $\mu=Q$, $Q/2$, and $2Q$. }
\label{tab:rc_fixed_scale}
\end{table}
\label{app:rc_other_mu}
\begin{figure}
    \centering
    \includegraphics[width=1.0\linewidth]{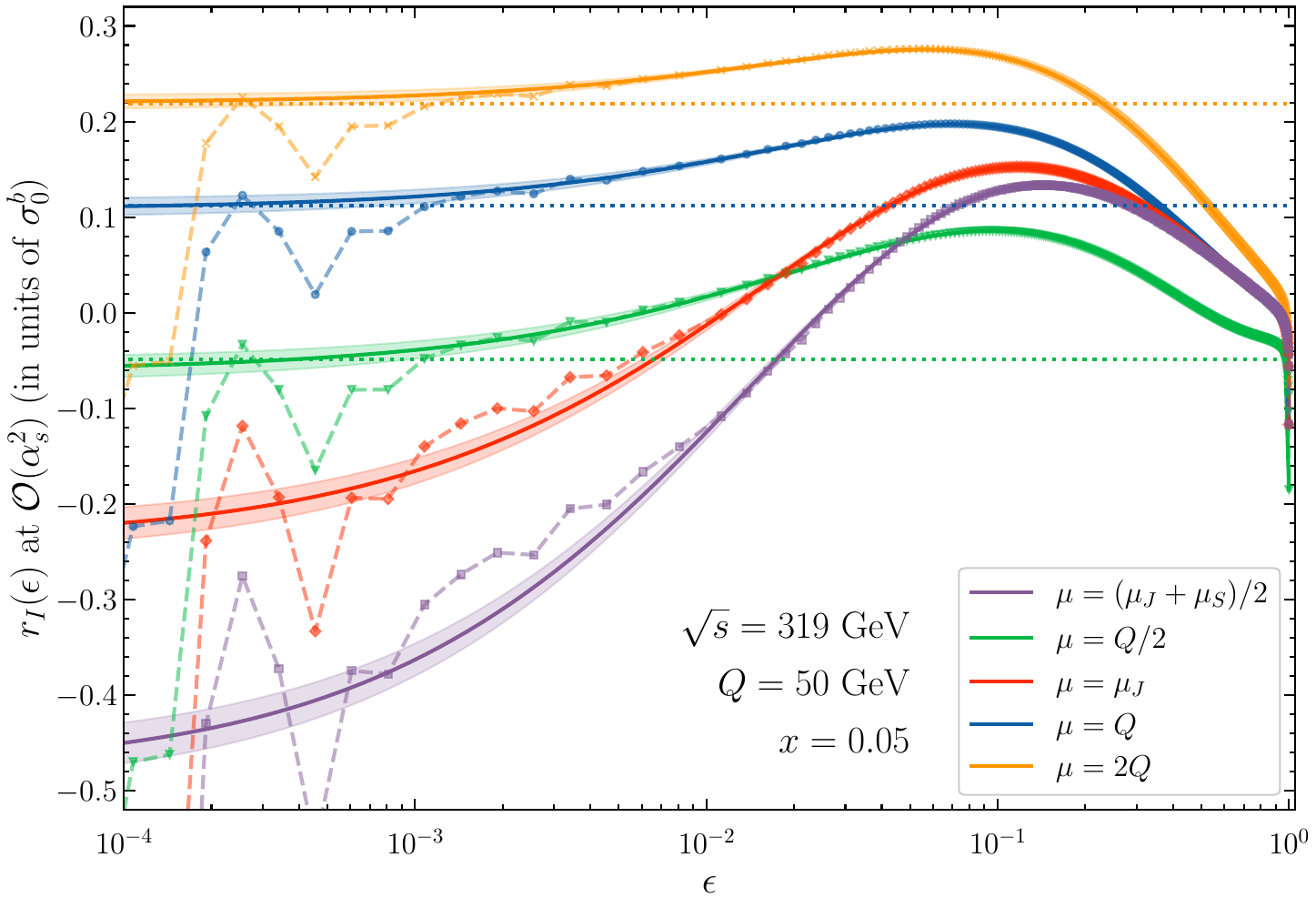}
    \vspace{-2em}
    \caption{
   Comparison of the $\mathcal{O}(\alpha_s^2)$ integral of the remainder function $r_I(\epsilon)$ with results from the fit/interpolation nonsingular functions. The dashed lines show numerical results from \texttt{NLOJet++}, the thin dotted horizontal lines are the $r_c$ values given in Table~\ref{tab:rc_fixed_scale}, and the solid lines are the predictions from the fit/interpolation functions in Eq.~\eqref{eq:ns_interp_fit}. The shaded bands are the $1\sigma$ confidence intervals of the fit functions.}
    \label{fig:rc1_NLO_fit}
\end{figure}

In Fig.~\ref{fig:rc1_NLO_fit}, we show $r_c$ analysis at $\mathcal{O}(\alpha_s^2)$ with the fit/interpolation nonsingular functions. This analysis checks the validity of the nonsingular cross sections derived from the fit/interpolation formula in Eq.~\eqref{eq:ns_interp_fit}, using the same model function in Eq.~\eqref{eq:fit_formula}, and assuming $\tau_\textrm{upper}=0.65\tau_\textrm{max}^b$. 
The figure shows $r_I(\epsilon)$ for the five scale variations of the nonsingular cross sections. For the fixed scales $\mu=Q$, $Q/2$, and $2Q$, we observe that $\lim_{\epsilon\to 0}r_I(\epsilon)$ accurately reproduces the analytic $r_c$ values in Table~\ref{tab:rc_fixed_scale} up to 1$\sigma$ confidence intervals shown as the bands.

%\acknowledgments

%This is the most common positions for acknowledgments. A macro is available to maintain the same layout and spelling of the heading.

%\paragraph{Note added.} This is also a good position for notes added after the paper has been written.

% The bibliography will probably be heavily edited during typesetting.
% We'll parse it and, using the arxiv number or the journal data, will
% query inspire, trying to verify the data (this will probalby spot
% eventual typos) and retrive the document DOI and eventual errata.
% We however suggest to always provide author, title and journal data:
% in short all the informations that clearly identify a document.

%\bibliography{DIS-N3LL.bib}

\end{document}